\documentclass[twocolumn,showpacs,preprintnumbers,amsmath,amssymb]{revtex4-1}
\usepackage{graphicx,color}
\usepackage{dcolumn}
\usepackage{bm}
\usepackage{longtable}
\usepackage{amsmath}
\usepackage{mathrsfs}
\usepackage{amsfonts}
\usepackage{textcomp}
\usepackage{esvect}
\usepackage{float}

\def\gsim {\mbox{\hbox{ \lower-.6ex\hbox{$>$}
\kern-1.12em \lower.5ex\hbox{$\sim$}\kern+.35em}}}
\def\lsim {\mbox{\hbox{ \lower-.6ex\hbox{$<$}
\kern-1.12em \lower.5ex\hbox{$\sim$}\kern+.35em}}}

\DeclareMathOperator{\arctanh}{arctanh}

\begin{document}


\title{Error analysis of linear optics measurements via 
       turn-by-turn beam position data in circular accelerators 
       \vspace{-0.0 cm}}

\author{A. Franchi\footnote[*]{\email{andrea.franchi@esrf.fr}}}
\affiliation{ESRF, Grenoble, France}

\date{\today}

\begin{abstract}\vspace{0.0 cm}
Many advanced techniques have been developed, tested and implemented 
in the last decades in almost all circular accelerators across the 
world to measure the linear optics. However, the 
greater availability and accuracy of beam diagnostics and the ever 
better correction of linear magnetic lattice imperfections (beta 
beating at $1\%$ level and coupling at 1\textperthousand) 
are reaching what seems to be the intrinsic accuracy and precision 
of different measurement techniques. This paper aims to highlight 
and quantify, when possible, the limitations of one standard method, 
the harmonic analysis of turn-by-turn beam position data. To this 
end, new analytic formulas for the evaluation of lattice parameters 
modified by focusing errors are derived. 
The unexpected conclusion of this study is that for the ESRF storage 
ring (and possibly for any third generation light source operating 
at ultra-low coupling and with similar diagnostics), measurement and 
correction of linear optics via orbit beam position data are to be 
preferred to the analysis of turn-by-turn data. 
\end{abstract}



\maketitle 

\section{Introduction and motivation}
Measurement and correction of focusing errors in circular 
accelerators is one of the top priorities in colliders and 
storage ring-based light sources to provide users with beam 
sizes and divergences as close as possible to the design 
values and to limit the possible detrimental effects on the 
beam lifetime caused by the integer and half-integer resonances. 
To this end, so many different techniques have been developed 
and successfully tested since decades that they already occupy 
entire chapters in textbooks~\cite{Zimmermann-book}. 

A brief and non-exhaustive historical overview may help 
understand the great advancements in this domain. Back to 
the early '80s, one of the first measurements of betatron 
phase advance and beta functions was documented in 
Ref.~\cite{ISR}. In the early '90s systematic measurements and 
correction were already carried out at the CERN Large Electron 
Positron (LEP) collider from the harmonic analysis of 
turn-by-turn (TBT) beam position monitor (BPM) data, reaching 
a beta beating of $10\%$~\cite{castro1,castro2}. In the first decades 
of the new century other colliders worldwide reported 
similar or even lower modulation~\cite{Colliders1,Colliders2,Colliders3}. 
With the advent of 3rd generation light sources, a 
major breakthrough was provided by the exploitation of orbit 
BPM data for the reconstruction of machine errors~\cite{LOCO1,LOCO2}, 
resulting in modulation of the beta functions from $10\%$ 
down to $1\%$ or even less in more recent facilities.

The ever increasing BPM resolution and computing power made 
the analysis and correction of linear optics (focusing error 
and betatron coupling) via measurements of the orbit response 
matrix (ORM) a routine task in basically all light sources 
worldwide. The inclusion of TBT features in the BPM electronics 
and the refinement of the data analysis via several 
model-independent techniques~\cite{SVD1,SVD2} provided these 
facilities with a viable alternative tool. However, systematic 
comparisons between these two techniques on the same machine  
appeared only recently (for the ALBA and SOLEIL storage 
rings)~\cite{Soleil,Alba}: 
While both approaches evaluated the measured beta beating 
in the $1\%-2\%$ range with respect to the ideal model, they 
differed of about the same quantity when compared against 
each other. Systematic numerical simulations on the same 
lattice showed indeed that the expected resolution of 
both approaches is of about 1\%~\cite{Alba,Alba-simulations}.

These observations fostered an intense debate (mostly oral, 
during workshops, conferences or informal meetings) on a 
series of questions: {\sl i)} Which is the best approach 
to measure the linear optics at $1\%$ level of beta beating? 
{\sl ii)} Why do the two methods predict different modulations 
of the optics functions and fail to converge toward the same model? 
{\sl iii)} Which is the ultimate resolution at which lattice 
errors can be measured and corrected? The community split in two 
main schools of thoughts. The ORM-oriented group argues that 
the higher BPM resolution of the orbit mode with respect to the 
TBT setup provides the most reliable observable (the ORM) and 
that the inferred model best reproducing it can be trusted 
for the evaluation (and correction) of the lattice parameters. 
The TBT-oriented school replies that TBT data can be trusted 
more in the evaluation of beta beating and betatron phase errors 
because these quantities are more direct observables, whereas their 
measurements result from a series of model-dependent fits when 
orbit data are used. 

This paper aims to help answer the above questions. 
Even though the scrutiny of TBT data is limited here to the  
harmonic analysis, the general final conclusions are expected 
to apply to other approaches, such as the BPM matrix, the 1-turn 
or N-turn matrix. Each aspect is quantified here by using 
the ESRF electron storage ring as example: Even though numbers 
may vary in other facilities, the overall considerations should 
apply to other machines with similar level of ultra-low coupling, 
diagnostics and beam stability. On the other hand, hadron 
machines are not expected to be subjected to the same conclusions, 
because their nonlinearities are weaker than in modern light 
sources, i.e. they have a larger area in the $x$-$y$ plane 
within the linear regime of the betatron motion compared to 
storage rings like the ESRF's. Moreover, analytic error estimates 
are derived here for the TBT analysis only. No equivalent results are 
obtained for the ORM approach, for which the only error study 
presently available is based on the same numerical parametric 
scans of Ref.~\cite{Alba-simulations}.
The scope of this comparison is also limited 
to the analysis of lattice errors without entering into the field 
of different correction schemes. The analysis of practical considerations 
which may prevent some facilities from using either technique 
is also out of the scope of this paper, which is organized as follows. 
After briefly reviewing the main physical and mathematical 
ingredients behind the two approaches in Sec.~\ref{today}, 
a more detailed discussion on the validity of approximations 
and assumptions proper of each method under typical 
measurement conditions is presented in Sec.~\ref{limits}. 
Cosequences for the present and future ESRF storage rings 
are eventually outlined in Sec.~\ref{sec:conclusion}.
All mathematical derivations have been put in separated appendices.

\section{Current approaches to the analysis of linear optics errors}
\label{today}
As mentioned in the Introduction, two main strategies are implemented 
in circular accelerators for the analysis of linear lattice errors. The 
first, which is routinely used in probably all synchrotron-based light 
sources, is based on the examination of the orbit response to a steering 
angle. The second focuses on the analysis of free betatron 
oscillations induced by a pulsed excitation and is the preferred one 
in hadron circular accelerators.

In the first approach, after introducing an orbit 
distortion via horizontal and vertical deflections, represented by two 
vectors $\vec{\Theta}_x=(\Theta_{x,1},\ \Theta_{x,2},\ ..., \Theta_{x,N_S})$ and 
$\vec{\Theta}_y=(\Theta_{y,1},\ \Theta_{y,2},\ ..., \Theta_{y,N_S})$, where 
$N_s$ is the number of available magnets, the horizontal and vertical 
orbits are recorded at $N_B$ BPMs $\vec{O}_x=(O_{x,1},\ O_{x,2},\ ...,
 O_{x,N_B})$ and $\vec{O}_y=(O_{y,1},\ O_{y,2},\ ..., O_{y,N_B})$. 
They can be written as
\begin{eqnarray}\hskip-0.0cm&&
\left(\begin{array}{c}\vec{O}_{x}\\ \vec{O}_{y}\end{array}\right)=\mathbf{ORM}
\left(\begin{array}{c}\vec{\Theta}_{x}\\ \vec{\Theta}_{y}\end{array}\right),\quad 
\mathbf{ORM}=
\left(\begin{array}{c c}\mathbf{O^{(xx)}} &\mathbf{O^{(xy)}} \\
                        \mathbf{O^{(yx)}} &\mathbf{O^{(yy)}} \end{array}
\right),\nonumber \\ \label{eq:ORM_01}
&&O^{(xx)}_{wj}=\frac{\partial O_{x,j}}{\partial \Theta_{x,w}}\ ,\quad
O^{(xy)}_{wj}=\frac{\partial O_{x,j}}{\partial \Theta_{y,w}}\ ,\quad 
1< j < N_B\ ,\\ 
&&O^{(yx)}_{wj}=\frac{\partial O_{y,j}}{\partial \Theta_{x,w}}\ ,\quad 
O^{(yy)}_{wj}=\frac{\partial O_{y,j}}{\partial \Theta_{y,w}}\ ,\quad 
1< w < N_S\nonumber
\end{eqnarray}
Optics codes such as MADX~\cite{madx} or AT~\cite{AT} can easily compute 
$\mathbf{ORM}$ for the ideal (or initial) lattice model and the 
difference between the measured and expected matrix may be written as 
\begin{eqnarray}
\mathbf{\delta ORM}&=&\mathbf{ORM^{(meas)}-ORM^{(ideal)}}\ .\label{eq:ORM_02}
\end{eqnarray}
The horizontal and vertical dispersion are also measured and their  
difference with respect to the ideal model is computed
\begin{eqnarray}
\vec{\delta D}_{x,y}=\vec{D}_{x,y}^{(meas)}-\vec{D}_{x,y}^{(ideal)}\ .
\end{eqnarray}
Both $\mathbf{\delta  ORM}$ and $\vec{\delta D}_{x,y}$ depend linearly 
on the linear lattice errors (i.e. from bending and quadrupole magnets). 
By sorting the elements of each ORM block sequentially in a vector, the 
dependence reads
\begin{eqnarray}
\left(\begin{array}{c}\delta\vec{O}^{(xx)}\\ 
      \delta\vec{O}^{(yy)}\\\delta\vec{D}_x\end{array}
\right)&=&\mathbf{M_{norm}}
\left(\begin{array}{c}\delta\vec{K}_{1}\\ \delta\vec{K}_{0}
\end{array}\right)\ , \label{eq:ORM_04}\\\label{eq:ORM_05}
\left(\begin{array}{c}\delta\vec{O}^{(xy)}\\ 
      \delta\vec{O}^{(yx)}\\ \delta\vec{D}_y\end{array}
\right)&=&\mathbf{M_{skew}}
\left(\begin{array}{c}\vec{\theta}^{(quad)}\\ \vec{\theta}^{(bend)}
\end{array}\right)\ .
\end{eqnarray}
$\delta\vec{K}_{1}$ and $\delta\vec{K}_{0}$ are the vectors containing 
the quadrupole and dipole errors, respectively, whereas 
$\vec{\theta}$ refers to the magnet tilts. The latter may be replaced 
in Eq.~\eqref{eq:ORM_05} by the corresponding skew multipolar components
\begin{eqnarray}
J_{1}=-K_{1}\sin{(2\theta^{(quad)})}\ ,\quad
J_{0}=-K_{0}\sin{(\theta^{(bend)})}\ .\quad
\end{eqnarray}
Throughout the paper, the MADX nomenclature for the multipolar expansion 
of magnetic fields is adopted,
\begin{equation}\label{eq:MADX}
-\Re\left[\sum_{n}{(K_{w,n-1}+iJ_{w,n-1})
            \frac{(x_w+iy_w)^n}{n!}}\right]\ ,
\end{equation}
with $K$ and $J$ referring to the integrated normal and skew 
magnetic strengths. The response matrices $\mathbf{M_{norm}}$ and 
$\mathbf{M_{skew}}$ can be computed by the optics codes. 
By pseudo-inverting the above system, for instance via singular value 
decomposition (SVD), effective models that best fit the measured ORM 
can be built. A unique model may not be extracted, since a trade-off 
between accuracy (i.e. large number of eigen-values in the 
decomposition) and reasonableness of the errors (i.e. low number of 
eigen-values to prevent numerical instabilities) shall be fixed on 
a subjective base. Moreover, the systems of Eqs.~\eqref{eq:ORM_04}-\eqref{eq:ORM_05} 
ignores contributions from the feed-down effects of quadrupoles and 
sextupoles induced by their misalignments and/or off-axis orbit at 
their locations. The closed orbit distortion resulting from this modelling 
renders the analysis more complex without adding values to the physical 
observables (betatron phase $\phi$ and amplitude $\beta$ 
at the BPMs) and are 
usually {\sl absorbed} by additional dipole errors (accounting for 
quadrupole misalignments) and quadrupole errors (representing the 
quadrupolar feed-down in sextupoles). In optics codes dipole errors 
induce a distortion of the reference orbit, though not of the closed 
one. Eqs.~\eqref{eq:ORM_04}-\eqref{eq:ORM_05} are the core  of the 
{\em Linear Optics from Closed Orbit} (LOCO) analysis~\cite{LOCO1,LOCO2}. 
Additional fit parameters may be included in the r.h.s. 
of the two equations, such as calibration factors and 
rolls of steerers and BPMs. Once the errors 
($\delta\vec{K}_{1},\ \delta\vec{K}_{0}$ and $\vec{\theta}$) are included 
into the lattice model, the optical parameters ($\beta$,  
$\phi$ and $D$) are computed by the optics codes and compared to 
the ones from the ideal model. Beta beating and phase advance errors 
are the most common figures of merit for focusing errors:
\begin{eqnarray}
\begin{aligned}
&\frac{\Delta\beta}{\beta}=\frac{\beta^{(meas)}-\beta^{(mod)}}{\beta^{(mod)}}
\\ &\delta\phi_{ij}=\Delta\phi_{ij}^{(meas)}-\Delta\phi_{ij}^{(mod)}\ , 
\quad \Delta\phi_{ij}=\phi_j-\phi_i
\end{aligned}\ ,\label{eq:ORM_06}
\end{eqnarray}
where both quantities are evaluated at the BPM locations, 
with $i$ and $j$ two different monitors, usually (though 
not necessarily~\cite{prstab_BPMselect}) consecutive. 
A consensus on a figure of merit for the evaluation of 
betatron coupling has not yet been reached. In hadron machines the 
amplitude of the difference resonance stop-band $|C|$ is widely 
used,~\cite{Zimmermann-book}
\begin{equation} \nonumber 
C=\hspace{-1mm}-\frac{1}{2\pi}\hspace{-1mm}\oint{\hspace{-1mm}ds_{_{\ }}
   \hspace{-.6mm}j(s)\sqrt{\beta_x(s)\beta_y(s)}
   e^{-i(\phi_x(s)-\phi_y(s))+i(s/R)\Delta_Q}}\ ,
\end{equation}
where $j(s)$ represents the distribution of the non-integrated 
skew quadrupole fields along the ring, $R$ is the machine 
radius, $s$ is the longitudinal coordinate, $\beta$ and 
$\phi$ are the Twiss parameters of 
the uncoupled lattice, and $\Delta_Q=Q_x-Q_y$ the fractional 
distance from the resonance of the set tunes. $|C|$ evaluated 
on the resonance ($\Delta_Q=0$) corresponds to the minimum 
separation experienced by the measured eigen-tunes 
$|\Delta Q_{min}|$~\cite{prstab_Tobias}. In lepton circular 
accelerators the ratio between the two transverse emittances 
$\epsilon_r=\epsilon_y/\epsilon_x$ is preferred, since any measurable 
value of $\epsilon_y$ is usually generated by betatron coupling 
and vertical dispersion. Both $|C|$ and $\epsilon_r$ are global 
parameters that prevent a detailed localisation and compensation 
of sources of coupling in light sources. To this end, in 
Ref.~\cite{prstab_esr_coupling} the coupling resonance driving 
terms (RDTs) were proposed as figure of merit along with vertical 
dispersion,\vskip -5mm
\begin{eqnarray}
f_{{\tiny\begin{array}{c} 1001 \\ 1010\end{array}},\ j}=
        \frac{\sum\limits_w^W J_{w,1}\sqrt{\beta_{w,x}\beta_{w,y}} 
        e^{i(\Delta\phi_{x,wj}\mp \Delta\phi_{y,wj})}}
        {4(1-e^{2\pi i(Q_u\mp Q_v)})}
+O(J_1^2)\ . \nonumber  \\ \label{eq:f}  
\end{eqnarray}
$J_{w,1},\ w=1,2,3 ...\ , W $ are the skew quadrupole integrated 
strengths present in the ring and originated by quadrupole tilts, 
sextupole misalignments, insertion devices, and corrector skew 
quadrupoles already powered. $\beta_w$ is the beta function at 
the source of coupling $w$, while $\Delta\phi_{wj}$ denotes its 
phase advance with respect to the BPM $j$. $Q_{u,v}$ 
are the measurable eigen-tunes, 
which are, in first approximation, equal to the set tunes:  
$Q_{u,v}=Q_{x,y}+ O(J_{w,1}^2)$. $\Delta\phi_{wj}$ is the phase 
advance between the source of coupling $w$ and the BPM $j$ 
where the RDTs are computed. These two RDTs can be evaluated (and 
minimized) at all BPMs from the model obtained after fitting the 
measured ORM and used to evaluate (and reduce) the vertical emittance 
along the ring, as shown in Ref.~\cite{prstab_esr_coupling}.\\

The second approach is based on the harmonic analysis of turn-by-turn (TBT)
free betatron oscillations induced by a pulsed magnet. Forced oscillations  
generated for example by an AC dipole are not discussed here. However, 
to the first order, and provided that the AC dipole driving 
frequency is sufficiently separated from the tune, the following analysis 
can be applied to both signals, either free or forced. At each BPM, 
the TBT signal can be decomposed in its main harmonics via a Fourier 
transform. The main harmonic is found at a frequency corresponding to 
the tune, whereas secondary harmonics appear at linear combinations 
of both tunes, $n_xQ_x+n_yQ_y$, with $n_{x,y}\in\mathbb{N}$, as reported in 
Fig.7 of Ref.~\cite{Andrea-arxiv}. Spectral lines in the horizontal and 
vertical planes are usually denoted as $H(n_x,n_y)$ and  $V(n_x,n_y)$, 
respectively. The tune lines at the BPM $j$ read
\begin{eqnarray}
\begin{aligned}\hskip -0.1cm
H(1,0)_{j,\beta}\hskip -0.5mm=\hskip -0.5mm\mathcal{C}_{x,j}\sqrt{2I_x\beta_{x,j}^{(meas)}}\hskip-0.5mm
      \cos{(2\pi NQ_x+\hskip -0.1cm\phi_{x,j}^{(meas)}\hskip -0.1cm+\hskip -0.1cm\psi_{x0})} \\
V(0,1)_{j,\beta}\hskip -0.5mm=\hskip -0.5mm\mathcal{C}_{y,j}\sqrt{2I_y\beta_{y,j}^{(meas)}}\hskip-0.5mm
      \cos{(2\pi NQ_y+\hskip -0.1cm\phi_{y,j}^{(meas)}\hskip -0.1cm+\hskip -0.1cm\psi_{y0})} 
\end{aligned}\hskip -0 cm .\nonumber
\end{eqnarray}
$\psi_0$ is an arbitrary initial phase equal for all BPMs (provided 
that the latter are perfectly synchronized in time), $N$ is the turn 
number, $\phi_j^{(meas)}$ is the BPM betatron phase 
and $2I$ is the invariant (i.e. the action) proportional to 
the strength of the pulsed excitation. The BPM calibration 
factor $\mathcal{C}_j$ is added to account for values potentially 
different from 1. By performing the harmonic analysis on the TBT 
signal normalized by the model beta function, the horizontal tune 
line reads
\begin{eqnarray}
\begin{aligned}
&\tilde{x}_j=\frac{x_j}{\sqrt{\beta_{x,j}^{(mod)}}}\longrightarrow
  \frac{H(1,0)_{j,\beta}}{\sqrt{\beta_{x,j}^{(mod)}}}=H(1,0)_j\ , \\
&\hskip -0.1cm H(1,0)_j\hskip -0.5mm=\hskip -0.5mm\mathcal{C}_{x,j}
       \sqrt{2I_x\frac{\beta_{x,j}^{(meas)}}{\beta_{x,j}^{(mod)}}}
              \cos{(2\pi NQ_x\hskip -0.5mm+\hskip -0.5mm\phi_{x,j}^{(meas)}
                    \hskip -0.1cm+\hskip -0.5mm\psi_{x0})}\ .
\end{aligned} \hskip -0.0 cm \label{eq:tunes1} 
\end{eqnarray} 
An equivalent expression applies to the vertical plane. At the ESRF 
the same filtered and interpolated Fast Fourier Transfort (FFT) of 
Ref.~\cite{tune} is used to extract amplitude and phase of this 
harmonic, namely\vskip -5mm
\begin{eqnarray}
\left\{\begin{aligned}
|H(1,0)_j|&=\frac{1}{2}\mathcal{C}_{x,j}\sqrt{2I_x
               \frac{\beta_{x,j}^{(meas)}}{\beta_{x,j}^{(mod)}}}\\
\Phi_{H(1,0)_j}&=\phi_{x,j}^{(meas)}+\psi_{x0}
\end{aligned} \right. \hskip 0.4 cm , \label{eq:tunes2} 
\end{eqnarray}
The invariant may be inferred by averaging the tune line 
amplitude over all BPMs, provided that their number and 
location are sufficient to cancel the contribution of the 
modulation of beta function and calibration factors,
\begin{eqnarray}
\sqrt{2I_x}\simeq2<|H(1,0)|>\ ,\quad \sqrt{2I_y}\simeq2<|V(0,1)|>
\hskip 0. cm .\qquad \label{eq:tunes3} 
\end{eqnarray}
The actual beta functions $\beta^{(meas)}$ can be extracted from 
Eqs.~\eqref{eq:tunes2}-\eqref{eq:tunes3}, according to 
\begin{eqnarray}
\begin{aligned}
\beta_{x,j}^{(meas)}\simeq\beta_{x,j}^{(mod)}
\left(\frac{|H(1,0)_j|}{<|H(1,0)|>}\right)^2\frac{1}{\mathcal{C}_{x,j}^2}\\
\beta_{y,j}^{(meas)}\simeq\beta_{y,j}^{(mod)}
\left(\frac{|V(0,1)_j|}{<|V(0,1)|>}\right)^2\frac{1}{\mathcal{C}_{y,j}^2}
\end{aligned}\hskip 0.5 cm . \quad\label{eq:BetaAmpli1} 
\end{eqnarray}
The uncomfortable dependence on the calibration factors 
in the above formulas motivated the search for an alternative way 
to extract $\beta^{(meas)}$ from the tune line. The BPM phase 
$\Phi$ of Eq.~\eqref{eq:tunes2} turns out to be independent of 
both $\mathcal{C}$, BPM rolls and betatron coupling. 
Because of this robustness, $\Phi$ was used in Ref.~\cite{castro1} 
to derive a different formula. The first observation is that 
the betatron phase advance $\Delta\phi_{ij}$ between two BPMs is 
equal to the difference of the tune line phases, namely
\begin{eqnarray}
\begin{aligned}
&\Delta\Phi_{H,ij}=\Phi_{H(1,0)_{j}}-\Phi_{H(1,0)_{i}}=\Delta\phi_{x,ij}\\
&\Delta\Phi_{V,ij}=\Phi_{V(0,1)_{j}}-\Phi_{V(0,1)_{i}}=\Delta\phi_{y,ij}
\end{aligned}\hskip 0.5 cm , \quad\label{eq:BetaPhase1} 
\end{eqnarray}
the initial phases $\psi_{x0,y0}$ of Eq.~\eqref{eq:tunes1} canceling 
out. By assuming that the region between three BPMs is free from 
unknown focusing errors, the following formula was derived in 
Ref.~\cite{castro1} to compute the beta function at the first 
BPM from the measured and model phase advances between three 
monitors:
\begin{eqnarray}
\beta_{1}^{(meas)}=\beta_{1}^{(mod)}
\frac{\cot{\Delta\phi_{12}^{(meas)}} -\cot{\Delta\phi_{13}^{(meas)}}}
     {\cot{\Delta\phi_{12}^{(mod)}}  -\cot{\Delta\phi_{13}^{(mod)}}}
\hskip 0.2 cm . \qquad\label{eq:BetaPhase2} 
\end{eqnarray}
The above relation applies to both transverse planes and is 
independent of any BPM calibration factor and roll. Even though 
originally conceived to work on three consecutive BPMs, it 
provides more robust results if several sets of triplets 
are used to apply Eq.~\eqref{eq:BetaPhase2} first, and the 
corresponding beta functions are properly averaged after, as shown 
in Ref.~\cite{prstab_BPMselect}.

As far as betatron coupling is concerned, the same RDTs of 
Eq.~\eqref{eq:f} can be measured independently only if the 
harmonic analysis is performed on the complex signal of the 
Courant-Snyder (C-S) coordinates $\tilde{x}-i\tilde{p}_x$ 
(and $\tilde{y}-i\tilde{p}_y$)~\cite{Andrea-arxiv}, whereas 
the harmonic analysis discussed here is carried out on the 
real signals $\tilde{x}$ ($\tilde{y}$) of Eq.~\eqref{eq:tunes1}. 
The momentum $\tilde{p}$ can be inferred by 
combining the position data of two BPMs~\cite{Caussyn} under 
the assumption that the invariant is constant between the two 
monitors. The presence of sextupoles and higher-order 
multipoles between two BPMs does alter the invariant, though 
this change is a higher order deformation that should not 
affect the linear analysis. On the other hand, the presence of 
coupling sources introduces a partial exchange of the two 
betatron invariants $2I_{x,y}$~\cite{prstab_coup}. This 
introduces a systematic error in the reconstruction of the 
momenta $\tilde{p}_{x,y}$, which is proportional to the coupling 
RDTs(i.e. to the unknown quantity to be measured), with the 
risk of corrupting the analysis of the betatron coupling. 
In Ref.~\cite{Andrea-arxiv} it is shown that combined coupling 
RDTs, $F_{xy}$ and $F_{yx}$ can be measured at each 
BPM $j$ from the two coupling harmonics of $\tilde{x}$ and $\tilde{y}$, 
i.e. $H(0,1)$ and $V(1,0)$, respectively:
\begin{eqnarray}\label{e:F_xy}
\left\{
\begin{aligned}
F_{xy,j}    &=f_{1001,j}   -f_{1010,j}^*\\
F_{yx,j}    &=f_{1001,j}^* -f_{1010,j}^* \\
|F_{xy,j}|  &=|H(0,1)_j|/(2|V(0,1)_j|)\mathcal{C}_{y,j}/\mathcal{C}_{x,j} \\
|F_{yx,j}|  &=|V(1,0)_j|/(2|H(1,0)_j|)\mathcal{C}_{x,j}/\mathcal{C}_{y,j} \\
q_{F_{xy},j}&=\hbox{arg}\{H(0,1)_j\}-\frac{3}{2}\pi-\hbox{arg}\{V(0,1)_j\} \\
q_{F_{yx},j}&=\hbox{arg}\{V(1,0)_j\}-\frac{3}{2}\pi-\hbox{arg}\{H(1,0)_j\}
\end{aligned}
\right.   \ , \hskip 1cm
\end{eqnarray}
where the possible dependence on the BPM calibration factors has 
been made explicit. In (hadron) machines where $|f_{1010}|\ll|f_{1001}|$,  
$F_{xy}\simeq F_{yx}^*\simeq f_{1001}$ and a calibration-independent 
formula to measure the amplitude of the RDT reads~\cite{Rogelio-thesis}
\begin{eqnarray}\label{e:f1001_hadron}
|f_{1001}|\simeq\frac{1}{2}\sqrt{\frac{|H(0,1)||V(1,0)|}{|H(1,0)||V(0,1)|}}
\ . \hskip 1cm
\end{eqnarray}

\section{Limits of the current approaches}
\label{limits}
\subsection{ORM analysis: model-dependent and time-consuming}
The retrieval of lattice parameters from the analysis 
of the closed orbit requires twice the employment of the computer 
lattice model: first in the fit of the measured ORM, 
Eqs.~\eqref{eq:ORM_04}-\eqref{eq:ORM_05}, then in computation of the 
new lattice parameters for the evaluation of beta beating and phase advance 
errors of Eq.~\eqref{eq:ORM_04}. Moreover, the analysis of orbit data 
is sensitive to BPM calibration factors, though effective coefficients 
can be inferred during the analysis of the ORM. 
This strong dependence on the initial lattice model, 
along with numerical issues related to possible degeneracies 
between fitting parameters, is considered by some as an intrinsic 
weakness of the ORM analysis.

Another drawback of the ORM analysis is its lengthy procedure for 
a single measurement and analysis. The acquisition typically foresees 
a sequence of current changes in orbit correctors and the retrieval of 
the corresponding orbit data. In the ESRF storage ring, this phase 
takes about 10 minutes for a partial ORM (32 out of 192 steerers), 
or 1 hour for a complete one. In larger machines such as the Large 
Hadron Collider (LHC) 
of CERN the time needed to scan the entire magnetic cycle makes this 
approach unsuitable for operational purposes. However, a new approach 
making use of alternating current steerers, fast BPM acquisition system 
(at 10 kHz) and harmonic analysis of orbit data was proved to 
obtain the same measurement with simultaneous magnet excitations at 
different frequencies, hence reducing dramatically the measurement 
time~\cite{AC-ORM}. Still, superconducting machines like the LHC 
may not benefit from this variation. The analysis too is quite time 
consuming, since the responses $\mathbf{M_{norm}}$ and $\mathbf{M_{skew}}$ 
of the ORM on the lattice errors ($\delta K$s and $\theta$s) in 
Eqs.~\eqref{eq:ORM_04}-\eqref{eq:ORM_05} is computed by simulating an 
ORM for each error: A heavy computation already for the ESRF storage 
ring with 256 quadrupoles and 64 dipoles, which can only become 
more lengthy in larger machines and future light sources.

\subsection{TBT analysis: error analysis of Eq.~\eqref{eq:BetaPhase2}}
Efforts have been made in the last three decades to conceive 
measurement techniques (as much as possible) independent of  
the initial computer model.  
Eq.~\eqref{eq:BetaPhase2} was derived in the 90s along with other handy 
expressions for the reconstruction of lattice parameters from TBT data 
in a way to be independent of the BPM calibration factors~\cite{castro1}.
Moreover, the dependence on the initial model was smartly limited 
optics functions only (i.e. of $\beta^{(meas)}$ starting from 
$\beta^{(mod)}$ and $\Delta\phi^{(mod)}$). The Model Independent Analysis 
(MIA) of Ref.~\cite{SVD1,MIA1,MIA2} and the more recent Independent 
Component Analysis (ICA) of Ref.~\cite{SVD2,ICA} proposed 
a statistical approach to extract the same lattice parameters with no 
{\sl a priori} knowledge of the initial model. All these 
advancements were successfully applied to many circular accelerators across 
the world. However, the autonomy from the initial model to fit and/or 
interpret the measured TBT data comes to the price of {\sl forcing} 
the description of the same data according to some hypothesis, 
assumptions or approximations, of which more will be said 
in Sec.~\ref{sec:high-order-tune}.

As mentioned in the previous section, Eq.~\eqref{eq:BetaPhase2} 
assumes that no quadrupole error is present between the three BPMs.
In Appendix~\ref{app:1} an extension of that formula
not requiring this hypothesis is derived up to the first order in the 
field errors  $\delta K_1$:
\begin{eqnarray}
&&\beta_{1}^{(meas)}\hskip -0.5mm=\hskip -0.5mm\beta_{1}^{(mod)}
\frac{\cot{\Delta\phi_{12}^{(meas)}}\hskip -0.5mm -\hskip -0.5mm\cot{\Delta\phi_{13}^{(meas)}}}
     {\cot{\Delta\phi_{12}^{(mod)}} \hskip -0.5mm -\hskip -0.5mm\cot{\Delta\phi_{13}^{(mod)}}
      +(\bar{h}_{12}\hskip -0.5mm -\hskip -0.5mm \bar{h}_{13})} \nonumber \\
&&\hskip 1.2cm+O(\delta K_1^2) \quad ,\label{eq:BetaPhase3} \\ \nonumber \\
&&\bar{h}_{ij}=\mp\frac{\displaystyle\sum_{i<w<j}{\beta_w^{(mod)}\delta K_{w,1}
          \sin^2{\Delta\phi_{wj}^{(mod)}}}}
         {\sin^2{\Delta\phi_{ij}^{(mod)}}}\label{eq:BetaPhase3B} 
\hskip 0.2 cm , 
\end{eqnarray} 
where the sum runs over all quadrupole errors between the BPMs $i$ 
and $j$. The sign depends on the plane: negative for $x$, positive 
for $y$.

If no error is present between the three BPMs, 
Eq.~\eqref{eq:BetaPhase2} is retrieved as expected. 
A special case where this equation still applies even in the 
presence of strong localized focusing errors is when 
$\bar{h}_{13}=\bar{h}_{12}\ne0$. More generally, it remains a 
robust approximation whenever the beating induced by the 
quadrupole errors between three BPMs has a minor impact 
on the cotangent of their phase advance, 
$|\bar{h}_{13}-\bar{h}_{12}|\ll |\cot{\Delta\phi_{13}^{(mod)}}-\cot{\Delta\phi_{12}^{(mod)}}|$, 
or when it is much smaller than the one generated 
by focusing glitches along the rest of the ring, see 
discussion after Eq.~\eqref{castro-10} of 
Appendix~\ref{app:1}. 

\begin{figure}
\rule{0mm}{0mm}
\centerline{\includegraphics[width=8.5cm]{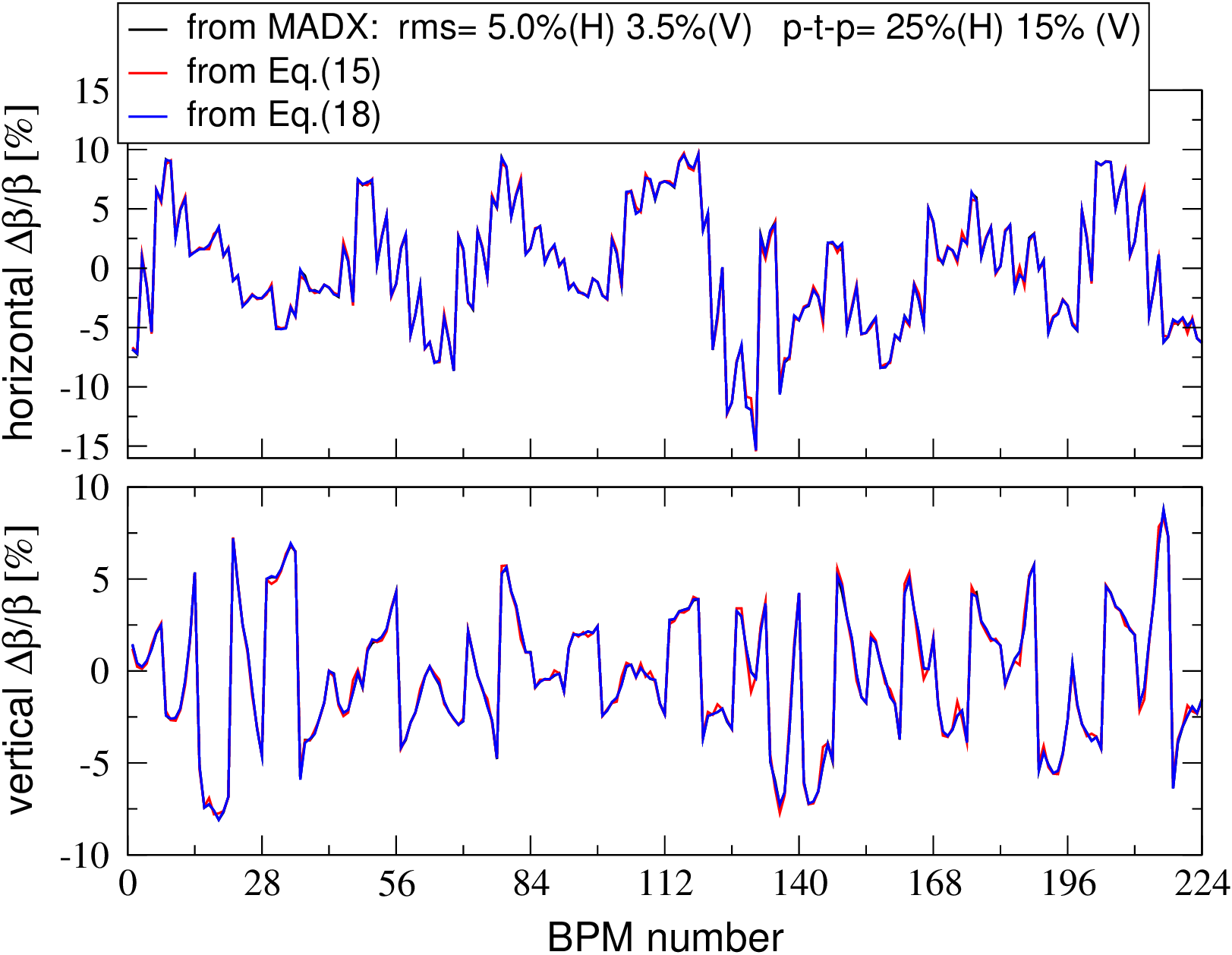}}
\rule{0mm}{1.5mm}
\centerline{\includegraphics[width=8.5cm]{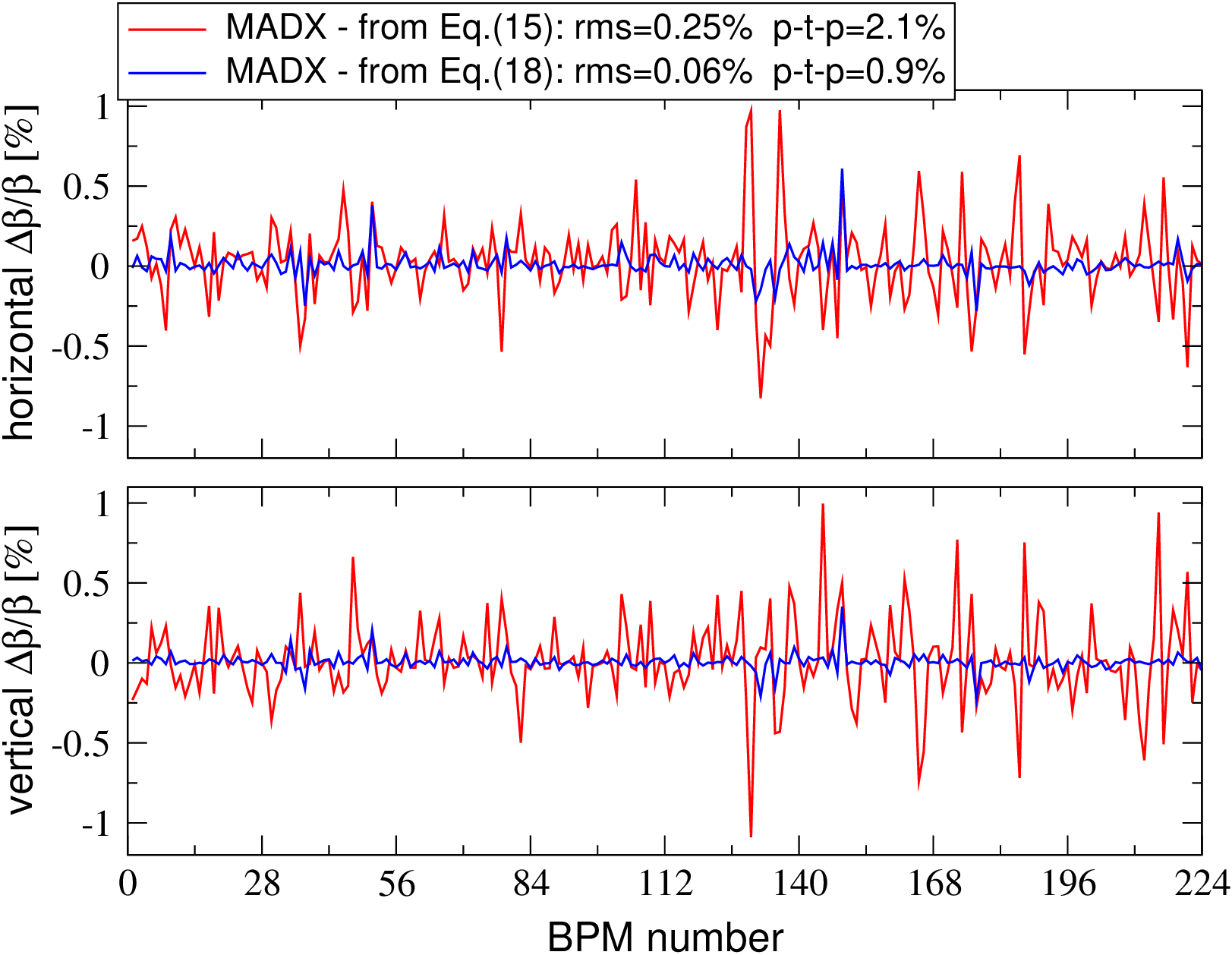}}
  \caption{\label{fig_beat1} (Color) Top: Beta beating induced by a typical 
           set of errors in the ESRF storage ring as computed by MADX and the 
           two formulas Eqs.~\eqref{eq:BetaPhase2} and \eqref{eq:BetaPhase3}. 
           {\sl rms} stands for ``root-mean square'', {\sl p-t-p} for 
           ``peak-to-peak''.
           Bottom: difference between the real beating (MADX) and one  
           obtained from two formulas: In this example, the new 
           Eq.~\eqref{eq:BetaPhase3} is four time more accurate than 
           Eq.~\eqref{eq:BetaPhase2}. A pity $\bar{h}_{ij}$ is not observable 
           and Eq.~\eqref{eq:BetaPhase3} may not be used for measurements! }
\rule{0mm}{3mm}
\end{figure}

In the upper 
plot of Fig.~\ref{fig_beat1} an example is shown with the beta 
beating computed for a typical set of linear lattice errors 
in the ESRF storage ring. The beating is first computed by 
MADX, then the BPM phase advances are used to evaluate it 
with the existing formulas of Eq.~\eqref{eq:BetaPhase2}, and 
eventually the $\bar{h}_{ij}$ are also calculated from the model 
and applied to Eq.~\eqref{eq:BetaPhase3}. The latter turns out to be 
more accurate than the former (see bottom plot of Fig.~\ref{fig_beat1}). 
Unfortunately, Eq.~\eqref{eq:BetaPhase3} is of no help in 
improving the measurement of the beta beating, since 
$\bar{h}_{ij}$ is not an observable. However, once the error 
model is built, it can be computed {\sl a posteriori} and used 
to estimate the accuracy (i.e. systematic error bars) of the 
direct measurement via Eq.~\eqref{eq:BetaPhase2}, which reads
\begin{eqnarray}
\frac{\delta\beta_{meas}}{\beta}\hskip -0.1cm=
\frac{\beta_{_{Eq.\eqref{eq:BetaPhase2}}}-
      \beta_{_{Eq.\eqref{eq:BetaPhase3}}}}
     {\beta_{_{Eq.\eqref{eq:BetaPhase3}}}}\hskip -0.1cm\simeq\hskip -0.1cm
\frac{\bar{h}_{13}-\bar{h}_{12}}
     {\cot{\Delta\phi_{12}^{(mod)}} -\cot{\Delta\phi_{13}^{(mod)}}}
     \ .\nonumber \\ \label{eq:BetaPhase4} 
\end{eqnarray}
An interesting feature of the above expression is that it depends on the ideal 
lattice parameters ($\phi^{(mod)}$ and $\beta^{(mod)}$) and the field 
deviations $\delta K_1$ only. Hence it does not require the evaluation of the 
new optics induced by the errors. This may accelerate the estimation of 
the systematic errors from the statistical analysis of various error 
distributions along large rings, as in Ref.~\cite{prstab_BPMselect}.

In the derivation of Eq.~\eqref{eq:BetaPhase3}, other 
handy formulas for the evaluation of C-S parameters 
modified by focusing errors have been obtained,
\begin{eqnarray}
&\hspace{-5mm}\left\{\begin{aligned}
\beta_{x_j}&\simeq\beta_{x,j}^{(mod)}
                     \left(1 +8\Im\left\{f_{2000,j}\right\}\right)\\
\alpha_{x,j}&\simeq\alpha_{x,j}^{(mod)}\left(1+8\Im\{f_{2000,j}\}\right)
             -8\Re\{f_{2000,j}\} \\
\Delta\phi_{x,ij}&\simeq\Delta\phi_{x,ij}^{(mod)}
                 \hspace{-.7mm}-\hspace{-.7mm}2h_{x,ij} 
                 +4\Re\left\{f_{2000,j}-f_{2000,i}\right\}\\
\end{aligned}\right. ,
\nonumber\\ &\hspace{8.7cm}\label{insertion-xy} \\\nonumber
&\hspace{-5mm}\left\{\begin{aligned}
\beta_{y_j}&\simeq\beta_{y,j}^{(mod)}
                     \left(1 +8\Im\left\{f_{0020,j}\right\}\right)\\
\alpha_{y,j}&\simeq\alpha_{y,j}^{(mod)}\left(1+8\Im\{f_{0020,j}\}\right)
             -8\Re\{f_{0020,j}\} \\
\Delta\phi_{y,ij}&\simeq\Delta\phi_{y,ij}^{(mod)}
                  \hspace{-.7mm}-\hspace{-.7mm}2h_{y,ij} 
                 +4\Re\left\{f_{0020,j}-f_{0020,i}\right\}\\
\end{aligned}\right. ,
\end{eqnarray}
where the focusing RDTs $f_{2000}$ and $f_{0020}$ are 
defined defined in Eqs.~\eqref{eq:def_f0020} and 
\eqref{RDT-2}, while explicit expressions for the 
detuning coefficients $h_{ij}$ can be found in 
Eq.~\eqref{phase-shift-9}. Even though the RDTs 
are complex quantities, only their real ($\Re$) and 
imaginary ($\Im$) parts enter in the above equations, 
whose remainders are proportional to $f^2$.

\subsection{TBT analysis: error analysis of Eq.~\eqref{eq:BetaAmpli1}}
In Appendix~\ref{app:1}, a more general version of 
Eq.~\eqref{eq:BetaAmpli1} is derived, namely 
\begin{eqnarray}
\left\{
\begin{aligned}
\beta_{x,j}^{(meas)}&=\beta_{x,j}^{(mod)}\left(\frac{|H(1,0)_j|}{<|H(1,0)|>}\right)^2
  \quad\times\\&\hskip 0.7cm
           [1+2(<\mathcal{E}_x>-\mathcal{E}_{x,j})+O(\mathcal{E}_x^2)]
  \quad\times\\&\hskip 0.7cm
           [1+64<|f_{2000}|^2>+O(|f_{2000}|^3)]\\
\beta_{y,j}^{(meas)}&=\beta_{y,j}^{(mod)}\left(\frac{|V(0,1)_j|}{<|V(0,1)|>}\right)^2
  \quad\times\\&\hskip 0.7cm
           [1+2(<\mathcal{E}_y>-\mathcal{E}_{y,j})+O(\mathcal{E}_y^2)]
  \quad\times\\&\hskip 0.7cm
           [1+64<|f_{0020}|^2>+O(|f_{0020}|^3)] \\
\mathcal{C}_{x,y}&=1+\mathcal{E}_{x,y}\quad,\quad 0\simeq\mathcal{E}_{x,y}\ll1
\end{aligned}\right.
,\hskip 0.7cm \label{beta-ampli-err1}
\end{eqnarray}
where $<|f|^2>$ and $<\mathcal{E}>$ represent the averaged values 
(over all BPMs) of the amplitudes of focusing RDTs squared and 
calibration errors, respectively. Unless these are determined 
by independent measurements, they cannot be disentangled and are 
not observable. A (rude) zero-order truncation is then needed in 
order to apply Eq.~\eqref{beta-ampli-err1} to real data, yielding to 
\begin{eqnarray}
\left\{
\begin{aligned}
&\beta_{x,j}=\beta_{x,j}^{(mod)}\left(\frac{|H(1,0)_j|}{<|H(1,0)|>}\right)^2
            +O(\mathcal{E}_x,|f_{2000}|^2)\\
&\beta_{y,j}=\beta_{y,j}^{(mod)}\left(\frac{|V(0,1)_j|}{<|V(0,1)|>}\right)^2
            +O(\mathcal{E}_y,|f_{0020}|^2)
\end{aligned}\right.
.\qquad \label{beta-ampli-err2}
\end{eqnarray}

\subsection{TBT analysis: Eq.~\eqref{eq:BetaAmpli1} Vs Eq.~\eqref{eq:BetaPhase2}}
Eq.~\eqref{beta-ampli-err2} shows that the 
error of Eq.~\eqref{eq:BetaAmpli1} is  proportional to 
the calibration error $\mathcal{E}$ and to square of the RDTs $|f|^2$. 
Since the latter are, to the first order, linearly dependent 
on the focusing errors, the error of Eq.~\eqref{eq:BetaAmpli1} 
scales with $\delta K_1^2$, whereas the uncertainty of 
Eq.~\eqref{eq:BetaPhase2} scales with $\delta K_1$, as indicated 
by Eq.~\eqref{eq:BetaPhase4}. From a purely theoretical 
point of view, hence, either formula is to be preferred to the 
other according to the largest source of uncertainty. If BPM 
calibration errors $\mathcal{E}$ are either unknown or 
expected to be larger than $\bar{h}_{ij}$ of Eq.~\eqref{eq:BetaPhase3B} 
(i.e. of the focusing errors between three BPMs), beta functions are 
better inferred by Eq.~\eqref{eq:BetaPhase2}. If the opposite is true, 
then Eq.~\eqref{eq:BetaAmpli1} is to be preferred. 

\begin{figure}
\rule{0mm}{0mm}
\centerline{\includegraphics[width=8.5cm]{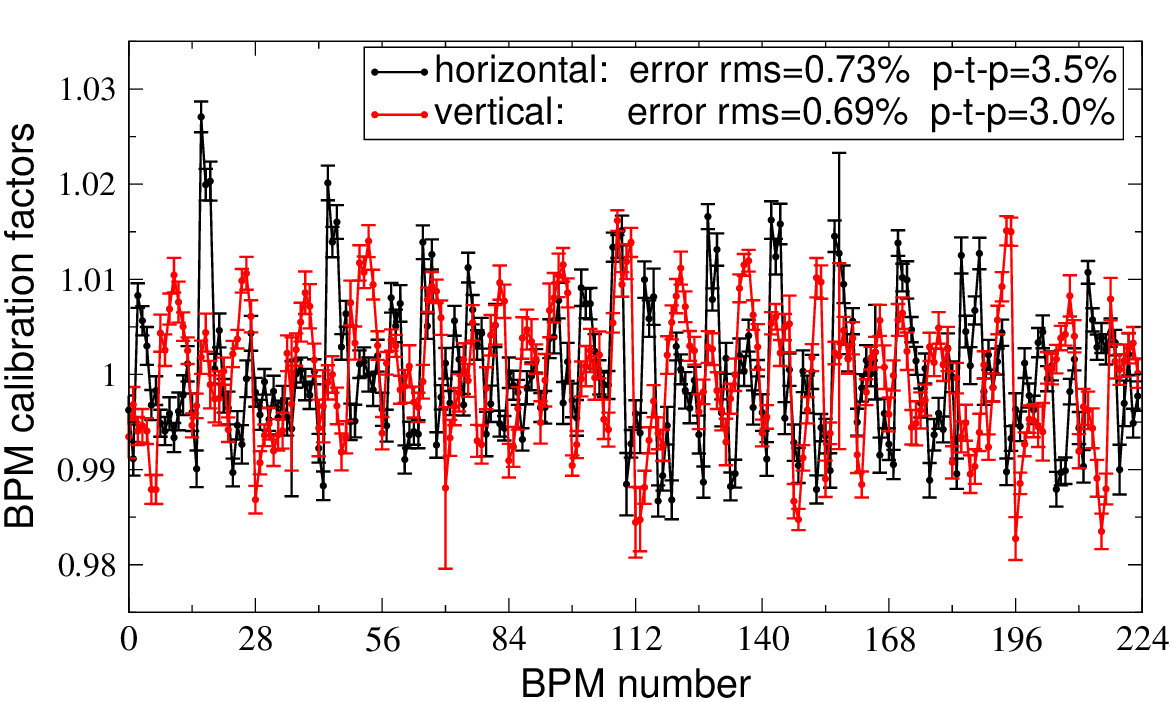}}
\vskip -3mm
\centerline{\includegraphics[width=8.5cm]{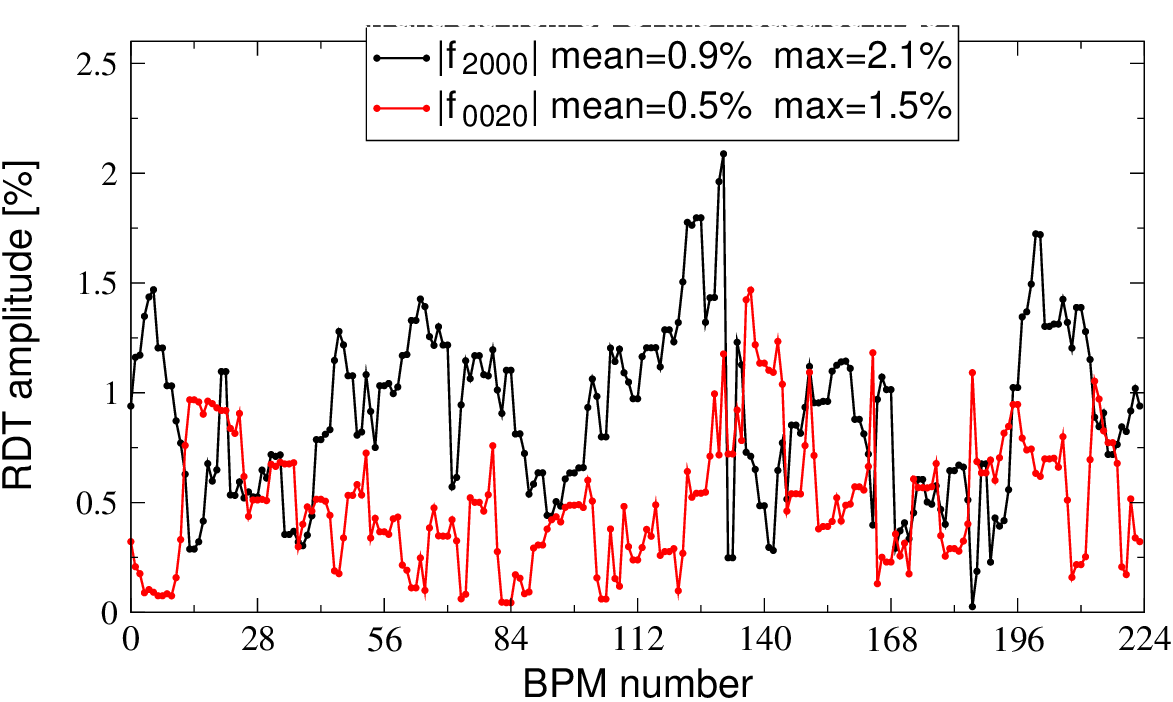}}
  \caption{\label{fig_beat2} (Color) Top: ESRF's effective BPM calibration 
factors $\mathcal{C}_{x,y}$ inferred from the analysis of the orbit 
response matrix, averaged over 32 measurements repeated during one 
whole year of operation. The error bars represent their standard 
deviation. Bottom : variation of the focusing RDT amplitudes along 
the ESRF storage ring for the same lattice error of Fig.~\ref{fig_beat1}.}
\rule{0mm}{3mm}
\end{figure}

As far as the ESRF storage ring is concerned, BPM calibration factors 
are routinely fitted from the analysis of measured ORM. In the top 
plot of Fig.~\ref{fig_beat2} the mean values over 32 measurement 
repeated during one whole year of operation are displayed along 
with the error bars representing their standard deviation. The 
rms (systematic) BPM gain error is of about $0.7\%$ ($2.7\%$ 
maximum), with rms (random) error bars below $0.1\%$. It is 
worthwhile noticing that these calibrations factors refers to 
Libera BPMs operating in slow (orbit) acquisition mode and that they 
may vary in the fast (TBT) acquisition mode. In the bottom plot of 
Fig.~\ref{fig_beat2} the variation of the RDT amplitudes along the 
ring is showed for the same lattice error of Fig.~\ref{fig_beat1}. 
With $<|f_{2000}|>$ of about $0.9\%$ ($2.1\%$ maximum),which is 
larger than $<|f_{0020}|>$, the RDT-related 
rms uncertainty $64|f|^2$ of Eq.~\eqref{beta-ampli-err1} is then 
of about $0.6\%$ ($2.8\%$ maximum).

The bottom plot of Fig.~\ref{fig_beat1} shows the error 
of Eq.~\eqref{eq:BetaPhase2} associated to $\delta K_1$ via the 
$\bar{h}_{ij}$: $0.25\%$ rms ($2.1\%$ peak-to-peak). 
Even admitting that the BPM calibration factors of 
Fig.~\ref{fig_beat2} are applicable to TBT data and can be 
used in Eq.~\eqref{eq:BetaAmpli1}, this formula is expected 
in any case to be less precise (residual rms error of $0.6\%$) 
than Eq.~\eqref{eq:BetaPhase2} (residual rms error of $0.25\%$). 
For storage rings with lower rms beta beating (which is of about
$5\%$ at the ESRF) the opposite might be true.

The above considerations are rather mathematical and other 
aspects are to be taken into account when selecting the best 
approach for the evaluation of the beta function. First, 
Eq.~\eqref{eq:BetaPhase2} relies on the perfect synchronization 
between all BPMs, i.e. BPM reporting on the same turn and on the 
same bunch. Any systematic delay or jitter (even of a tiny fraction 
of revolution frequency) between the BPM data acquisition, would 
result in an artificial BPM phase advance error, since the 
initial arbitrary phase $\psi_{0}$ of Eq.~\eqref{eq:tunes1} 
would be no longer the same for all BPMs and, hence, not 
canceled out in Eq.~\eqref{eq:BetaPhase1}. 
\begin{eqnarray}
\begin{aligned}
&\Delta\Phi_{ij}=\Delta\phi_{ij}+\delta\phi_{ij}^{(tim)}
\end{aligned}\hskip 0.5 cm , \quad\label{eq:BetaComp1} 
\end{eqnarray}
where $\delta\phi_{ij}^{(tim)}$ is the phase error introduced 
by the monitors  relative delay. The BPMs in the 
ESRF storage ring, for instance, are synchronized within 
about 0.1 $\mu$s (peak-to-peak) over a revolution time of 
2.82 $\mu$s. Eq.~\eqref{eq:BetaAmpli1} 
does not suffer from such a constraint and would work even 
with BPMs reporting on different turns. Second, the presence 
of trigonometric functions in the denominator of 
Eq.~\eqref{eq:BetaPhase1} requires that BPMs are separated 
by a phase advance away from either zero or multiples of $\pi$, 
to prevent the cotangent from becoming infinite. If this is 
the case, the triplet can be defined by using BPMs further 
downstream, as indicated in Ref.~\cite{prstab_BPMselect}. This 
may come to the price of increasing the number of sources of 
focusing errors between the BPMs, i.e. $\bar{h}_{ij}$, and 
in turn the measurement error of Eq.~\eqref{eq:BetaPhase4}. 
Eq.~\eqref{eq:BetaAmpli1} is not affected by the relative 
position of the BPMs.

\subsection{ORM and TBT analysis: impact of BPM resolution and ultra-low coupling}
\label{subsec:resolution}
Ideally, ORM and TBT should be generated by exciting the beam (with 
orbit correctors and dipole kickers, respectively) so to optimize 
the signal-to-noise ratio, while remaining in the linear regime of 
the betatron motion. In this section rough estimates of the beam excitation 
amplitudes ensuring sufficient resolution are provided for the 
ESRF storage ring equipped with commercial Libera {\sl Brilliance} 
BPMs and operating with ultra-low coupling, i.e. with a ratio 
between the betatron equilibrium emittances $E_x/E_y\simeq1$\textperthousand. 
The question whether these amplitudes remain in the linear regime 
or not is addressed in the next sections. 

ORMs are measured by using the BPMs in the slow-mode acquisition 
(10 Hz), which ensures a resolution of about 10 nm. Specialists 
prefer to define the BPM resolution as {\sl integrated noise spectrum} 
or {\sl integrated rms noise}~\cite{BPMnoise1,BPMnoise2}, denoted as 
measured uncertainty for frequencies integrated from 0.1 Hz or 
1 Hz up to a specified bandwidth and represented by a spatial 
resolution per square root of the bandwidth (the noise being 
typically expected to be white so that the measurement error 
would decrease with the square root of the bandwidth). For the 
Libera {\sl Brilliance} BPMs, the typical value in slow-mode 
acquisition is 10 nm/$\sqrt{\hbox{Hz}}$. Since the resolution 
depends on the beam intensity too, ORMs are measured at 30mA, 
which ensure sufficient beam signal while remaining low enough 
to prevent beam induced effects on the orbit motion. The rms orbit 
distortion is of 200-250 $\mu$m in the plane of the steerer, 
2-5 $\mu$m in the other one because of the ultra-low coupling 
achieved in the machine. Despite this low orbit distortion in 
the orthogonal plane, data remain more than two orders of 
magnitude above the noise floor (10 nm), permitting a reliable 
coupling measurement. 

\begin{figure}
\rule{0mm}{0mm}
\centerline{\includegraphics[width=8.5cm]{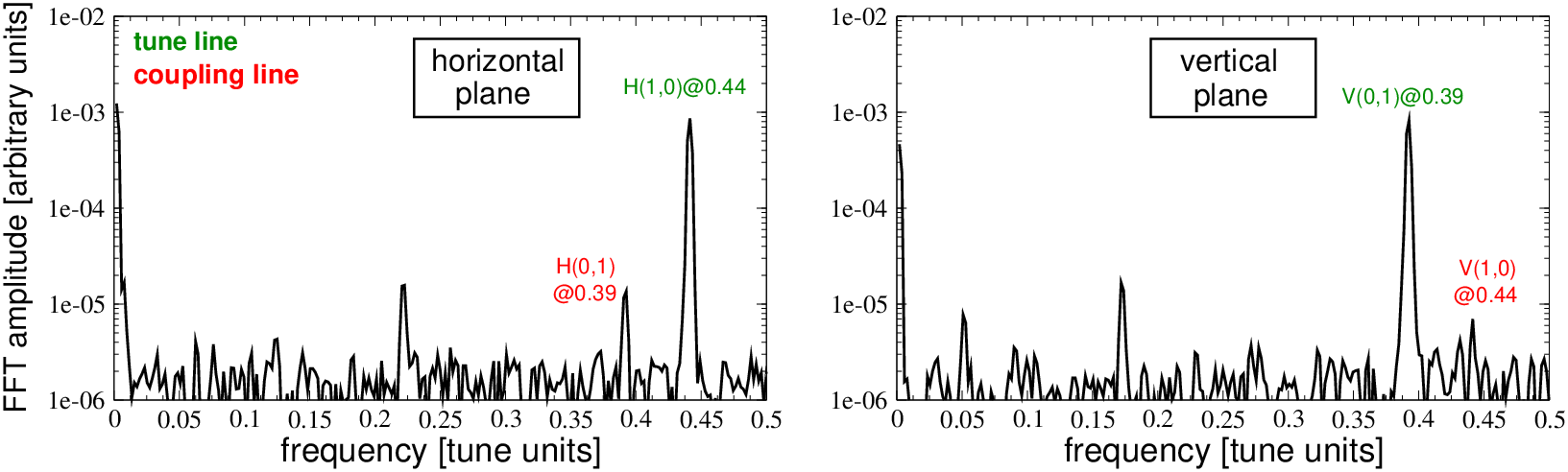}}
\vskip 2mm
\centerline{\includegraphics[width=8.5cm]{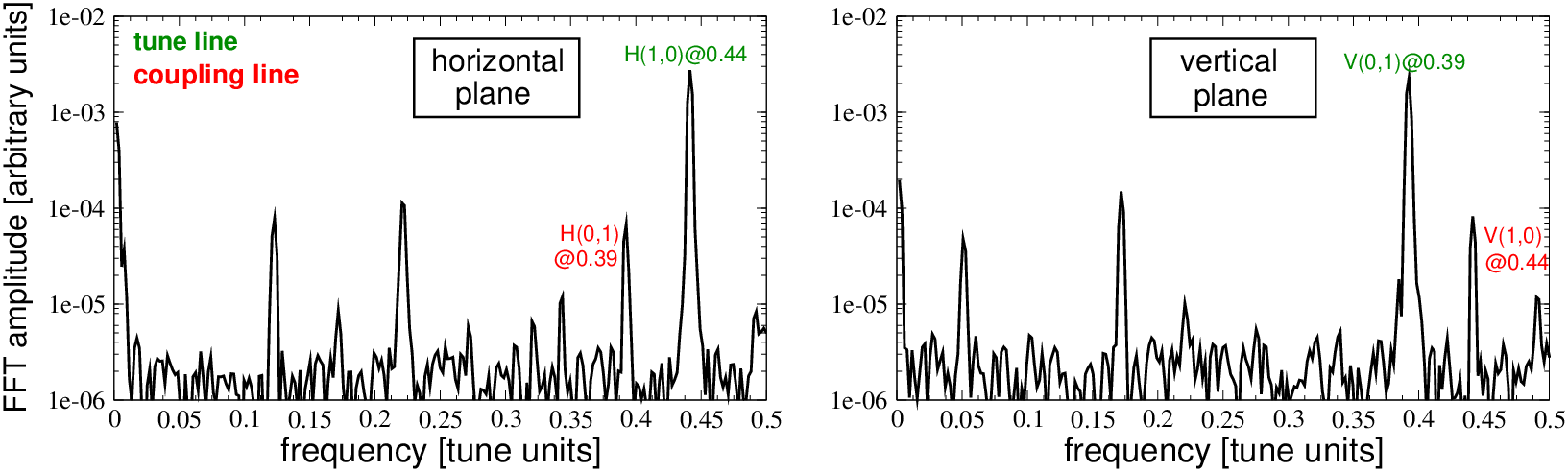}}
  \caption{\label{fig_fftTbT} (Color) Examples of spectra from measured TBT 
    BPM data of the ESRF storage ring. Because of the ultra-low coupling, 
    data with an initial excitation of 1 mm ($\beta_x=35$ m) 
    and 0.3 mm ($\beta_x=3$ m) result in coupling lines close to the background 
    noise (top plots). In order to limit the latter to about 2\% of the 
    coupling lines, the initial oscillations need to be tripled (bottom plots).
    }
\rule{0mm}{0mm}
\end{figure}

TBT data are acquired by switching the BPMs into the 
$\sim$355kHz acquisition mode, whose expected resolution is 
in the $\mu$m range. Two independent evaluations of the 
noise floor with beam (one recording TBT data of the unperturbed 
beam, the other via SVD of the BPM matrix) indicate that 
the actual noise floor is of about 10 $\mu$m. This accounts also 
for the natural beam motion which can be corrected by a fast orbit 
correction scheme during operation but not during the acquisition of 
TBT data (the method requires free oscillations). The resolution of 
TBT data is then three orders of magnitude worse than in the ORM 
measurement. This has a dramatic consequence in the minimum 
beam excitation necessary for a robust evaluation of coupling via 
TBT data. This is performed by analyzing the coupling lines of
the TBT spectrum, $H(0,1)$ and $V(1,0)$ of Eq.~\eqref{e:F_xy}, 
whose amplitudes in real units [m] read
\begin{eqnarray}\label{e:resTBT1}
\left\{ \begin{aligned}
|H(0,1)|_{[m]}&=2\sqrt{\frac{\beta_x}{\beta_y}}\times|V(0,1)|_{[m]}\times|F_{xy}|\\
|V(1,0)|_{[m]}&=2\sqrt{\frac{\beta_y}{\beta_x}}\times|H(1,0)|_{[m]}\times|F_{yx}|
\end{aligned} \right.   \ . \hskip 1cm
\end{eqnarray}
The beta functions at the BPMs are such that both square roots 
in the above expressions range between 0.4 and 2.8: 
$\sqrt{\beta_{x,y}/\beta_{y,x}}\sim 1$ can be then assumed. Because 
of the ultra-low coupling, the amplitudes of combined RDTs $|F|$ 
are of the order of $10^{-2}$. An excitation in both planes of 
1 mm ($|H(1,0)|_{[m]}\sim|V(0,1)|_{[m]}\sim10^{-3}$ m, since 
these are by far the largest harmonics of the TBT signal) 
would result then in coupling lines 
$|H(0,1)|_{[m]}\sim|V(1,0)|_{[m]}\sim10\ \mu$m, which is of the same 
order of magnitude than the noise floor (see upper plots of 
Fig.~\ref{fig_fftTbT}). 

In conclusion, even though a tune line of $\sim1$ mm would suffice 
for a reliable analysis of focusing errors, the evaluation of 
ultra-low coupling via the secondary harmonics would be rather 
inaccurate. Hence, unless the BPM resolution in TBT mode and 
the natural beam stability are  
significantly improved, an initial oscillation of several mm is 
necessary for a robust and complete study of the linear lattice 
errors with TBT data (see lower plots of 
Fig.~\ref{fig_fftTbT}). These numbers may of course be different 
in other 3rd generation (and more recent) light sources, but 
the orders of magnitude are expected to be similar as long as 
they operate with ultra-low coupling and comparable BPM spectral 
background noise. 

\subsection{TBT analysis: Can higher-order terms may be really neglected?}
\label{sec:high-order-tune}
The harmonic analysis of TBT data discussed here, as well as other 
techniques analyzing the BPM matrix, assumes that the tune line (or 
the {\sl betatron mode}) is exclusively generated by quadrupolar terms, 
independent of the initial oscillation amplitude.

In Appendix~\ref{app:2A} more general expressions for the tune line 
amplitude and BPM phase advance in the nonlinear (amplitude dependent) 
regime are derived. The result is
\begin{eqnarray} 
&&\left\{\begin{aligned}                            \nonumber
&\Delta\Phi_{H,ij}=\Delta\phi_{x,ij}^{(mod)}+\hbox{arg}\{B_{x,i}-B_{x,j}\} 
                -2h_{1100,ij}\\ &\hskip 12mm -4h_{2200,ij}(2I_x)-2h_{1111,ij}(2I_y)\\
&\Delta\Phi_{V,ij}=\Delta\phi_{y,ij}^{(mod)}+\hbox{arg}\{B_{y,i}-B_{y,j}\} 
                -2h_{0011,ij}\\ &\hskip 12mm -2h_{1111,ij}(2I_x)-4h_{0022,ij}(2I_y) 
\end{aligned}\right. , \\ &&\left\{\begin{aligned}  \label{eq:nonlinT1}
&|H(1,0)_j|=\frac{\sqrt{2I_x}}{2}|B_{x,j}| \\
&|V(0,1)_j|=\frac{\sqrt{2I_y}}{2}|B_{y,j}| 
\end{aligned}\right. \qquad, \\ &&\left\{\begin{aligned}\nonumber
&\hspace{-.5mm}B_{x,j}\hspace{-.9mm}=\hspace{-.9mm}1\hspace{-.5mm}+\hspace{-.5mm}
         i4f^*_{2000,j}\hspace{-.9mm}+\hspace{-.5mm}
         iF_{xx,j}(K_2^2,K_3,I_{x,y})\hspace{-.9mm}+\hspace{-.9mm}
         T_{H,j}(K_2^2,I_{x,y})\\ 
&\hspace{-.5mm}B_{y,j}\hspace{-.9mm}=\hspace{-.9mm}1\hspace{-.5mm}+\hspace{-.5mm}
         4if^*_{0020,j}\hspace{-.9mm}+\hspace{-.5mm}
         iF_{yy,j}(K_2^2,K_3,I_{x,y})\hspace{-.9mm}+\hspace{-.9mm}
         T_{V,j}(K_2^2,I_{x,y})
\end{aligned}\right. .
\end{eqnarray}
The functions $F$ and the octupolar-like amplitude dependent detuning 
terms $h$ are proportional to octupolar fields ($\propto K_3$) 
and to quadratic functions of sextupole strengths ($\propto K_2^2$), 
whereas $T$ scales quadratically with $K_2$. 

The above expressions indicate that when the initial 
oscillation amplitude $(2I)$ is {\sl too large}, the betatron 
BPM phase advance $\Delta\phi_{ij}$ is no longer measurable 
from the difference of the tune line phases $\Delta\Phi_{ij}$, 
since (octupolar-like) amplitude dependent focusing 
terms {\sl corrupt} the tune line. 
The same is true for the invariant itself $(2I)$, which 
is no longer measurable from the tune line amplitude.  
In fact, the latter is not anymore constant along the ring 
and its modulation depends on the invariant itself via the 
functions $F$ and $T$ of Eq.~\eqref{eq:nonlinT1}.

\begin{figure}[!t]
\rule{0mm}{0mm}
\centerline{\includegraphics[width=8.5cm,height=3.5cm]{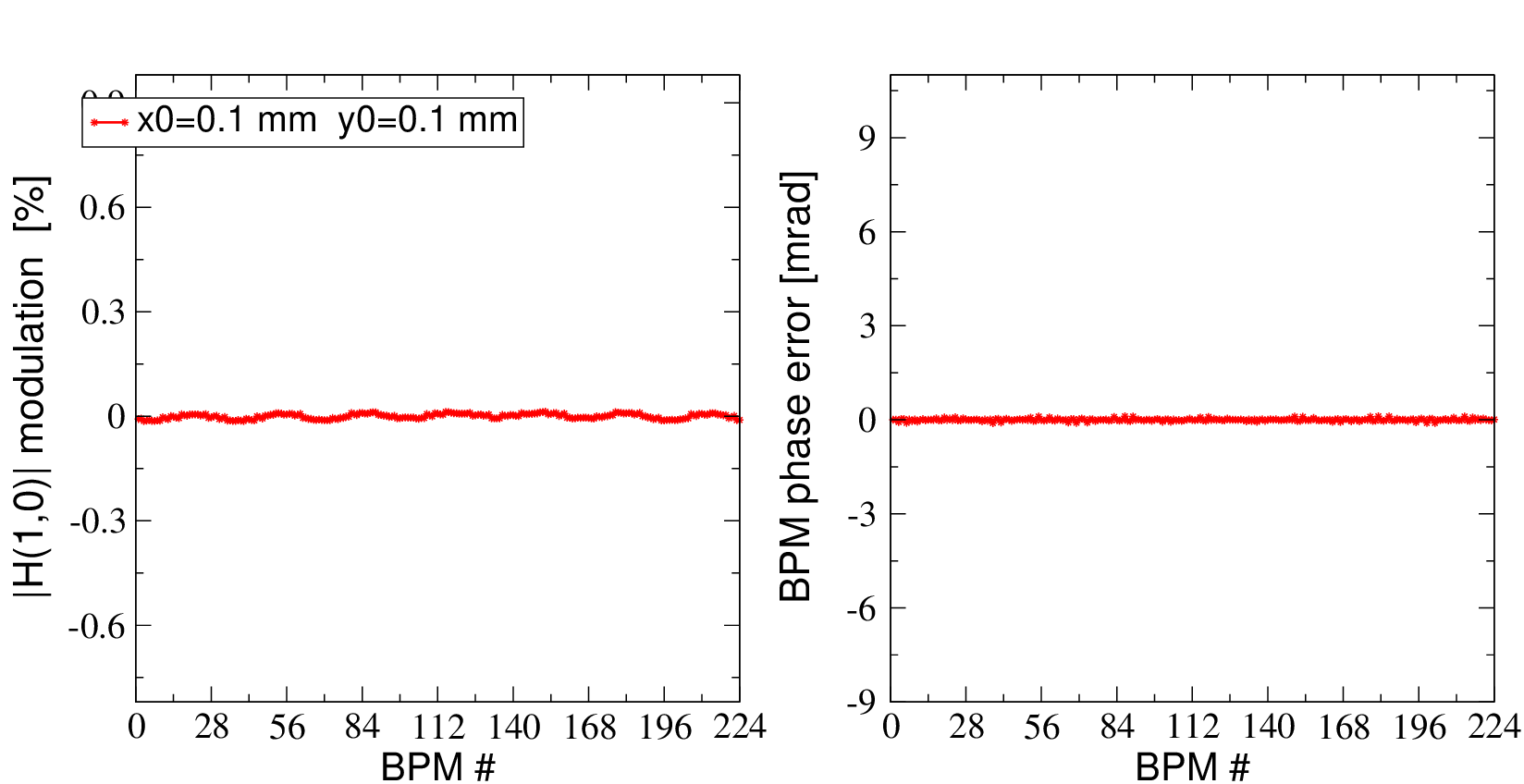}}
\vskip -2mm
\centerline{\includegraphics[width=8.5cm,height=3.5cm]{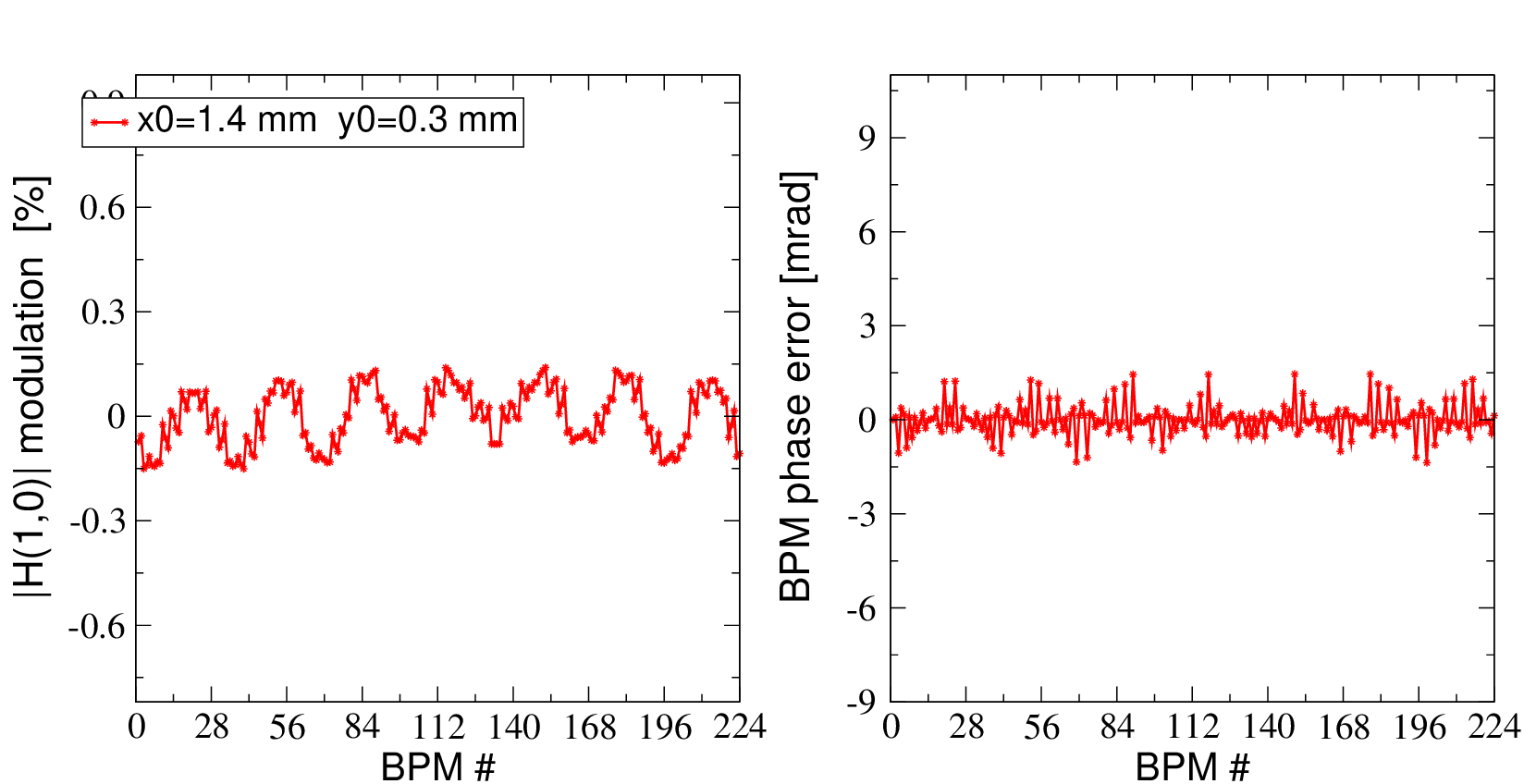}}
\vskip -2mm
\centerline{\includegraphics[width=8.5cm,height=3.5cm]{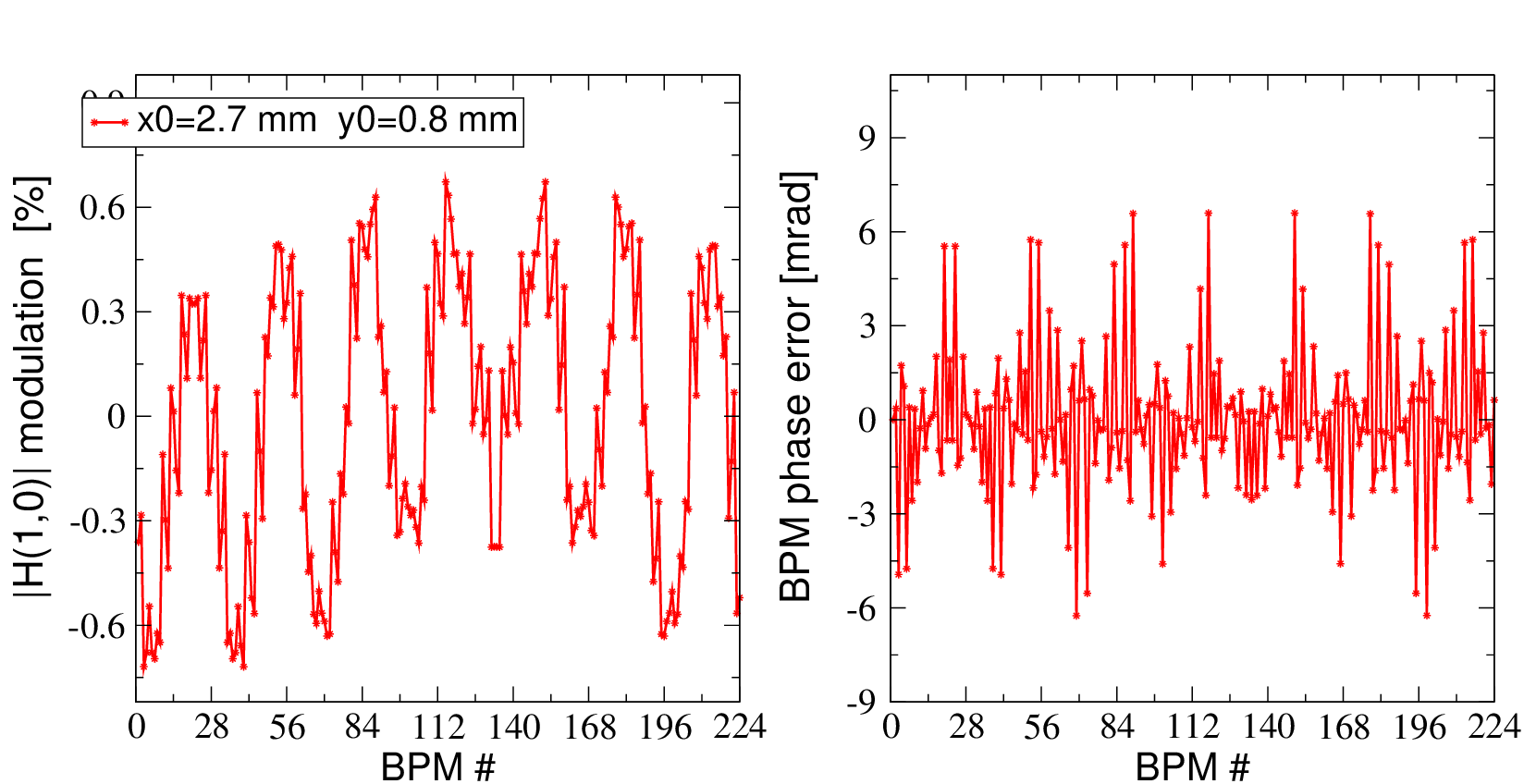}}
\vskip -2mm
\centerline{\includegraphics[width=8.5cm,height=3.5cm]{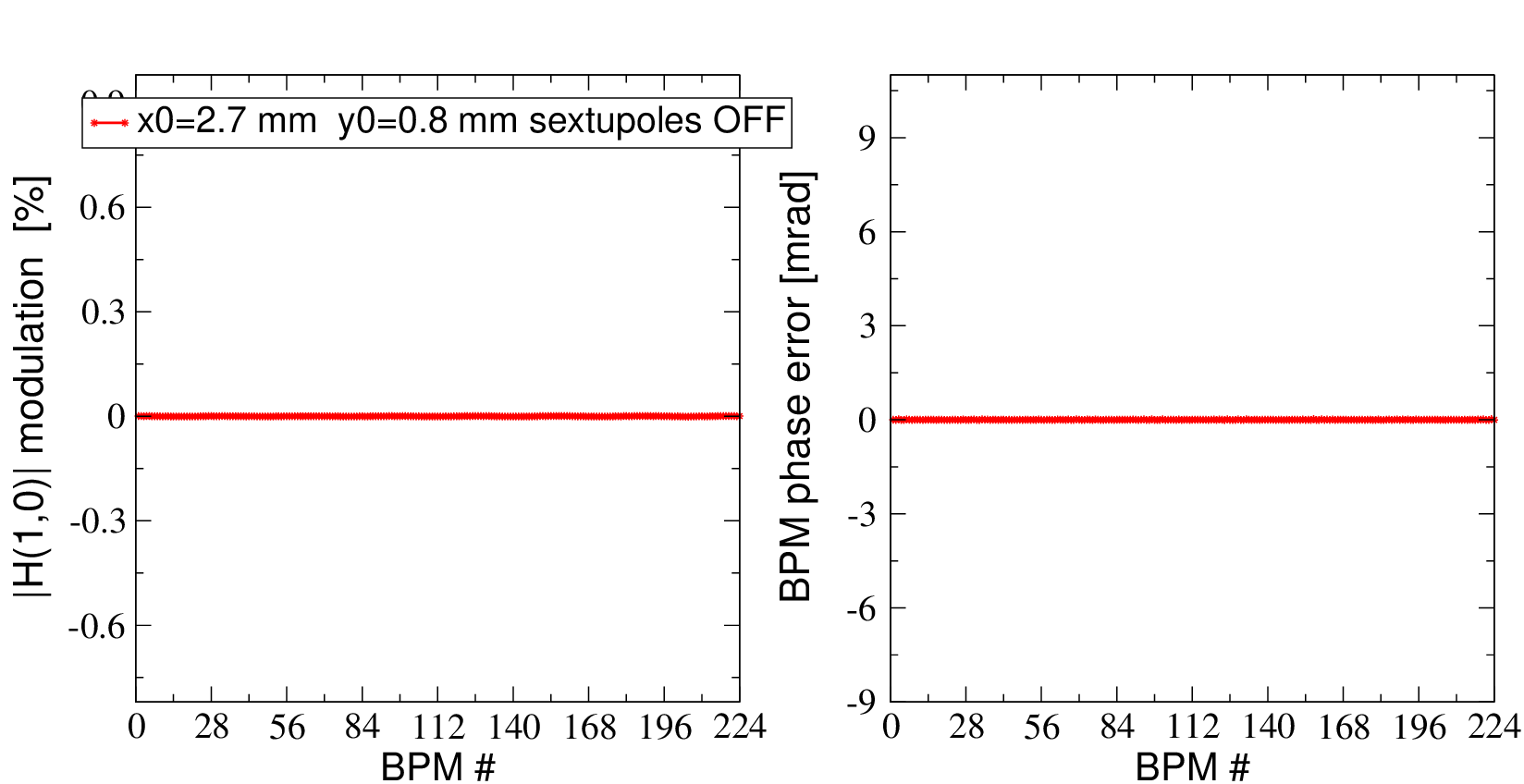}}
  \caption{\label{fig_NonLin1} (Color) Modulation of the horizontal 
          tune line amplitude
          (left) and error of the BPM phase advance $\Delta\Phi_{x,ij}$ 
          inferred from the tune line phase with respect to the 
          betatron BPM phase advance $\Delta\phi_{x,ij}$ (right) obtained 
          from single particle tracking simulations with different initial 
          conditions. The large BPM phase error observed at 
          $(x_0,y_0)=(2.7,0.8)$ mm (third row) disappears when nonlinear 
          magnets are removed from the lattice model of the ESRF storage 
          ring (last row). For the vertical tune line (not 
          shown) amplitude modulation and phase errors are about a factor 
          two and three lower, respectively.}
\rule{0mm}{3mm}
\end{figure}

There are two ways to estimate the maximum acceptable 
initial oscillation preventing nonlinearities from 
polluting the linear analysis of the tune line. The 
first is to evaluate the explicit expressions for $B$ and 
$h$ from the nonlinear lattice model. The second is 
to perform the harmonic analysis of single-particle tracking. 
To this end, it is enough to compare the tune line 
modulation and the deviation of the BPM phase advance 
$\Delta\Phi_{ij}$ from the betatron phase $\Delta\phi_{ij}$ 
for an ideal model with no focusing error. By repeating 
this test for several initial conditions, a threshold 
can be set to ensure a certain limit to the pollution. 
In Fig.~\ref{fig_NonLin1} the results of four different 
tests are reported for the horizontal tune line. The impact 
on the vertical tune line is weaker because of the 
stronger horizontal focusing and chromaticity which 
result in a stronger impact of nonlinearities in that 
plane. A single particle has been tracked 
through the ideal lattice (i.e. with no focusing error, 
 $f_{2000}=f_{0020}=0$ along the ring) 
of the ESRF storage ring for 1024 turns. The positions 
recorded at the 224 BPMs are Fourier analyzed and the tune 
line amplitude is used to evaluate the modulation of $|H(1,0)|$ 
(left plots) along the ring due to the nonlinear terms 
$F$ and $T$ of Eq.~\eqref{eq:nonlinT1}.  
The tune line phase is used to compute 
the BPM phase advance $\Delta\Phi_{ij}$ and its difference 
with the betatron phase advance $\Delta\phi_{ij}$ is 
reported in the right plots. The test is first run for 
an extremely low initial excitation of 100 $\mu$m in both 
planes (uppermost plot), then $(x_0,y_0)=(1.4,0.3)$ mm (second plot) 
and $(x_0,y_0)=(2.7,0.8)$ mm (third plot). 
Even though 
the tune line modulation is relatively modest, below 0.3$\%$ 
rms, an sizable deviation from the betatron phase advance 
of 2.6 mrad rms (6.6 mrad maximum), roughly corresponding to an 
rms {\sl artificial} beta beating of 1\%, is observed in the 
last case ($(x_0,y_0)=(2.7,0.8)$ mm). This means that 
the linear analysis carried out in Ref.~\cite{Andrea-arxiv}, 
which fitted quadrupolar errors from similar TBT data to best match 
$\Delta\phi_{ij}$ and $\Delta\phi_{ij}$ up to 0.9 mrad rms, was 
indeed erroneous, since a great part of the initial deviation 
came from the nonlinear terms of Eq.~\eqref{eq:nonlinT1} and 
not from quadrupolar errors. The confirmation that such deviations 
from the betatron parameters stem from nonlinear magnets is 
given in the lower most plot, where both the invariant and 
the betatron phase advance are retrieved with the same 
initial conditions $(x_0,y_0)=(2.7,0.8)$ mm after turning off 
all sextupoles and octupoles in the lattice model. 

The results of Fig.~\ref{fig_NonLin1} have important consequences 
on the possibility of using TBT data for a complete linear 
analysis of the ESRF storage ring. The above simulations indeed 
suggest to limit the initial oscillation amplitude to about 
1.4 mm horizontally and 0.3 mm vertically in order to limit the 
nonlinear contribution to the BPM phase to less than 0.5 mrad rms. 
On the other hand, in Sec.~\ref{subsec:resolution} it was shown 
that at this level of excitation the coupling analysis via 
the spectral lines $H(0,1)$ and $V(1,0)$ becomes inaccurate 
when the machine operates (as it has been doing since 2010) 
with ultra-low coupling.

\subsection{TBT analysis: which  model BPM phase advance?}

A possible way out to this dilemma can be found in replacing
as reference betatron phase advance the one computed from the 
lattice $\Delta\phi_{ij}$ (\verb ptc_twiss  command in MADX or 
\verb linopt  function in AT) with the BPM phase advance 
$\Delta\Phi_{ij}^{(SPT)}$ inferred from the harmonic analysis 
of single-particle tracking simulations with the same 
oscillation amplitudes of the measured data. The linear 
analysis could be then performed on the difference between 
measured and model BPM phase advances, namely
\begin{eqnarray}\nonumber
\delta\Delta\phi_{ij}(\delta K_1)\simeq&& 
\Delta\Phi_{ij}^{(meas)}(\delta K_1,K_2,K_3,I_{x,y})\ -\\
&&\hskip 0mm\Delta\Phi_{ij}^{(SPT)}(K_2,K_3,I_{x,y})
\quad ,\label{e:PhaseModel1}
\end{eqnarray}
since in first approximation (i.e. by ignoring sextupole 
and octupole errors) the amplitude dependent nonlinear 
terms $B$ and $h_{pprr}$ of Eq.~\eqref{eq:nonlinT1} would cancel 
out and the difference would depend on quadrupole errors 
only via $\delta\Delta\phi_{ij}$. 

\begin{figure}
\rule{0mm}{0mm}
\centerline{\includegraphics[width=8.5cm,height=5.5cm]{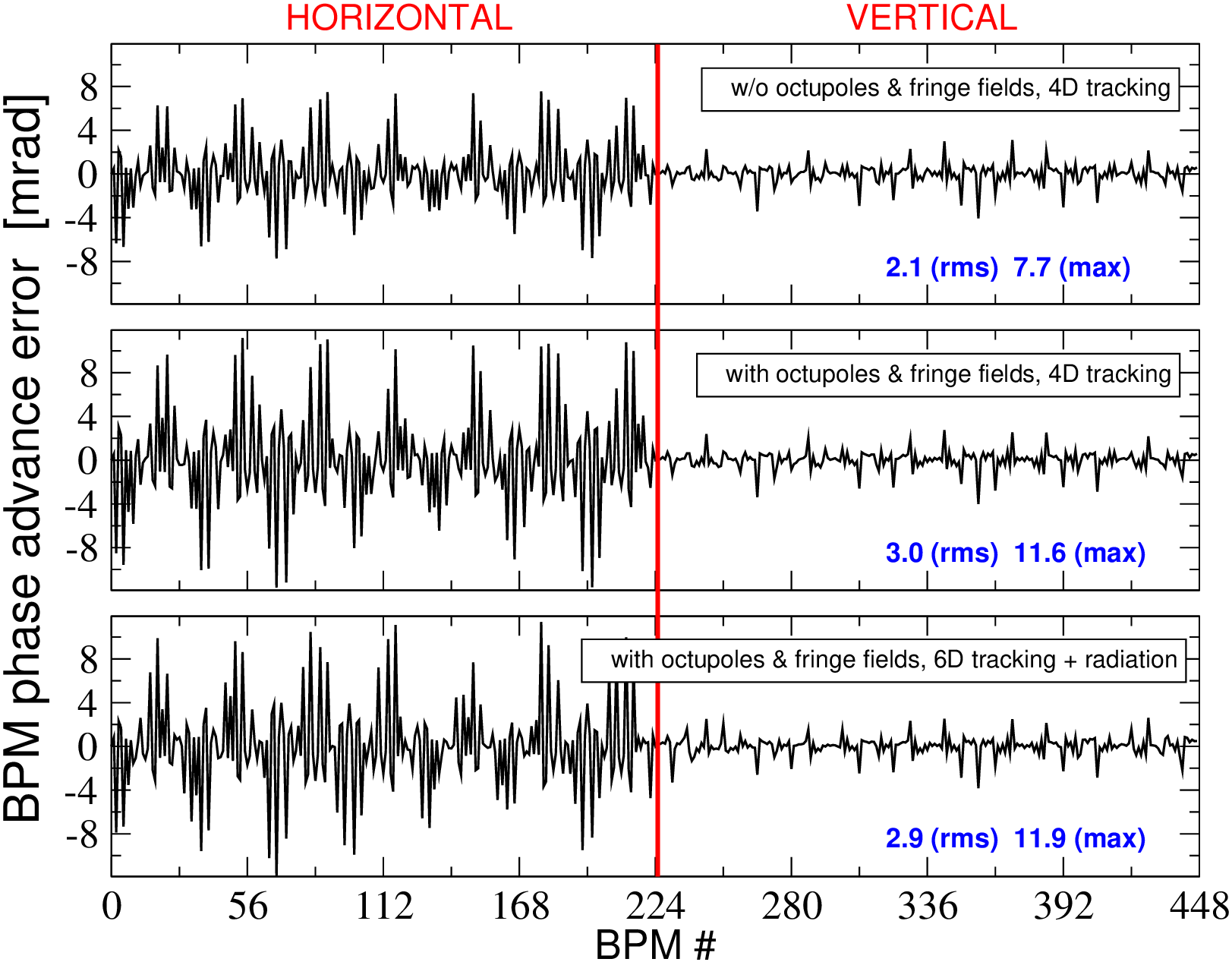}}
  \caption{\label{fig_NonLin3} (Color) BPM phase advance error 
           $\Delta\Phi_{ij}^{(SPT)}-\Delta\phi_{ij}$ evaluated from 
           single-particle simulations with initial conditions 
           $(x_0,y_0)=(2.7,0.8)$ mm against tracking parameters and additional 
           nonlinearities on top of the sextupoles: 4D tracking with ideal
           sextupole setting (top), including the same nonlinear lattice error 
           model (sextupole errors, octupolar field component in quadrupoles 
           and fringe fields) of Ref.~\cite{Andrea-arxiv} (center), and 
           6D tracking including radiation effects (bottom).}
\rule{0mm}{3mm}
\end{figure}

By doing so, however, the 
quality of the linear analysis would become dependent 
on the nonlinear lattice setting and model, as well as on 
tracking parameters, since they all affect 
$\Delta\Phi_{ij}^{(SPT)}$. In Fig.~\ref{fig_NonLin3} the 
results of three numerical tests are reported for the same 
(linear and nonlinear) optics of Fig.~\ref{fig_beat1}. The
center plot shows how actually the inclusion of nonlinear magnetic 
errors in the model does indeed influence $\Delta\Phi_{ij}^{(SPT)}$. 
A weak dependence on the inclusion of longitudinal tracking with 
radiation effects can be also observed in the lower most plot. 

\begin{figure}
\rule{0mm}{0mm}
\centerline{\includegraphics[width=8.5cm,height=5.5cm]{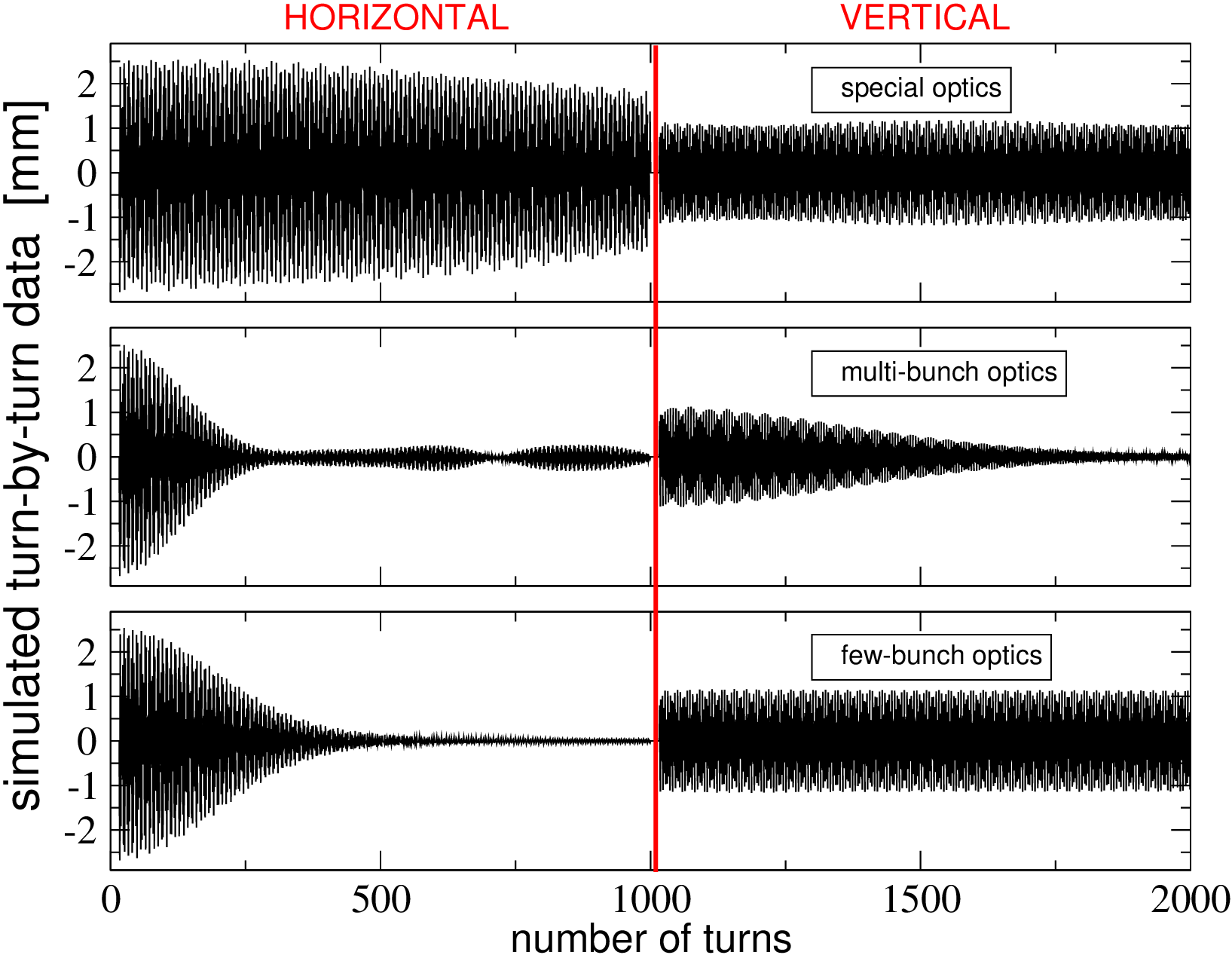}}
\vskip 2mm
\centerline{\includegraphics[width=8.5cm,height=5.5cm]{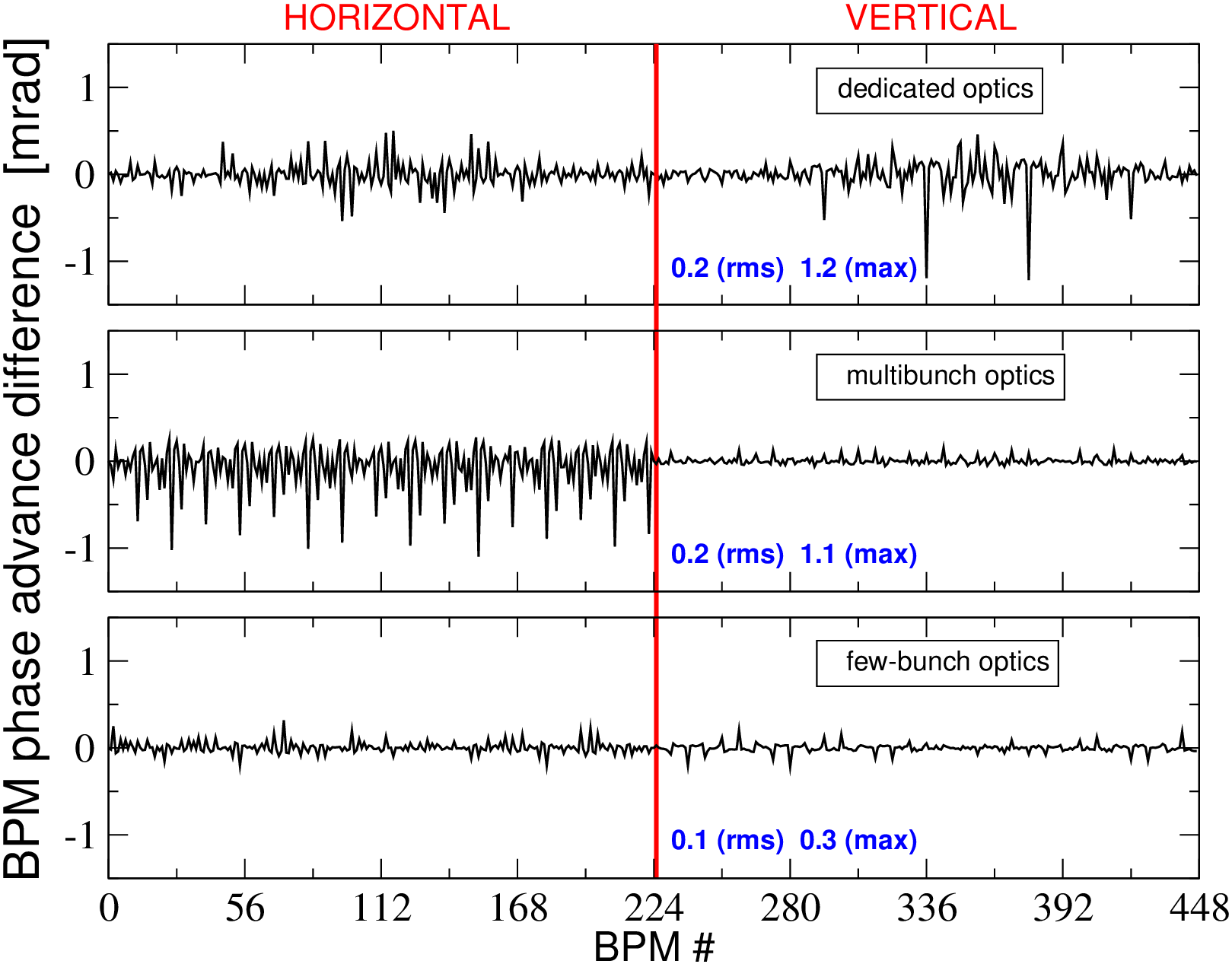}}
  \caption{\label{fig_NonLin2} (Color) Top: TBT beam centroid oscillation 
           obtained from multi-particle 4D tracking (Gaussian distribution, 
           $5\times10^4$ particles, $E_x=4$ nm, $E_y=4$ pm, $\sigma_p=0.1\%$). Three 
           various sextupole settings of the ESRF storage ring are 
           tested: low chromaticity and weak detuning with amplitude ({\sl 
           special optics}), low chromaticity and strong detuning 
           ({\sl multi-bunch optics}), and high vertical chromaticity and weak 
           detuning ({\sl few-bunch optics}). Bottom: corresponding BPM 
           phase advance deviation between the harmonic analysis of 
           a multi-particle TBT signals with respect its single-particle 
           counterpart, $\Delta\Phi_{ij}^{(MPT)}-\Delta\phi_{ij}^{(SPT)}$.}
\rule{0mm}{3mm}
\end{figure}

It can be argued that  the beam itself is not a single 
particle and that its multi-particle nature and its finite rms 
emittances ($E_x=4$ nm, $E_y=4$ pm, $\sigma_p=0.1\%$ for the 
ESRF electron beam) would require an even more realistic approach 
to account for the damping of the TBT signal of its centroid 
(i.e. decoherence) due to chromaticity, amplitude dependent 
detuning~\cite{dechoerence1,dechoerence2,dechoerence3} and 
possibly radiation effects. To this end, multi-particle tracking 
simulations and the harmonic analysis on the TBT motion 
of the beam centroid could be performed to infer the 
corresponding BPM phase advance
$\Delta\Phi_{ij}^{(MPT)}$.

In the top plots of Fig.~\ref{fig_NonLin2} the simulated 
TBT signal at one BPM is 
shown for three different sextupole settings of the ESRF storage ring:
with low chromaticity and detuning with amplitude ({\sl special 
optics}, typically used for TBT studies), with low chromaticity but 
large detuning ({\sl multi-bunch optics}), and high vertical 
chromaticity and low detuning ({\sl few-bunch optics}). Tracking 
is performed in the transverse plane only (4D) with frozen 
longitudinal motion and no radiation effects. The decoherence 
is much more visible in the horizontal plane because of the much 
larger horizontal emittance compared to the vertical plane. Simulations 
were run with no betatron coupling. In the bottom plots of 
Fig.~\ref{fig_NonLin2} the BPM phase difference between multi-particle 
and single-particle BPM phase advance, 
$\Delta\Phi_{ij}^{(MPT)}-\Delta\phi_{ij}^{(SPT)}$, 
is plotted along the ring : Deviations are more pronounced in 
the horizontal plane (as expected from the stronger decoherence) 
though they are a mere $10\%$ of $\Delta\phi_{ij}^{(SPT)}$ of 
Fig.~\ref{fig_NonLin3}.

\begin{figure}
\rule{0mm}{0mm}
\centerline{\includegraphics[width=8.5cm,height=5.5cm]{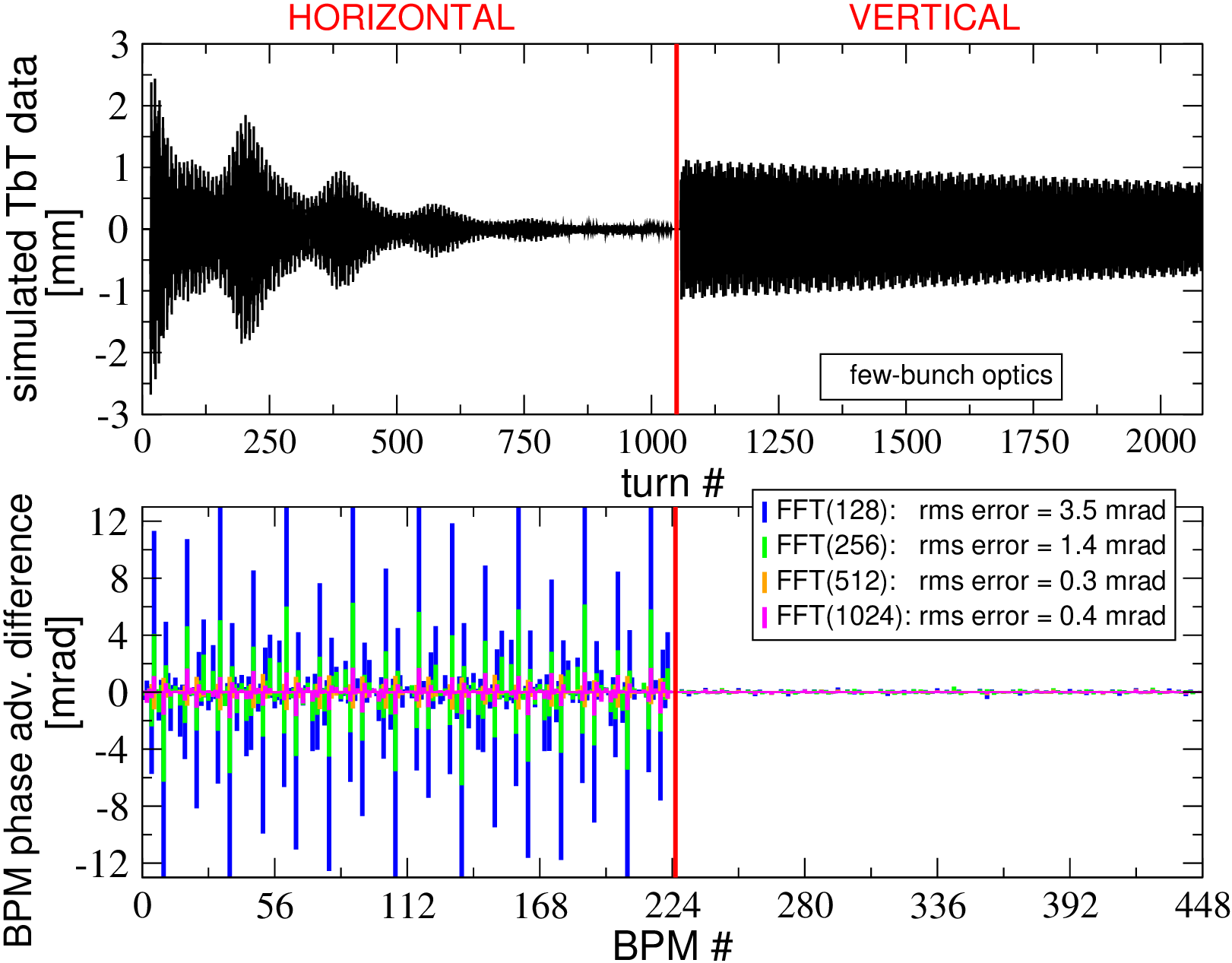}}
\rule{0mm}{0mm}
\centerline{\includegraphics[width=8.5cm,height=5.5cm]{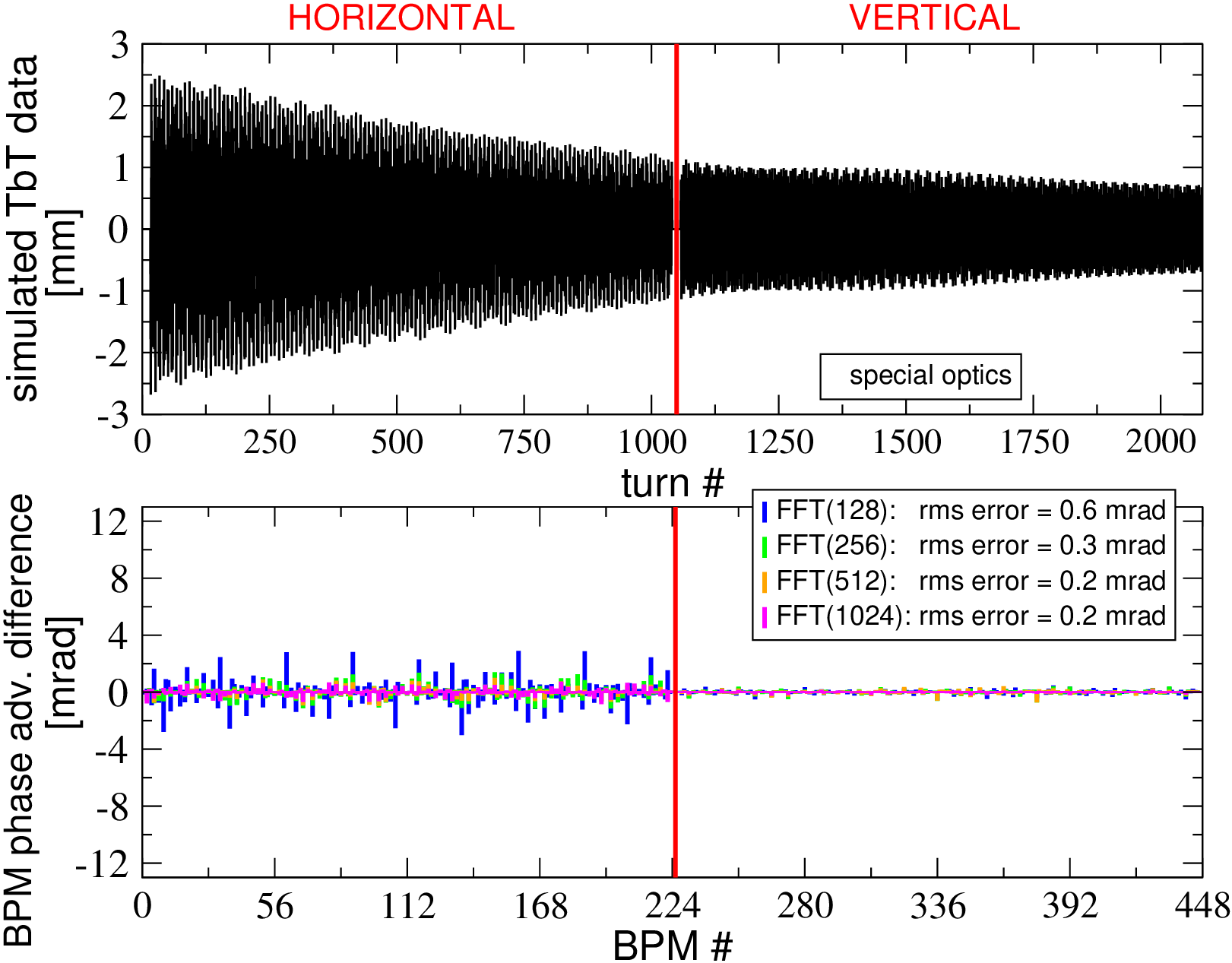}}
  \caption{\label{fig_NonLin4} (Color) BPM phase advance error 
            $\Delta\Phi_{ij}^{(MPT)}-\Delta\phi_{ij}^{(SPT)}$ evaluated 
            against the number of turns used in the harmonic analysis for 
            two different sextupole settings, with strong decoherence 
            (top) and negligible attenuation (bottom). The same multi-particle 
            simulations of Fig.~\ref{fig_NonLin2} is used, this time with a 
            full 6D tracking including radiation effects.}
\rule{0mm}{0mm}
\end{figure}

Since the spectral resolution of the 
harmonic analysis depends on the number of turns suitable 
to be Fourier-analyzed, the quality of the linear analysis 
is expected to increase with the number of turns. 
This effect is displayed in Fig.~\ref{fig_NonLin4}, where  
the difference $\Delta\Phi_{ij}^{(MPT)}-\Delta\phi_{ij}^{(SPT)}$ 
is showed for two different nonlinear optics against the 
number of turns used for the FFT. Differently from Fig.~\ref{fig_NonLin2}, 
tracking is here performed in all planes (6D) including 
radiation effects. In both cases, when 512 or more 
turns are analyzed multi-particle effects appear to account 
for a mere fraction of mrad. 

These and other multi-particle simulations confirm that 
the multi-particle effects are negligible compared to 
$\Delta\phi_{ij}^{(SPT)}$ and that the latter can be 
effectively used for a linear analysis via 
Eq.~\eqref{e:PhaseModel1}, provided that a solid nonlinear 
lattice model is available.

\subsection{TBT analysis: Can beta beating and ultra-low coupling be evaluated with 
           separate measurements?}
It can be argued that TBT analysis of focusing errors and 
betatron coupling at the ESRF storage ring could 
be carried out with two separate 
measurements: one at low excitation amplitude for the 
evaluation and correction of beta beating only (with no or 
limited {\sl pollution} of the tune lines by nonlinear 
terms), and a second with large oscillation to enhance the 
coupling spectral lines well above the noise floor. While 
the first measurement is perfectly feasible, two mains 
obstacles prevent the second from being viable. 

First, at large amplitudes nonlinear terms affect the
coupling spectral lines too. In Appendix~\ref{app:2B} 
analytic expressions for the coupling lines of the real 
signals  $\tilde{x}$ and $\tilde{y}$ including the 
leading amplitude dependent terms are derived:
\begin{eqnarray}\hspace{-1mm}
\left\{\begin{aligned}
&\hspace{-1mm}
 |H(0,1)_j|\hspace{-.8mm}=\hspace{-1mm}\left|F_{xy,j}(J_1)+
        T_{xy,j}(J_3,K_3,K_2^2,J_1,I_{x,y})\right|\hspace{-.8mm}\sqrt{2I_y}\\
&\hspace{-1mm}
 |V(1,0)_j|\hspace{-.8mm}=\hspace{-1mm}\left|F_{yx,j}(J_1)+
        T_{yx,j}(J_3,K_3,K_2^2,J_1,I_{x,y})\right|\hspace{-.8mm}\sqrt{2I_x}\\
&\hspace{-1mm}
 \hbox{arg}\left\{H(0,1)_j\right\}=\phi_{x,j}+\psi_{x0}+
                  \hbox{arg}\left\{F_{xy,j}+T_{xy,j}\right\}-\frac{\pi}{2}\\
&\hspace{-1mm}
 \hbox{arg}\left\{V(1,0)_j\right\}=\phi_{y,j}+\psi_{y0}+
                  \hbox{arg}\left\{F_{yx,j}+T_{yx,j}\right\}-\frac{\pi}{2}
\end{aligned}\right. \hspace{-2mm},
\nonumber \\ \label{eq:NLcoup4T}
\end{eqnarray}
where the betatron coupling terms $F_{xy}$ and $F_{yx}$ are 
the same of Eq.~\eqref{e:F_xy}. The complex nonlinear amplitude 
dependent coupling functions $T_{xy}$ and $T_{yx}$, defined in 
Eq.~\eqref{eq:NLcoup2}, scale linearly with the skew 
octupole gradient $J_3$, as well as with the products $K_3J_1$ 
(cross product between normal octupole and skew quadrupole 
strengths) and $K_2^2J_1$ (cross product between normal 
sextupole and skew quadrupole fields). Hence, even in 
the absence of physical octupoles, normal sextupoles 
excite $T_{xy}$ and $T_{yx}$ via betatron coupling. Since both 
$F_{xy}$ and $F_{yx}$ scale linearly with $J_1$ too, the overall 
amplitude dependent modulation of the coupling lines scales 
quadratically with the sextupole fields, i.e. with the same 
order of magnitude of the tune line modulation $B_{x,y}$ 
and $h_{pprr}$ of Eq.~\eqref{eq:nonlinT1}. The analysis 
of the coupling lines to evaluate betatron coupling 
at large amplitudes would then be corrupted by the machine 
nonlinearities in the same way the study of focusing 
errors from the tune lines would be.

Second, the tune lines are used to 
extract the coupling RDTs $f_{1001}$ and $f_{1010}$ (or 
their combined functions $F_{xy}$ and $F_{yx}$) via 
Eqs.~\eqref{e:F_xy}-\eqref{e:f1001_hadron}. If a large 
excitation is imparted to generate measurable coupling 
lines, the nonlinear terms contributing to the tune line 
amplitudes and phase, $B_{x,y}$ of Eq.~\eqref{eq:nonlinT1}, 
would corrupt the evaluation of the coupling RDTs, since 
$|H(1,0)|\ne\sqrt{2I_x}$, $|V(0,1)|\ne\sqrt{2I_y}$, 
$\hbox{arg}\{H(1,0)\}\ne\phi_x+\psi_{x0}$ and 
$\hbox{arg}\{V(0,1)\}\ne\phi_y+\psi_{y0}$.

\begin{figure}
\rule{0mm}{0mm}
\centerline{\includegraphics[width=8.5cm]{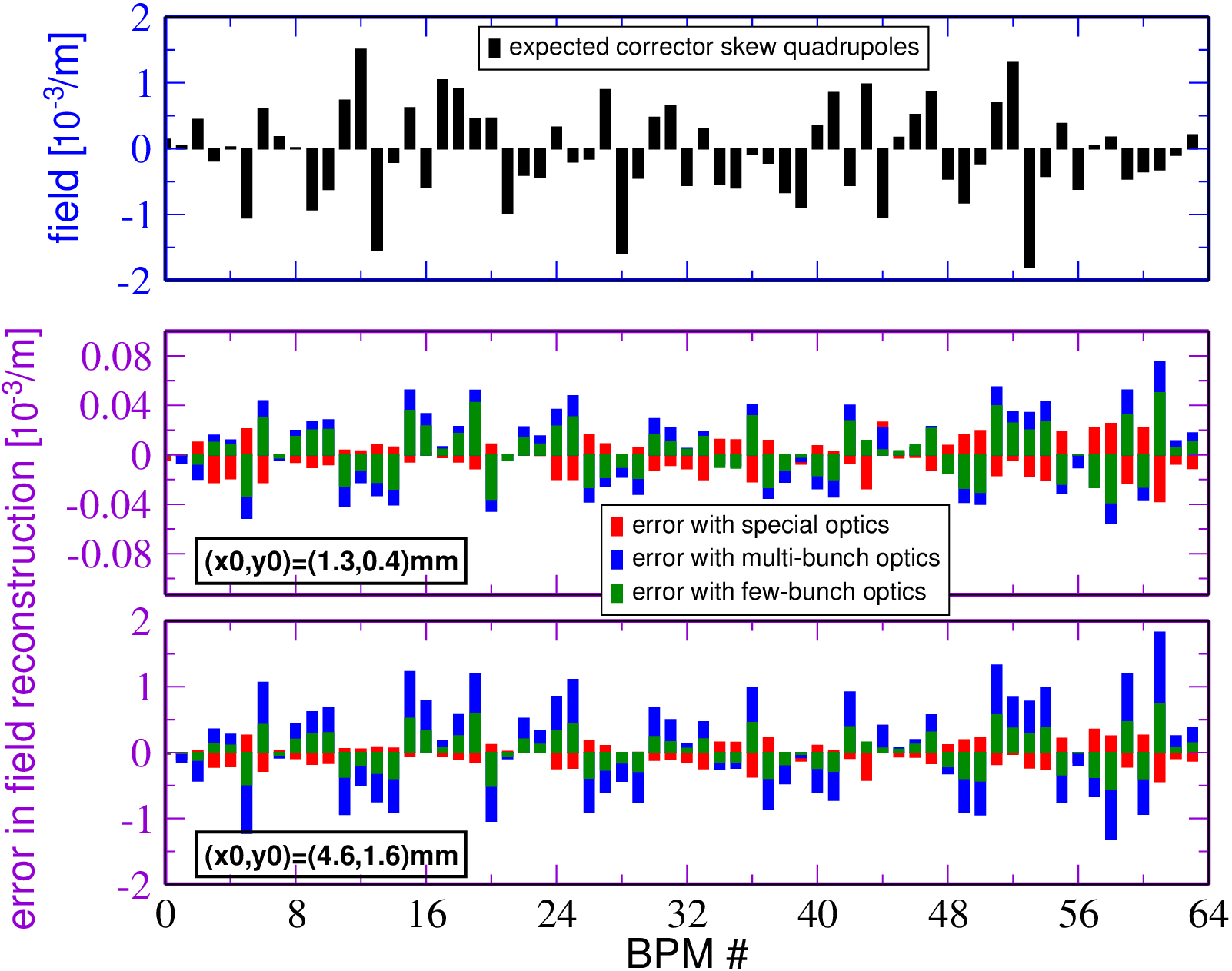}}
  \caption{\label{fig_coup1} (Color) Example of simulated betatron 
    coupling analysis corrupted by nonlinear terms. The coupling lines 
    $H(0,1)$ and $V(1,0)$ of simulated single-particle TBT BPM data with 
    the lattice of the ESRF storage ring comprising 64 skew quadrupoles 
    (top plot) are assumed to be excited by linear coupling functions 
    $F_{xy}$ and $F_{yx}$ only. They are inferred via Eq.~\eqref{e:F_xy} 
    and the skew quadrupole strengths retrieved by pseudo-inverting 
    the system via Eq.~\eqref{eq:f}. At low amplitude the difference 
    between set and reconstructed skew quadrupole fields is of a few 
    percent (center plot), while at large initial excitation it increases 
    to 25\% and 100\% depending on the sextupole settings (bottom plot). When 
    sextupoles are turned off in the model (which does not comprise 
    octupoles), the errors remains well below 0.5\% (not shown), 
    irrespective of the initial conditions.}
\rule{0mm}{3mm}
\end{figure}

An example of betatron coupling analysis corrupted by 
nonlinear terms is shown in Fig.~\ref{fig_coup1}. The 
harmonic decomposition is carried out on simulated 
single-particle TBT BPM data with the lattice of the 
ESRF storage ring comprising 64 skew quadrupoles 
distributed along the ring, whose normalized integrated 
strengths $J_1$ are reported in the upper plot. By assuming that 
the coupling lines $H(0,1)$ and $V(1,0)$ are generated only 
by betatron coupling terms $F_{xy}$ and $F_{yx}$, respectively, 
the latter are inferred and used to extract the strengths 
of the 64 skew quadrupoles by pseudo-inverting the system of 
Eqs.~\eqref{e:F_xy} and~\eqref{eq:f}. This exercise has 
been repeated for three different sextupoles settings and at 
diverse initial excitations. The errors remain in the 
few percent level when the initial displacement is of about 
1 mm (center plot), though this amplitude is too low for 
real measurements. However, when simulating oscillation 
amplitudes sufficient to generate measurable coupling 
spectral lines, the rms error ranges from 25\% to 
100\% depending on the nonlinear setting (bottom 
plot). When turning off all sextupoles in the lattice (along with 
any other nonlinearities) the rms error remains well below 0.5\% 
for any initial condition. 


\section{Conclusion}
\label{sec:conclusion}

Analytic formulas for the evaluation of linear lattice parameters 
from either turn-by-turn beam position data or an error lattice model 
have been derived and used to perform an error analysis. This 
study also presented a procedure for the estimation of detrimental 
effects of nonlinear terms stemming from sextupoles and higher order 
multipole magnets. These may result in a wrong evaluation of the 
BPM phase advance and hence of the focusing errors. Preliminary 
single-particle simulations however would suffice to properly 
account for such nonlinear terms. It has been also shown how 
beam decoherence does not corrupt the evaluation of the BPM phase 
advance, the error found in multi-particle simulations being 
the same determined by single-particle tracking (within a $10\%$ 
uncertainty). 

The elements presented in this paper indicate that for the ESRF 
electron storage ring operating with ultra-low coupling and making 
use of the Libera {\sl Brilliance} BPMs, the analysis of the linear 
lattice errors (focusing and coupling) is limited in its accuracy 
and precision by several factors. As a rule of thumb, 
3 mrad of rms BPM phase advance error correspond to about $1\%$ 
of rms beta beating. The ORM analysis is to be preferred to the 
harmonic study of TBT data for several reasons:
\begin{itemize}
\item The ORM analysis of focusing errors and coupling requires 
      a maximum beam excitation of about 250~$\mu$m well within 
      the linear regime of the betatron motion.
\item Because of the natural beam motion (vibrations), the 
      worse BPM resolution when operating in TBT 
      mode and the need of measuring an ultra-low coupling, the 
      harmonic analysis requires a minimum beam excitation of some 
      mm, reaching a region of the betatron motion where magnetic 
      nonlinearities reduce the measurement accuracy in the range 
      of 2-6 mrad for the rms BPM phase advance error and of 
      $1\%-2\%$ in the evaluation of the beta beating (depending on 
      the sextupole settings). If nonlinear 
      terms are not taken into account, the inferred quadrupolar 
      errors would wrongly account for sextupolar and octupolar 
      contributions to the betatron motion.  
\item Formulas for the evaluation of beta beating from TBT data 
      are affected by intrinsic errors at the level of $\sim 0.3\%$ 
      rms ($\sim2\%$ peak to peak) if the BPM phase advance is used 
      (and a perferct synchronization between the monitors is assumed),
      whereas if the tune line amplitude is used (and BPM calibration 
      factors inferred from orbit data can be trusted) the uncertainty 
      is of $\sim 0.6\%$ rms ($\sim3\%$ maximum). No error estimate 
      for the ORM analysis has been performed so far.
\end{itemize}

The above numbers may clearly vary in other ring-based light 
sources, though recent comparisons between ORM and TBT analysis 
of linear lattice errors in other facilities report similar 
uncertainties~\cite{Soleil,Alba}.

The same considerations made here for the FFT-based analysis of 
TBT data should apply the the techniques of 
Refs.~\cite{SVD1,SVD2,MIA1,MIA2,ICA}. Indeed the nonlinearities of 
Eq.~\eqref{eq:nonlinT1} affecting the tune lines are expected 
to alter the {\sl betatron modes} which are used for the 
evaluation of focusing errors. In these schemes, in fact, it 
is assumed that any deviation from the ideal betatron modes, 
i.e. those oscillating at the frequency of the linear tunes, 
would stem from quadrupolar errors, whereas Eq.~\eqref{eq:nonlinT1} 
suggests that at {\sl large} amplitudes (needed for the 
analysis of ultra-low coupling) the impact of nonlinear 
effects needs to be assessed. Hence, as for the harmonic 
analysis, if nonlinearities are ignored, the 
inferred quadrupolar errors would wrongly account for 
sextupolar and octupolar contributions to the betatron 
motion. 

A last consideration worth to be made concerns the new 
ESRF storage ring under design (and possibly any new 
light source with sub-nm natural horizontal emittance). The 
new lattice design is expected to provide a natural emittance 
of about 130 pm (4 nm today) and features beta functions 
globally much smaller than in the existing machine. Today  
the lowest beta function at the BPMs is of about 5.6 m. 
In the new machine two BPMs (out of ten) per cell are 
located in regions with $\beta_x=1.1$ m ($\beta_y=3.3$ m), 
and $\beta_x=1.9$ m  ($\beta_y=2.3$ m) in other two monitors. 
In the future storage ring the BPM electronics remains 
based on the existing Libera {\sl Brilliance} hardware 
(additional Libera {\sl Spark} modules will be installed to 
cover the increased number of monitors). This means that in 
order to preserve today's spectral resolution, the initial 
beam excitation, i.e. the invariant, shall be larger by about 
a factor two, in a machine which is by far more nonlinear than 
the existing one. This casts even stronger concerns on the 
possibility of measuring and correcting linear optics (focusing 
errors and betatron coupling) via TBT data in the 
upcoming storage ring. 

The above conclusions are expected not to apply to hadron 
circular accelerators with less aggressive 
focusing lattices and larger regions of the betatron 
phase space, i.e. the invariants, within the linear optics 
regime.

\section{Acknowledgment}
I am deeply indebted with Rogelio Tom\'as for inspiring 
this work. I am also grateful to him and Reine Versteegen 
for reading the original manuscript and for providing precious 
comments and suggestions during its preparation.





\appendix
\begin{widetext}
\section{Linear lattice parameters with focusing errors}
\label{app:1}
In Appendix C of Ref.~\cite{Andrea-arxiv} a non-truncated expression for the 
tune lines $H(1,0)$ and $V(0,1)$ is derived assuming ultra low coupling, i.e. 
that coupling RDTs are negligible compared to the ones excited by focusing 
errors, $f_{2000}$ and $f_{0020}$. A second, though not less important, 
assumption is that the impact on the tune lines from nonlinear RDTs is 
negligible, i.e. that the oscillation amplitudes ($2I_{x,y}$) are low enough 
to prevent octupolar-like RDT $f_{3100}$, $f_{2011}$, $f_{0031}$ and 
$f_{1120}$ from contributing to $H(1,0)$ and $V(0,1)$, See Table V-VII of 
Ref.~\cite{Bartolini1}. Note that the spectral lines reported there 
refer to the complex signals $h_x=\tilde{x}-i\tilde{p}_x$ and 
$h_y=\tilde{y}-i\tilde{p}_y$, for which the above octupolar-like RDTs 
excite the lines $H_h(-1,0)$ and $V_h(0,-1)$, hence introducing   
amplitude-dependent focusing errors. Since the harmonic analysis 
is performed here on the real signals  $\tilde{x}$ and $\tilde{y}$, those 
lines {\sl pollute} the tune peaks, since $H(1,0)=1/2[H_h(1,0)+H_h^*(-1,0)]$ 
and $V(0,1)=1/2[V_h(0,1)+V_h^*(0,-1)]$. A more detailed discussion is 
made in the Appendices C and D of Ref.~\cite{Andrea-arxiv}. A third 
condition is that Hamiltonian octupolar-like terms $h_{2200}(2I_x)^2$, 
$h_{1111}(2I_x)(2I_y)$ and $h_{0022}(2I_y)^2$ can be neglected, as 
they would introduce amplitude-dependent detuning and shifts of the 
betatron phase unrelated to linear lattice errors. It is worthwhile 
reminding that such nonlinear resonant and detuning terms are generated 
by octupole magnets (to the first order) as well as by sextupoles 
(to the second order) and are in general much stronger in light 
sources than in hadron machines (because of the higher natural 
chromaticity). Unless specified, throughout this appendix only ideal 
BPMs with calibration factors $\mathcal{C}_{x,y}=1$ are considered. 
Under these three important assumptions the tune lines at a 
generic BPM $j$ read
\begin{eqnarray}
\left\{
\begin{aligned}
&H(1,0)_j=\frac{1}{2}\Big[\cosh{(4|f_{2000,j}|)}
                    +i\sinh{(4|f_{2000,j}|)}e^{-iq_{2000,j}}\Big] 
      \sqrt{2I_x}e^{i(2\pi NQ_x+\phi_{x,j}^{(mod)}+\psi_{x0})}\\
&V(0,1)_j=\frac{1}{2}\Big[\cosh{(4|f_{0020,j}|)}
                    +i\sinh{(4|f_{0020,j}|)}e^{-iq_{0020,j}}\Big]
      \sqrt{2I_y}e^{i(2\pi NQ_y+\phi_{y,j}^{(mod)}+\psi_{y0})}
\end{aligned}\right. \quad ,\qquad\label{tune-beat}
\end{eqnarray}
where 
\begin{eqnarray}
\left\{
\begin{aligned}
&f_{2000,j}=\frac{\sum\limits_w^W \beta_{x,w}^{(mod)}\delta K_{w,1}
                e^{2i\Delta\phi_{x,wj}^{(mod)}}}
        {8(1-e^{4\pi iQ_x})}+O(\delta K_{1}^2)\\ 
&q_{2000,j}=\hbox{arg}\left\{f_{2000,j}\right\}
\end{aligned}\right. \quad , \qquad
\left\{
\begin{aligned}
&f_{0020,j} =-\frac{\sum\limits_w^W \beta_{y,w}^{(mod)}\delta K_{w,1}
                e^{2i\Delta\phi_{y,wj}^{(mod)}}}
        {8(1-e^{4\pi iQ_y})}+O(\delta K_{1}^2) \\
&q_{0020,j} = \hbox{arg}\left\{f_{0020,j}\right\} 
\end{aligned}\right.\ , \qquad  \label{eq:def_f0020}
\end{eqnarray}
with $\beta^{(mod)}$ and $\Delta\phi^{(mod)}$ are the the 
Courant-Snyder (C-S) parameters 
of the ideal lattice (i.e. without focusing errors $\delta K_{w,1}$). 
$O(\delta K_{1}^2)$ denotes the remainder proportional to the square of 
the focusing errors. The above sums run over all $W$ quadrupole 
errors along the ring, and $\Delta\phi_{wj}$ denotes the phase advance 
between the magnet $w$ and the BPM $j$. Eq.~\eqref{tune-beat} may be 
rewritten as
\begin{eqnarray}
H(1,0)_j&=&\frac{1}{2}\sqrt{2I_x}A_{f,j}
           e^{i(2\pi NQ_x+\phi_{x,j}^{(mod)}+\theta_{f,j})}
\  ,\qquad\label{tunex-beat2}\\
A_{f,j}&=&\left(1 +2\sinh{(4|f_{2000,j}|)}\Big[ 
                \sinh{(4|f_{2000,j}|)}+\cosh{(4|f_{2000,j}|)}\sin{q_{2000,j}} \Big]\right)^{1/2} 
       ,\ \label{tunex-beat3} \\
\theta_{f,j}&=&
             \tan^{-1}\left\{\frac{\sinh{(4|f_{2000,j}|)}\cos{q_{2000,j}}}
             {\cosh{(4|f_{2000,j}|)}+\sinh{(4|f_{2000,j}|)}\sin{q_{2000,j}}}
             \right\}+\psi_{x0}\ .\label{tunex-beat4}
\end{eqnarray}
The $s$-dependent term $A_{f}$ represents the phase space deformation 
induced by focusing errors not included in the model. With the ideal 
lattice the $s$-dependent phase space ellipses of the Cartesian 
coordinates are mapped into circles of constant radius $\sqrt{2I_x}$ when 
moving in the C-S coordinates. With lattice errors not included in the 
model, the C-S transformation with the ideal C-S parameters will map the 
initial ellipses in other ellipses whose semi-axis depend on $A_{f}$. Only 
when those errors are included in the model $f_{2000}=0$ and 
$A_{f}=1$ along the ring, and the phase space circles are retrieved 
with the new C-S parameters and transformation. \\

{\bf Analytic formulas for the beta beating:} 
By comparing Eq.~\eqref{tunex-beat2} and Eq.~\eqref{eq:tunes2} (assuming 
an ideal BPM calibration factor, $\mathcal{C}=1$), the measured 
$\beta$ can be interpreted as the initial model $\beta^{(mod)}$ 
modified by the focusing errors via the RDTs so to have  
tune line amplitude constant along the ring and equal to $\sqrt{2I}$. This 
is equivalent to say that at the BPM $j$ 
$\beta_{x,j}=\beta_{x,j}^{(mod)}A_{f,j}^2$, i.e. 
\begin{eqnarray}
\left\{
\begin{aligned}
&\beta_{x,j}=\beta_{x,j}^{(mod)}\left\{1 +2\sinh{(4|f_{2000,j}|)}
            \Big[\sinh{(4|f_{2000,j}|)}+\cosh{(4|f_{2000,j}|)}\sin{q_{2000,j}}
            \Big]\right\}\\
&\beta_{y,j}=\beta_{y,j}^{(mod)}\left\{1 +2\sinh{(4|f_{0020,j}|)}
            \Big[\sinh{(4|f_{0020,j}|)}+\cosh{(4|f_{0020,j}|)}\sin{q_{0020,j}}
            \Big]\right\}
\end{aligned}\right.\quad , \label{beta-beat1}
\end{eqnarray}
where the expression for the vertical plane follows from the same 
interpretation of $V(0,1)$ in Eq.~\eqref{tune-beat}. The beta beating 
then reads
\begin{eqnarray}
\left\{
\begin{aligned}
&\frac{\Delta\beta_{x,j}}{\beta_{x,j}}=2\sinh{(4|f_{2000,j}|)}
            \Big[\sinh{(4|f_{2000,j}|)}+\cosh{(4|f_{2000,j}|)}\sin{q_{2000,j}}
            \Big]\\
&\frac{\Delta\beta_{y,j}}{\beta_{y,j}}=2\sinh{(4|f_{0020,j}|)}
            \Big[\sinh{(4|f_{0020,j}|)}+\cosh{(4|f_{0020,j}|)}\sin{q_{0020,j}}
            \Big]
\end{aligned}\right.\quad . \label{beta-beat2}
\end{eqnarray}
To the first order in the RDTs the hyperbolic functions can be 
truncated to their leading terms,
\begin{eqnarray}
\left\{
\begin{aligned}
&\beta_{x,j}=\beta_{x,j}^{(mod)}
           \left(1 +8\Im\left\{f_{2000,j}\right\}\right)+O(|f_{2000}|^2)\\
&\beta_{y,j}=\beta_{y,j}^{(mod)}
           \left(1 +8\Im\left\{f_{0020,j}\right\}\right)+O(|f_{0020}|^2)
\end{aligned}\right.\quad ,\quad
\left\{
\begin{aligned}
&\left(\frac{\Delta\beta_x}{\beta_x}\right)_j
           = 8\Im\left\{f_{2000,j}\right\}+O(|f_{2000}|^2)\\
&\left(\frac{\Delta\beta_y}{\beta_y}\right)_j
           = 8\Im\left\{f_{0020,j}\right\}+O(|f_{0020}|^2)
\end{aligned}\right.
\quad .\quad\label{beta-beat2B}
\end{eqnarray}
$O(|f|^2)$ denotes the remainder proportional to the square of the RDT 
amplitude. To the first order in $\delta K_1$ the RDTs can be substituted 
by Eq.~\eqref{eq:def_f0020},  yielding
\begin{eqnarray}
\left\{
\begin{aligned}
&\left(\frac{\Delta\beta_x}{\beta_x}\right)^{(1)}_j
           \simeq+\frac{1}{2\sin{(2\pi Q_x)}}
             \sum\limits_w^W \beta_{x,w}^{(mod)}\delta K_{w,1}
                           \cos{(2|\Delta\phi_{x,wj}^{(mod)}|-2\pi Q_x)}\\
&\left(\frac{\Delta\beta_y}{\beta_y}\right)^{(1)}_j
           \simeq-\frac{1}{2\sin{(2\pi Q_y)}}
             \sum\limits_w^W \beta_{y,w}^{(mod)}\delta K_{w,1}
                           \cos{(2|\Delta\phi_{y,wj}^{(mod)}|-2\pi Q_y)}
\end{aligned}\right.\hskip -10mm , \hskip 10mm\label{beta-beat3}
\end{eqnarray}
which are the standard textbook formulas. \\

{\bf Analytic formulas for the phase shift.} 
The phase space deformation introduced by $A_f$ in Eq.~\eqref{tunex-beat2} 
and the resulting beta beating are accompanied by local jumps (or shifts) 
of the betatron phase with respect the ideal one. These are generated by 
the phase space deformation induced by the RDTs via the {\sl s}-dependent 
term $\theta_f$ in Eq.~\eqref{tunex-beat2} and by detuning Hamiltonian 
coefficient $h_{1100}$ ($h_{0011}$ in the vertical plane) which does not 
alter the phase space topology. The tune $Q_x$ in Eq.~\eqref{tunex-beat2} 
is indeed equal to the ideal one minus the derivative of all additional 
phase-independent Hamiltonian terms,
\begin{eqnarray}
Q_x&=&Q_x^{(mod)}-\frac{1}{2\pi}\frac{\partial<H>_{\phi}}{\partial I_x}
   =Q_x^{(mod)}-\frac{1}{2\pi}\frac{\partial h_{1100}(2I_x)}{\partial I_x} 
    +O(I_x)\simeq Q_x^{(mod)}-\frac{1}{\pi}h_{1100}
\qquad ,\label{phase-shift-1}\\
h_{1100}&=&-\frac{1}{4}\sum_{w=1}^W{\beta_{x,w}^{(mod)}\delta K_{w,1}}+
        O(\delta K_{1}^2)
\qquad , \qquad \label{phase-shift-2}
\end{eqnarray}
where the remainder $O(I_x)$ includes amplitude-dependent octupolar-like 
detuning not discussed here, and $O(\delta K_{1}^2)$ denotes the second 
order contribution to detuning from quadrupole errors (as well as from 
coupling) which is neglected in the following derivation though it can 
be computed as shown in Ref.~\cite{Andrea-arxiv}. In the vertical plane 
the following relations apply 
\begin{eqnarray}
Q_y&=&\simeq Q_y^{(mod)}-\frac{1}{\pi}h_{0011}
\qquad ,\qquad
h_{0011}=+\frac{1}{4}\sum_{w=1}^W{\beta_{y,w}^{(mod)}\delta K_{1,w}}+
        O(\delta K_{1}^2)
\qquad . \qquad \label{phase-shift-3}
\end{eqnarray}
The betatron phase computed by any optics code refers always to the 
origin ({\sl s}=0). When comparing the betatron phases with and without 
lattice errors, it shall be noted that with errors the initial phase is 
not zero with respect to the ideal case, namely
\begin{eqnarray}
\left\{
\begin{aligned}
\phi_{x,s=0}&= \theta_{f_{2000},s=0}\\
\phi_{x,j}&=\phi_{x,j}^{(mod)}-2h_{1100,j} + \theta_{f_{2000},j}-\phi_{x,s=0}
\end{aligned}\right.
\quad , \quad\left\{
\begin{aligned}
\phi_{y,s=0}&= \theta_{f_{0020},s=0}\\
\phi_{y,j}&=\phi_{y,j}^{(mod)}-2h_{0011,j} + \theta_{f_{0020},j}-\phi_{y,s=0}
\end{aligned}\right.
\quad .\quad \label{phase-shift-4}
\end{eqnarray}
The {\sl s}-dependent terms $h_{1100,j}$ and $h_{0011,j}$ include the 
focusing errors from the origin ($s$=0) and the BPM $j$: 
\begin{eqnarray}
h_{1100,j}=-\frac{1}{2}\sum_{w=1}^{W<j}{\beta_{x,w}^{(mod)}\delta K_{w,1}}
           +O(\delta K_1^2)\quad ,\quad 
h_{0011,j}=+\frac{1}{2}\sum_{w=1}^{W<j}{\beta_{y,w}^{(mod)}\delta K_{w,1}}
           +O(\delta K_1^2)\quad .\quad\label{phase-shift-5}
\end{eqnarray}
Note that even if the final detuning is zero (in practice two or more 
dedicated quadrupole families are trimmed so to have the desired ideal 
tunes), $h_{1100,j}$ and $h_{0011,j}$ are in general nonzero along the 
ring. By manipulating Eq.~\eqref{phase-shift-4} the phase shift 
then reads 
\begin{eqnarray}
\left\{
\begin{aligned}
\phi_{x,j}-\phi_{x,j}^{(mod)}&=-2h_{1100,j} 
             +\theta_{f_{2000},j}-\theta_{f_{2000},s=0}\\
\phi_{y,j}-\phi_{y,j}^{(mod)}&=-2h_{0011,j} 
             +\theta_{f_{0020},j}-\theta_{f_{0020},s=0}
\end{aligned}\right.
\quad .\quad \label{phase-shift-6}
\end{eqnarray}
The truncation to the first order in the RDT of $\theta_f$ from 
Eq.~\eqref{tunex-beat4} reads
\begin{eqnarray}\displaystyle
\theta_{f_{2000},j}\simeq 
               \tan^{-1}\left\{\frac{4|f_{2000,j}|\cos{q_{2000,j}}}
             {1+4|f_{2000,j}|\sin{q_{2000,j}}}\right\}
           \simeq\tan^{-1}{(4|f_{2000,j}|\cos{q_{2000,j}})}
           \simeq 4|f_{2000,j}|\cos{q_{2000,j}}\simeq 4\Re\{f_{2000,j}\}
\quad .\qquad \label{phase-shift-7}
\end{eqnarray}
The equivalent approximation in the vertical plane yields to 
$\theta_{f_{0020},j}\simeq4\Re\{f_{0020,j}\}$. Eq.~\eqref{phase-shift-6} 
then simplifies to 
\begin{eqnarray}
\left\{
\begin{aligned}
\phi_{x,j}-\phi_{x,j}^{(mod)}&=-2h_{1100,j} 
             +4\Re\left\{f_{2000,j}-f_{2000,s=0}\right\}+O(|f_{2000}|^2)\\
\phi_{y,j}-\phi_{y,j}^{(mod)}&=-2h_{0011,j} 
             +4\Re\left\{f_{0020,j}-f_{0020,s=0}\right\}+O(|f_{0020}|^2)
\end{aligned}\right.
\quad .\quad \label{phase-shift-8}
\end{eqnarray}
The shift of the BPM phase advance can be computed from the above 
expressions
\begin{eqnarray}
\left\{
\begin{aligned}
\Delta\phi_{x,ij}&=\Delta\phi_{x,ij}^{(mod)}-2h_{1100,ij} 
             +4\Re\left\{f_{2000,j}-f_{2000,i}\right\}+O(|f_{2000}|^2)\\
\Delta\phi_{y,ij}&=\Delta\phi_{y,ij}^{(mod)}-2h_{0011,ij} 
             +4\Re\left\{f_{0020,j}-f_{0020,i}\right\}+O(|f_{0020}|^2)
\end{aligned}\right.
\ ,\  
\left\{
\begin{aligned}
h_{1100,ij}&=-\frac{1}{2}\sum_{i<w<j}{\beta_{x,w}^{(mod)}\delta K_{w,1}}
             +O(\delta K_1^2) \\
h_{0011,ij}&=+\frac{1}{2}\sum_{i<w<j}{\beta_{y,w}^{(mod)}\delta K_{w,1}}
             +O(\delta K_1^2)
\end{aligned}\right. \hskip -1mm ,\nonumber\\ \label{phase-shift-9}
\end{eqnarray}
where the above sums extend over the focusing errors between the 
two BPMs $i$ and $j$ only. Since the latter monitor is donwstream
the former, i.e. $s_j>s_i$, the above sum is well defined.
Explicit expressions truncated to the first 
order in $\delta K_1$ similar to Eq.~\eqref{beta-beat3} can be retrieved 
after substituting the RDTs in the above equations with 
Eq.~\eqref{eq:def_f0020}. \\

{\bf Analytic formulas for the alpha shift.} 
The last C-S parameters to be evaluated is $\alpha=-1/2\beta'$, where 
the derivative is with respect to $s$. The beta function is the one of 
Eq.~\eqref{beta-beat1}. The derivative can be written as 
\begin{eqnarray}
\beta'(\beta^{(mod)},|f|,q)=
     \frac{\partial\beta}{\partial\beta^{(mod)}}\beta'^{(mod)}+
     \frac{\partial\beta}{\partial|f|}|f|'+
     \frac{\partial\beta}{\partial q}q'\qquad,\label{alpha-shift-1}
\end{eqnarray}
Two approximations are made here to simplify the mathematical derivation. 
The first is that $0\simeq|f|'\ll\beta'^{(mod)},q'$ and corresponds to 
the fact that the variation along the ring of $|f|$ can be neglected, 
this being much smaller than the one of oscillating functions $\beta$ 
and $q$. The second is that $q\simeq 2\phi^{(mod)}$, see 
Eq.~\eqref{eq:def_f0020}. Both are actually 
exact conditions along regions free of focusing errors, as proved in 
Ref.~\cite{Andrea-thesis}. From Eq.~\eqref{beta-beat1} and the above 
relation, $\alpha$ reads
\begin{eqnarray}
\alpha_{x,j}&\simeq&-\frac{1}{2}\beta_{x,j}'^{(mod)}\left\{1 +2\sinh{(4|f_{2000,j}|)}
            \Big[\sinh{(4|f_{2000,j}|)}+\cosh{(4|f_{2000,j}|)}\sin{q_{2000,j}}
            \Big]\right\} \nonumber \\
            &&-\beta_{x,j}^{(mod)}\sinh{(4|f_{2000,j}|)}\cosh{(4|f_{2000,j}|)}
              \cos{q_{2000,j}}q_{2000,j}'
\qquad.\label{alpha-shift-2}
\end{eqnarray}
Since $q'\simeq(2\phi)'=2/\beta^{(mod)}$ and 
$\frac{1}{2}\beta'^{(mod)}=\alpha^{(mod)}$, the above expression becomes 
\begin{eqnarray}
\alpha_{x,j}&\simeq&\alpha_{x,j}^{(mod)}\left\{1 +2\sinh{(4|f_{2000,j}|)}
            \Big[\sinh{(4|f_{2000,j}|)}+\cosh{(4|f_{2000,j}|)}\sin{q_{2000,j}}
            \Big]\right\}
            -\sinh{(8|f_{2000,j}|)}\cos{q_{2000,j}}
\quad,\nonumber 
\end{eqnarray}
and the alpha shifts $\Delta\alpha=\alpha-\alpha^{(mod)}$ read 
\begin{eqnarray}
\left\{
\begin{aligned}
\Delta \alpha_{x,j}&\simeq\alpha_{x,j}^{(mod)}2\sinh{(4|f_{2000,j}|)}
            \Big[\sinh{(4|f_{2000,j}|)}+\cosh{(4|f_{2000,j}|)}\sin{q_{2000,j}}
            \Big]-\sinh{(8|f_{2000,j}|)}\cos{q_{2000,j}}\\
\Delta \alpha_{y,j}&\simeq\alpha_{y,j}^{(mod)}2\sinh{(4|f_{0020,j}|)}
            \Big[\sinh{(4|f_{0020,j}|)}+\cosh{(4|f_{0020,j}|)}\sin{q_{0020,j}}
            \Big]-\sinh{(8|f_{0020,j}|)}\cos{q_{0020,j}}
\end{aligned}\right.
.\qquad\label{alpha-shift-4}
\end{eqnarray}
By keeping only the leading terms from the hyperbolic functions, the 
following first-order expression is retrieved
\begin{eqnarray}
\left\{
\begin{aligned}
\alpha_{x,j}&\simeq\alpha_{x,j}^{(mod)}\left(1+8\Im\{f_{2000,j}\}\right)
       -8\Re\{f_{2000,j}\}+O(|f_{2000}|^2)\\
\alpha_{y,j}&\simeq\alpha_{y,j}^{(mod)}\left(1+8\Im\{f_{0020,j}\}\right)
       -8\Re\{f_{0020,j}\}+O(|f_{0020}|^2)
\end{aligned}\right.
.\qquad\label{alpha-shift-5}
\end{eqnarray}
Explicit expressions truncated to the first order in $\delta K_1$ similar 
to Eq.~\eqref{beta-beat3} can be retrieved after substituting 
the RDTs in the above equations with Eq.~\eqref{eq:def_f0020}. \\

{\bf Improved formula to evaluate the beta beating from BPM phase advance.}
Eq.~\eqref{eq:BetaPhase2} was derived in Ref.~\cite{castro1} under
the assumption that the region between the three BPMs is free 
of unknown focusing errors. Here a more general formula is 
derived, which does not requires this condition. The only 
approximation made is a series of truncations to the first 
order in $\delta K_1$, and hence in the RDTs $f_{2000}$ and 
$f_{0020}$. The starting point are 
Eqs.~\eqref{beta-beat2B},~\eqref{phase-shift-9},~\eqref{alpha-shift-5}
\begin{eqnarray}
\left\{
\begin{aligned}
\beta_j&\simeq\beta_j^{(mod)}
                     \left(1 +8\Im\left\{f_j\right\}\right)\\
\Delta\phi_{ij}&\simeq\Delta\phi_{ij}^{(mod)}-2h_{ij} 
                 +4\Re\left\{f_j-f_i\right\}\\
\alpha_{j}&\simeq\alpha_{j}^{(mod)}\left(1+8\Im\{f_j\}\right)-8\Re\{f_j\}
\end{aligned}\right.
\quad,\quad
\left\{
\begin{aligned}
\frac{1}{\beta_1}\left(\cot{\Delta\phi_{12}}+\alpha_1\right)\\
\frac{1}{\beta_1}\left(\cot{\Delta\phi_{13}}+\alpha_1\right)
\end{aligned}\right.
\quad,\quad\label{castro-1}
\end{eqnarray}
where $h_{ij}$ and $f_j$ are the detuning term of Eq.~\eqref{phase-shift-9} 
and the RDT, respectively, corresponding to each plane, whose 
subscript is omitted here for the sake of notation (the derivation 
is the same). Before making explicit the two expressions in the 
above second bracket, the term $\Re\left\{f_j-f_i\right\}$ needs to 
be evaluated first. From Eq.(4.2)-(4.3) of Ref.~\cite{Andrea-thesis} 
$f_j$ can be rewritten as function of $f_i$ according to
\begin{eqnarray}
\left\{
\begin{aligned}
&\hat{h}_{ij}=f_je^{-i2\phi_{j}^{(mod)}}-f_ie^{-i2\phi_{i}^{(mod)}}
\quad\Rightarrow\quad 
f_j=\hat{h}_{ij}e^{i2\phi_{j}^{(mod)}}+f_ie^{i2\Delta\phi_{ij}^{(mod)}} \\
&\hat{h}_{ij}=\mp\frac{1}{8}\sum_{i<w<j}{\beta_{w}^{(mod)}\delta K_{w,1}}e^{-i2\phi_{w}^{(mod)}}
\end{aligned}\right.
\quad,\quad\label{castro-2}
\end{eqnarray}
where again the sum extends over the focusing errors between the 
two BPMs $i$ and $j$ only, while the sign is negative for $x$, positive 
for $y$. The label $i$ shall not be confused with the imaginary 
unit in the above exponential terms  $i=\sqrt{-1}$. Hence
\begin{eqnarray}
\Re\left\{f_j-f_i\right\}&=&\Re\left\{\hat{h}_{ij}e^{i2\phi_j^{(mod)}}
        +f_i\left(e^{i2\Delta\phi_{ij}^{(mod)}}-1\right)\right\} \nonumber\\
 &=&\Re\left\{\hat{h}_{ij}e^{i2\phi_j^{(mod)}}\right\} +
    |f_i|\Re\left\{e^{i(2\Delta\phi_{ij}^{(mod)}+q_i)}-e^{iq_i}\right\}
    \hskip 2cm q_i\hbox{ is the phase of }f_i \nonumber\\
 &=&\Re\left\{\hat{h}_{ij}e^{i2\phi_j^{(mod)}}\right\} +
    |f_i|\left\{\cos{(2\Delta\phi_{ij}^{(mod)}+q_i)}-\cos{q_i}\right\} \nonumber\\
 &=&\Re\left\{\hat{h}_{ij}e^{i2\phi_j^{(mod)}}\right\} +
    |f_i|\left\{\cos{q_i}\left[\cos{2\Delta\phi_{ij}^{(mod)}}-1\right]
              -\sin{q_i}\sin{2\Delta\phi_{ij}^{(mod)}}\right\} \nonumber\\
 &=&\Re\left\{\hat{h}_{ij}e^{i2\phi_j^{(mod)}}\right\} +
    \Re\left\{f_i\right\}\left[-2\sin^2{\Delta\phi_{ij}^{(mod)}}\right]
   -\Im\left\{f_i\right\}2\sin{\Delta\phi_{ij}^{(mod)}}\cos{\Delta\phi_{ij}^{(mod)}} 
\quad.\quad\label{castro-3}
\end{eqnarray}
By making use the  Taylor expansion of $\cot{(x+\epsilon)}$ with 
$\epsilon\ll x$, the BPM phase advance $\Delta\phi_{ij}$ of 
Eq.~\eqref{castro-1} can be approximated to
\begin{eqnarray}
\cot{(x+\epsilon)}=\cot{x}-\frac{\epsilon}{\sin^2{x}}+O(\epsilon^2)
\quad\Rightarrow\quad
\cot{\Delta\phi_{ij}}\simeq\cot{\Delta\phi_{ij}^{(mod)}}
           +\frac{2h_{ij}-4\Re\left\{f_j-f_i\right\}}{\sin^2{\Delta\phi_{ij}^{(mod)}}}
\quad.\quad\label{castro-4}
\end{eqnarray}
By substituting $\Re\left\{f_j-f_i\right\}$ with Eq.~\eqref{castro-3}, 
the above expression reads
\begin{eqnarray}
\cot{\Delta\phi_{ij}}\simeq\cot{\Delta\phi_{ij}^{(mod)}}
            \left(1+8\Im\left\{f_i\right\}\right)
           +\bar{h}_{ij} + 8\Re\left\{f_i\right\}           
\quad,\quad
\bar{h}_{ij}=\frac{2h_{ij}-4\Re\left\{\hat{h}_{ij}e^{i2\phi_j^{(mod)}}\right\}}
            {\sin^2{\Delta\phi_{ij}^{(mod)}}}
\quad.\quad\label{castro-5}
\end{eqnarray}
$\bar{h}_{ij}$ can be further made explicit via Eqs.~\eqref{phase-shift-2} 
and~\eqref{castro-2}, namely
\begin{eqnarray}
\bar{h}_{ij}&=&\mp\frac{1}{2\sin^2{\Delta\phi_{ij}^{(mod)}}}
        \sum_{i<w<j}{\beta_w^{(mod)}\delta K_{w,1}
        \left[1-\Re\left\{e^{i2(\phi_{j}^{(mod)}-\phi_{w}^{(mod)})}
        \right\}\right]}
        +O(\delta K_1^2)\nonumber\\
 &=&\mp\frac{1}{2\sin^2{\Delta\phi_{ij}^{(mod)}}}
         \sum_{i<w<j}{\beta_w^{(mod)}\delta K_{w,1}
         \left[1-\cos{2\Delta\phi_{wj}^{(mod)}}\right]}
         +O(\delta K_1^2)\nonumber\\
 &=&\mp\frac{1}{2\sin^2{\Delta\phi_{ij}^{(mod)}}}
        \sum_{i<w<j}{\beta_w^{(mod)}\delta K_{w,1}2\sin^2{\Delta\phi_{wj}^{(mod)}}}
        +O(\delta K_1^2)\nonumber\\
 &=&\mp\frac{1}{\sin^2{\Delta\phi_{ij}^{(mod)}}}
        \sum_{i<w<j}{\beta_w^{(mod)}\delta K_{w,1}\sin^2{\Delta\phi_{wj}^{(mod)}}}
        +O(\delta K_1^2)
\quad,\quad\label{castro-6}
\end{eqnarray}
where the sign depends on the plane and the above sum extends over 
the focusing errors between the two BPMs $i$ and $j$, while  
$\Delta\phi_{wj}^{(mod)}$ denotes the phase advance between the 
BPM $j$ and the source of error $w$ of the ideal (or initial) lattice 
model.
\begin{eqnarray}
\left\{
\begin{aligned}
&\bar{h}_{x,ij}=-\frac{1}{\sin^2{\Delta\phi_{x,ij}^{(mod)}}}
        \sum_{i<w<j}{\beta_{x,w}^{(mod)}\delta K_{w,1}\sin^2{\Delta\phi_{x,wj}^{(mod)}}}
        +O(\delta K_1^2)\\
&\bar{h}_{y,ij}=+\frac{1}{\sin^2{\Delta\phi_{y,ij}^{(mod)}}}
        \sum_{i<w<j}{\beta_{y,w}^{(mod)}\delta K_{w,1}\sin^2{\Delta\phi_{y,wj}^{(mod)}}}
        +O(\delta K_1^2)
\end{aligned}\right.
\quad,\quad\label{castro-7}
\end{eqnarray}
All the ingredients are now ready to make explicit the quantities 
in the most right block of Eq.~\eqref{castro-1}, by noting that 
\begin{eqnarray}
\frac{1}{\beta_1}\left(\cot{\Delta\phi_{12}}+\alpha_1\right)&\simeq&
\frac{\cot{\Delta\phi_{12}^{(mod)}}\left(1+8\Im\left\{f_1\right\}\right)
      +\bar{h}_{12} + 8\Re\left\{f_1\right\}
      +\alpha_{1}^{(mod)}\left(1+8\Im\{f_1\}\right)-8\Re\{f_1\}}
    {\beta_1^{(mod)}\left(1+8\Im\left\{f_1\right\}\right)}\nonumber\\
&\simeq&\frac{1}{\beta_1^{(mod)}}\left(\cot{\Delta\phi_{12}^{(mod)}}
        +\alpha_1^{(mod)}\right)
       +\frac{\bar{h}_{12}}{\beta_1^{(mod)}\left(1+8\Im\left\{f_1\right\}\right)}
\label{castro-7B}\\\nonumber
&\simeq&\frac{1}{\beta_1^{(mod)}}\left(\cot{\Delta\phi_{12}^{(mod)}}
        +\alpha_1^{(mod)}\right)
       +\frac{\bar{h}_{12}}{\beta_1^{(mod)}}+O(\delta K_1^2) \quad,\\
\frac{1}{\beta_1}\left(\cot{\Delta\phi_{13}}+\alpha_1\right)&\simeq&
        \frac{1}{\beta_1^{(mod)}}\left(\cot{\Delta\phi_{13}^{(mod)}}
        +\alpha_1^{(mod)}\right)
       +\frac{\bar{h}_{13}}{\beta_1^{(mod)}}+O(\delta K_1^2) 
\quad.\quad\label{castro-8}
\end{eqnarray}
The difference between the above expressions reads
\begin{eqnarray}
\frac{1}{\beta_1}\left(\cot{\Delta\phi_{13}}-\cot{\Delta\phi_{12}}\right)
   \simeq\frac{1}{\beta_1^{(mod)}}\left[(\cot{\Delta\phi_{13}^{(mod)}}
          -\cot{\Delta\phi_{12}^{(mod)}})+(\bar{h}_{13}-\bar{h}_{12})\right]
     +O(\delta K_1^2) 
\quad,\quad\label{castro-9}
\end{eqnarray}
which is equivalent to
\begin{eqnarray}
\beta_1\simeq\beta_1^{(mod)}\frac{\cot{\Delta\phi_{13}}-\cot{\Delta\phi_{12}}}
     {(\cot{\Delta\phi_{13}^{(mod)}}-\cot{\Delta\phi_{12}^{(mod)}})
      +(\bar{h}_{13}-\bar{h}_{12})}  +O(\delta K_1^2) 
\quad,\quad\label{castro-10}
\end{eqnarray}
with $\bar{h}_{ij}$ defined in Eq.~\eqref{castro-7}. 
Eq.~\eqref{eq:BetaPhase3} is then demonstrated. When no source of 
focusing error is present between the three BPMs, 
$\bar{h}_{13}=\bar{h}_{12}=0$ and Eq.~\eqref{eq:BetaPhase2} is retrieved. 
A special case where Eq.~\eqref{eq:BetaPhase2} still 
applies even in the presence of strong localized focusing errors 
is when $\bar{h}_{13}=\bar{h}_{12}\ne0$. More generally, 
Eq.~\eqref{eq:BetaPhase2} remains a robust approximation 
whenever $|\bar{h}_{13}-\bar{h}_{12}|\ll |\cot{\Delta\phi_{13}^{(mod)}}-\cot{\Delta\phi_{12}^{(mod)}}|$, 
or when the beating induced by any quadrupole errors between 
two BPMs  is much smaller than the one generated by focusing 
glitches along the rest of the ring, i.e. 
$|\bar{h}_{12}|\ll |\beta_1^{(mod)}\left(1+8\Im\left\{f_1\right\}\right)|$
in Eq.~\eqref{castro-7B}, the RDT $f_1$ being generated by all 
sources of error, see Eq.~\eqref{eq:def_f0020}.\\

{\bf Interpreting an insertion optics as a closed RDT bump.} 
Eqs.~\eqref{beta-beat2B},~\eqref{phase-shift-9},~\eqref{alpha-shift-5}
provide an interesting interpretation of an insertion optics, i.e. 
of a local modification of the linear optics confined between two 
points $i$ and $j$, with no change outside. 
\begin{eqnarray}
\left\{
\begin{aligned}
\beta_i&\simeq\beta_i^{(mod)}
                     \left(1 +8\Im\left\{f_i\right\}\right)\\
\alpha_{i}&\simeq\alpha_{i}^{(mod)}\left(1+8\Im\{f_i\}\right)-8\Re\{f_i\} 
\end{aligned}\right.
\quad,\quad
\left\{
\begin{aligned}
\beta_j&\simeq\beta_j^{(mod)}
                     \left(1 +8\Im\left\{f_j\right\}\right)\\
\alpha_{j}&\simeq\alpha_{j}^{(mod)}\left(1+8\Im\{f_j\}\right)-8\Re\{f_j\} \\
\Delta\phi_{ij}&\simeq\Delta\phi_{ij}^{(mod)}-2h_{ij} 
                 +4\Re\left\{f_j-f_i\right\}\\
\end{aligned}\right.
\quad.\quad\label{insertion-1}
\end{eqnarray}
If the insertion is perfectly matched to the rest of the machines, 
$\beta_i=\beta_j$ and $\alpha_{i}=\alpha_{j}$ (in general the same is 
true for the dispersion function and its derivative, not discussed here). 
This implies that: $(i)$ $f_i=f_j=0$, $(ii)$ the RDTs are zero outside the 
two locations $i$ and $j$, and $(iii)$ the phase advance of the whole 
insertion is $\Delta\phi_{ij}=\Delta\phi_{ij}^{(mod)}-2h_{ij}$, 
with $h_{ij}$ in general nonzero. This in turn implies 
that the above equations are actually exact, since the remainders 
proportional to $|f|^2$ is also zero at the insertion ends. An 
example of matched insertion optics introduced in the lattice of the 
ESRF storage ring is showed in Fig.~\ref{fig_insertion1}, along 
with the amplitude of the two RDTs. As predicted by 
Eq.~\eqref{insertion-1}, the RDTs are zero at the ends and outside 
the insertion region, with a closed bump inside. \\
\begin{figure}
\rule{0mm}{0mm}
\centerline{\includegraphics[width=11.0cm]{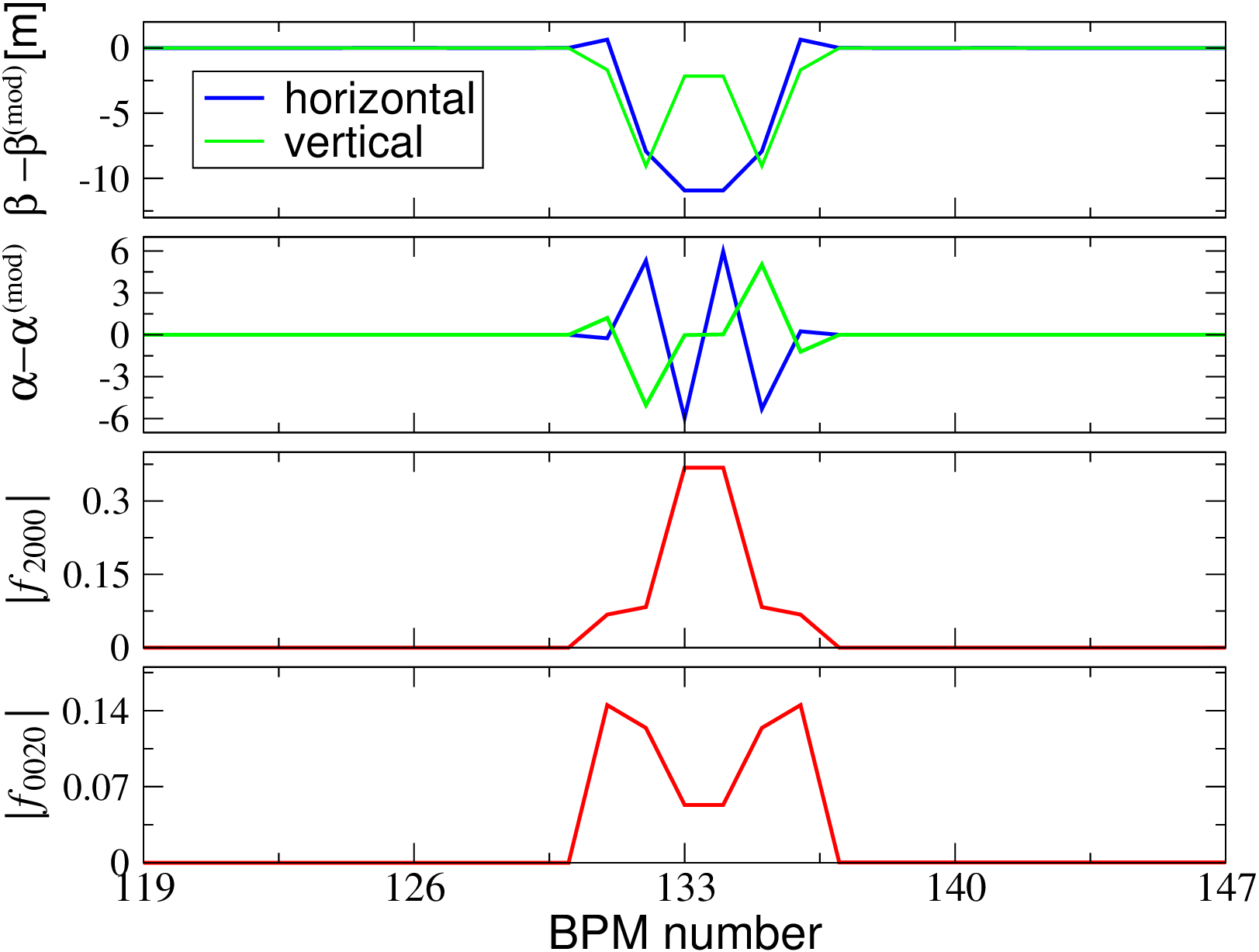}}
  \caption{\label{fig_insertion1} (Color) Difference between the  
    C-S parameters computed by MADX at the BPMs with and without 
    an example of insertion optics introduced in the lattice of the ESRF 
    storage ring (upper plots). The amplitude of the two corresponding 
    RDTs is plotted in the lower plots, which is zero outside the 
    insertion region. The perfectly matched insertion corresponds 
    hence to a closed RDT bump.}
\rule{0mm}{0mm}
\end{figure}

{\bf Accurate evaluation of the focusing errors RDTs.} 
Eqs.~\eqref{beta-beat2B},~\eqref{phase-shift-8},~\eqref{alpha-shift-5}
compute the C-S parameters modified by focusing errors (or insertion 
optics) via the corresponding RDTs $f_{2000}$ and $f_{0020}$. They represent 
already an approximation, linear in RDTs, since terms proportional to 
higher powers ($f^2,\ f^3,\ ...$) are neglected. For light sources such as 
the ESRF storage ring, with a typical rms beating of about 3-5$\%$ (see 
top plot of Fig.~\ref{fig_beat1}) and ultra-low coupling 
(the emittance ratio $\epsilon_y/\epsilon_x$ is about 1\textperthousand) 
this approximation is already rather robust and is expected to be of 
no concern for more recent machines with the same coupling level and 
an RMS beating lower than $1\%$. It remains to asses how reliable is 
the computation of the RDT from the lattice formula (linear in 
$\delta K_1$) of Eq.~\eqref{eq:def_f0020}. In Fig.~\ref{fig_RDT1} the 
differences between the C-S parameters computed from an error model 
(the same of Fig.~\ref{fig_beat1}) and from the ideal lattice of the 
ESRF storage ring are displayed. By taking as reference the values 
computed by MADX (red curves), an overall good agreement is observed 
when the lattice formula, Eq.~\eqref{eq:def_f0020}, is used to 
evaluate the RDTs (green curves), though some local and global 
discrepancies can be observed: up to 1 m for $\beta$, 0.5 for $\alpha$, 
and 5 mrad for the betatron phases $\phi$, see Fig.~\ref{fig_RDT2}.

\begin{figure}
\rule{0mm}{0mm}
\centerline{\includegraphics[width=14.0cm,height=10cm]{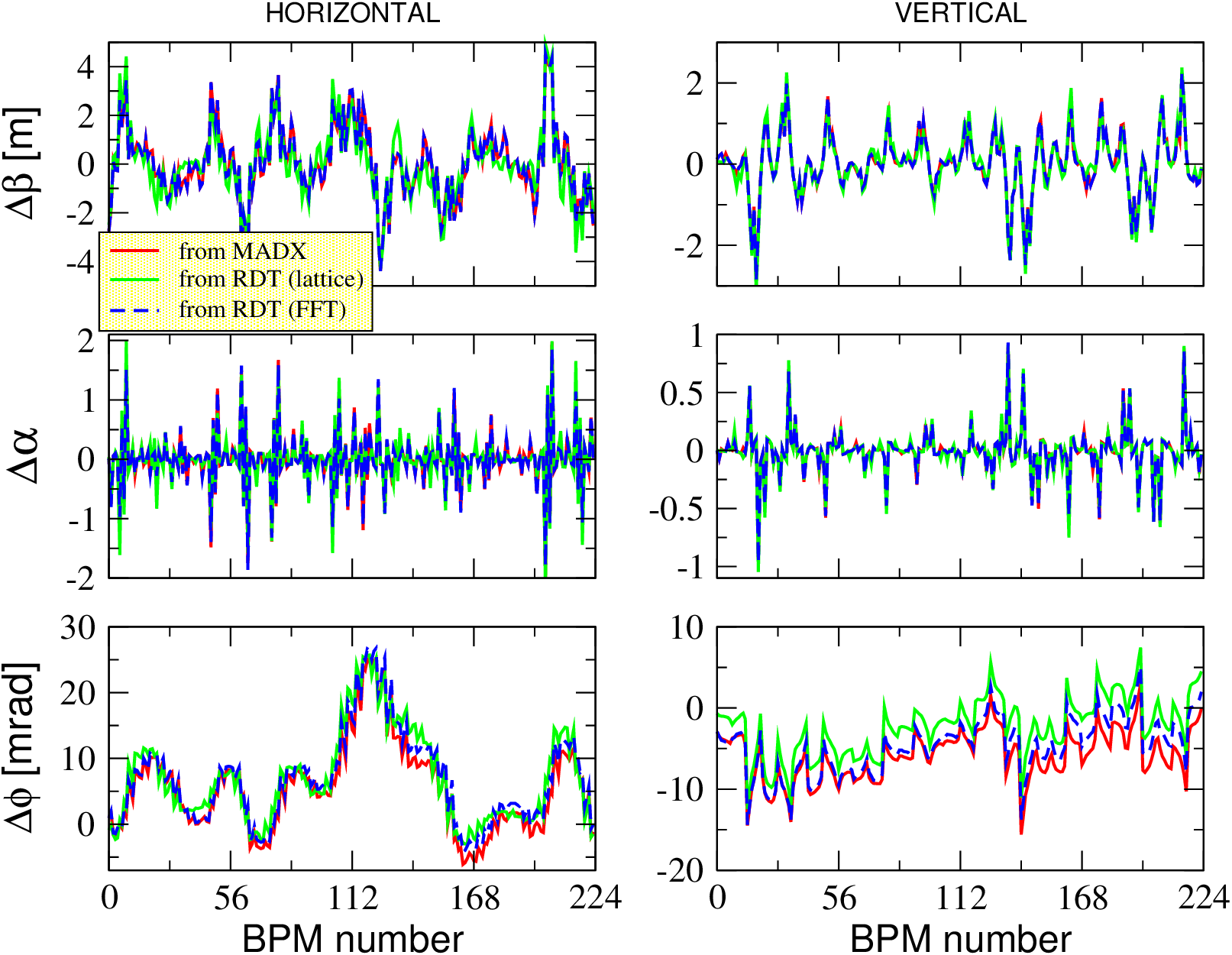}}
  \caption{\label{fig_RDT1} (Color) Differences between the C-S 
    parameters evaluated from an error model (the same of 
    Fig.~\ref{fig_beat1}) and from the ideal lattice of the ESRF storage 
    ring. These differences are computed by MADX (solid red line) and from 
    Eq.~\eqref{beta-beat2B} (for $\beta$), Eq.~\eqref{alpha-shift-5} (for 
    $\alpha$) and Eq.~\eqref{phase-shift-8} (for $\phi$). The RDTs in those 
    formulas are computed by the lattice formula (linear in $\delta K_1$) of 
    Eq.~\eqref{eq:def_f0020} (solid green line) and from Eq.~\eqref{RDT-2} 
    (FFT of simulated single particle tracking data, dashed blue line).}
\rule{0mm}{0mm}
\rule{0mm}{0mm}
\centerline{\includegraphics[width=14.0cm,height=10cm]{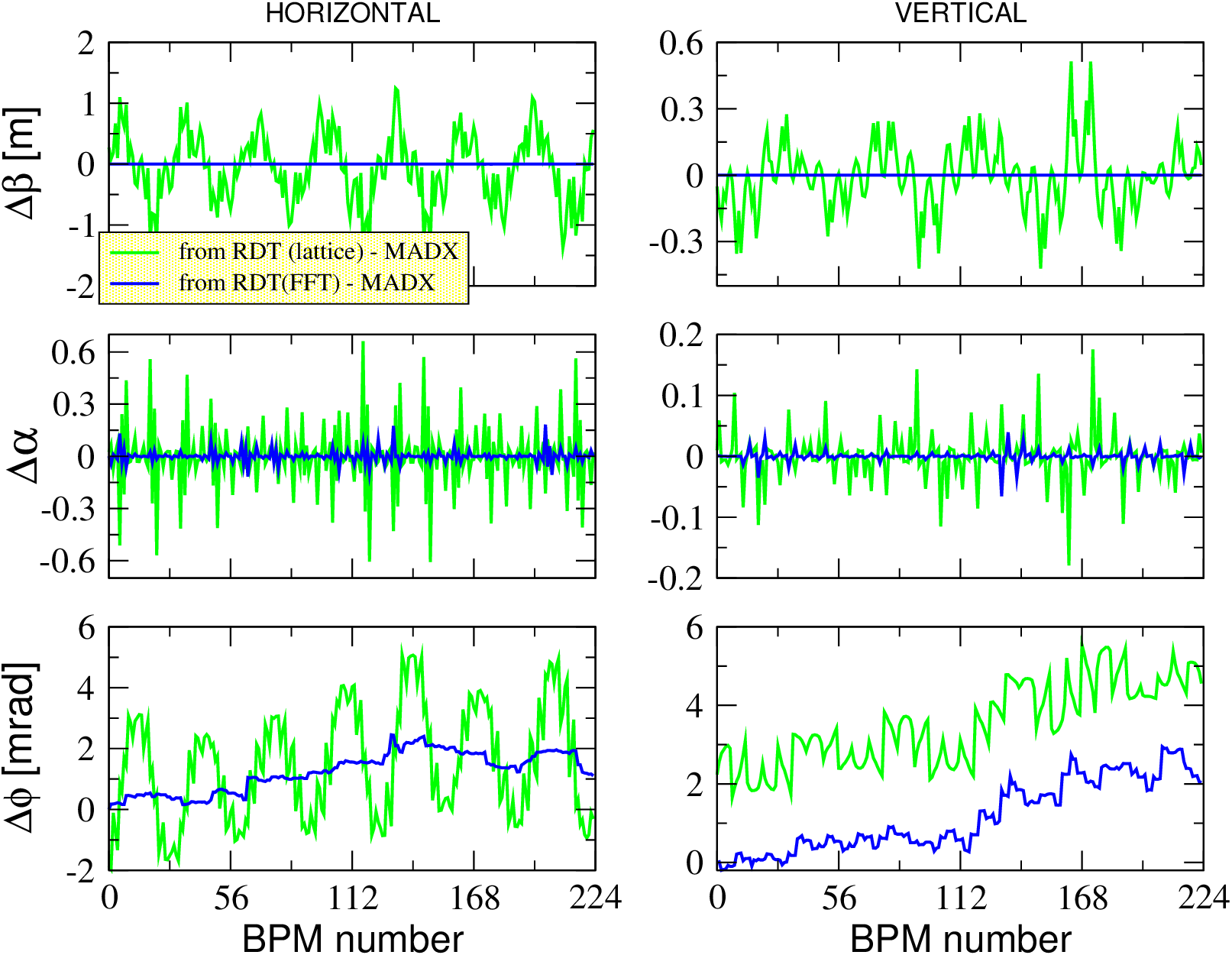}}
  \caption{\label{fig_RDT2} (Color). Deviations between the change of the 
    C-S parameters computed by MADX and the one predicted by 
    Eq.~\eqref{beta-beat2B} (for $\beta$), Eq.~\eqref{alpha-shift-5} (for 
    $\alpha$) and Eq.~\eqref{phase-shift-8} (for $\phi$). The green line 
    is obtained when computing the RDTs via Eq.~\eqref{eq:def_f0020}, 
    whereas the application of Eq.~\eqref{RDT-2} results in the blue curve, 
    which is much more accurate.}
\rule{0mm}{0mm}
\end{figure}
Fortunately a more accurate way to compute the RDTs exists, though it 
requires several computational steps. First, single particle 
tracking is to be performed, with focusing errors included in the 
lattice and with non-zero initial conditions in both planes, small 
enough so to remain in the linear regime (a few tens of $\mu$m). 
Turn-by-turn position $(x,p_x)$ and momentum $(y,p_y)$ shall be 
recorded at the BPMs. The complex C-S variables $h_x=\tilde{x}-i\tilde{p}_x$ 
and $h_y=\tilde{y}-i\tilde{p}_y$ are then computed at each BPM, 
where the C-S parameters used to convert the Cartesian 
coordinates are the ones of the ideal model (i.e. without the focusing 
errors used for tracking). As showed in Appendix C of 
Ref.~\cite{Andrea-arxiv}, from the FFT of $h_x$ at each BPM two main 
harmonics can be extracted, one at the tune frequency $H_h(1,0)$ 
and the other at  its opposite $H_h(-1,0)$ (the same applies to 
the vertical plane): 
\begin{eqnarray}
h_{x}&=&\cosh{(4|f_{2000}|)}\ \zeta_{x,-} -
       i\sinh{(4|f_{2000}|)}\ e^{iq_{2000}}\ \zeta_{x,+} \ ,  
	\nonumber \\ 
&& \hskip 1.5cm \Uparrow  \hskip3.6 cm  \Uparrow \nonumber \\
&& \hskip 1.2cm H_h(1,0)  \hskip2.2cm H_h(-1,0)\label{RDT-1}\\
h_{y}&=&\cosh{(4|f_{0020}|)}\ \zeta_{y,-} -
       i\sinh{(4|f_{0020}|)}\ e^{iq_{0020}}\ \zeta_{y,+} \ ,  
	\nonumber \\ 
&& \hskip 1.5cm \Uparrow  \hskip3.6 cm  \Uparrow \nonumber \\
&& \hskip 1.2cm V_h(0,1)  \hskip2.2cm V_h(0,-1)\nonumber
\end{eqnarray}
where $\zeta_{\pm}=\sqrt{2I}e^{\mp i(2\pi NQ+\phi+\psi_{0})}$. The RDT phase 
$q$ and amplitude $|f|$ can then be inferred from the phase and 
amplitude of the four lines, according to
\begin{eqnarray}
\left\{
\begin{aligned}
&q_{2000}=\Phi_{H_h(-1,0)}+\Phi_{H_h(1,0)}+\frac{\pi}{2} \\
&q_{0020}=\Phi_{V_h(0,-1)}+\Phi_{V_h(0,1)}+\frac{\pi}{2} 
\end{aligned}\right.
\ ,\  
\left\{
\begin{aligned}
&|f_{2000}|=\frac{1}{4}\arctanh{g_x}
           =\frac{1}{8}\left[\ln{(1+g_x)}-\ln{(1-g_x)}\right]
           \ ,\  g_x=\frac{|H_h(-1,0)|}{|H_h(1,0)|}\\
&|f_{0020}|=\frac{1}{4}\arctanh{g_y}
           =\frac{1}{8}\left[\ln{(1+g_y)}-\ln{(1-g_y)}\right]
           \ ,\  g_y=\frac{|V_h(0,-1)|}{|V_h(0,1)|}
\end{aligned}\right. \ .
\nonumber \\ \label{RDT-2}
\end{eqnarray}
By using the RDTs inferred from the above FFT formulas in 
Eqs.~\eqref{beta-beat2B},~\eqref{phase-shift-8},~\eqref{alpha-shift-5}, 
the agreement with the 
C-S parameters computed by MADX is greatly improved, as showed 
by the blue curves of Figs.~\ref{fig_RDT1} and~\ref{fig_RDT2}: 
The accuracy is better than 1 mm for $\beta$, 0.15 for $\alpha$, 
and up 3 mrad for $\phi$. When computing the BPM phase advance 
error, the deviation drops to about 0.1 mrad.
The accumulation of inaccuracy for $\phi$ along the ring (bottom 
plot of Fig.~\ref{fig_RDT2}) can be attributed to higher-order 
terms in the computation of $h_{ij}$ and nonlinear terms put in the 
remainder $O(|f|^2)$ of Eq.~\eqref{phase-shift-9}.



{\bf Evaluating the beta beating from the tune line amplitude.} 
From Eqs.~\eqref{tunex-beat2}-\eqref{tunex-beat3} the tune 
line amplitude at the BPM $j$ reads
\begin{eqnarray}
|H(1,0)_j|=\frac{\mathcal{C}_{x,j}}{2}\sqrt{2I_x}A_{f,j}
\quad,\quad \mathcal{C}_{x,j}=1+\mathcal{E}_{x,j}
\quad,\quad 0\simeq\mathcal{E}_{x,j}\ll 1
\qquad ,\label{beta-tuneline1}
\end{eqnarray}
where the BPM calibration factor is included and represented by 
a small calibration error $\mathcal{E}_{x,j}$ for a later 
perturbative expansion. By averaging over all BPMs the following 
expression for the invariant is obtained
\begin{eqnarray}
<|H(1,0)|>=\frac{<\mathcal{C}_x>}{2}\sqrt{2I_x}<A_f>
\qquad\Rightarrow\quad 
\sqrt{2I_x}=\frac{2<|H(1,0)|>}{<\mathcal{C}_x><A_f>}
\qquad ,\label{beta-tuneline2}
\end{eqnarray}
$\mathcal{C}$ and $A_f$ being uncorrelated quantities. 
On the other hand, $A_{f,j}$ can be written as the ratio between 
the real beta and the ideal one, see Eq.~\eqref{eq:tunes2}, 
\begin{eqnarray}
A_{f,j}=\sqrt{\frac{\beta_{x,j}}{\beta_{x,j}^{(mod)}}}
\qquad\Rightarrow\quad 
|H(1,0)_j|=\frac{\mathcal{C}_{x,j}}{2}\sqrt{2I_x}
           \sqrt{\frac{\beta_{x,j}}{\beta_{x,j}^{(mod)}}}
\qquad\Rightarrow\quad 
\beta_{x,j}=\beta_{x,j}^{(mod)}\frac{2|H(1,0)_j|^2}
           {\mathcal{C}_{x,j}^2(2I_x)}
\quad .\quad \label{beta-tuneline3}
\end{eqnarray}
By replacing $(2I_x)$ with the expression of Eq.~\eqref{beta-tuneline2},
the above expression reads
\begin{eqnarray}
\beta_{x,j}=\beta_{x,j}^{(mod)}\left(\frac{|H(1,0)_j|}{<|H(1,0)|>}\right)^2
            \left(\frac{<\mathcal{C}_x>}{\mathcal{C}_{x,j}}\right)^2
            <A_f>^2
\quad .\quad \label{beta-tuneline4}
\end{eqnarray}
The last two terms in the r.h.s of the above equation can be approximated 
by
\begin{eqnarray}
\left\{
\begin{aligned}
&\left(\frac{<\mathcal{C}_x>}{\mathcal{C}_{x,j}}\right)^2=
\left(\frac{1+<\mathcal{E}_x>}{1+\mathcal{E}_{x,j}}\right)^2\simeq
\left(1+<\mathcal{E}_x>-\mathcal{E}_{x,j}+O(\mathcal{E}_x^2)\right)^2\simeq
1+2(<\mathcal{E}_x>-\mathcal{E}_{x,j})+O(\mathcal{E}_x^2) \\
&<A_f>^2\simeq<1+32|f_{2000}|^2+8|f_{2000}|\sin{q_{2000}}+O(|f_{2000}|^3)>^2
        \simeq 1+64<|f_{2000}|^2>+O(|f_{2000}|^3)
\end{aligned}\right. 
\quad .\quad \label{beta-tuneline5}
\end{eqnarray}
Eq.~\eqref{beta-tuneline4} then becomes
\begin{eqnarray}
\left\{
\begin{aligned}
&\beta_{x,j}=\beta_{x,j}^{(mod)}\left(\frac{|H(1,0)_j|}{<|H(1,0)|>}\right)^2
            [1+2(<\mathcal{E}_x>-\mathcal{E}_{x,j})+O(\mathcal{E}_x^2)]
            [1+64<|f_{2000}|^2>+O(|f_{2000}|^3)]\\
&\beta_{y,j}=\beta_{y,j}^{(mod)}\left(\frac{|V(0,1)_j|}{<|V(0,1)|>}\right)^2
            [1+2(<\mathcal{E}_y>-\mathcal{E}_{y,j})+O(\mathcal{E}_y^2)]
            [1+64<|f_{0020}|^2>+O(|f_{0020}|^3)]
\end{aligned}\right.
\quad ,\quad \label{beta-tuneline6}
\end{eqnarray}
where the expression for the vertical plane is obtained with the 
same derivation. $<\mathcal{E}>$ and $<|f|^2>$ represent the 
averaged values (over all BPMs) of the calibration errors and of the 
amplitudes of the RDTs, respectively . Unless these are determined 
by independent measurements, they cannot be disentangled and are 
not observable. A (rude) zero-order truncation is then needed in 
order to apply Eq.~\eqref{beta-tuneline6} to real data, yielding to 
\begin{eqnarray}
\left\{
\begin{aligned}
&\beta_{x,j}=\beta_{x,j}^{(mod)}\left(\frac{|H(1,0)_j|}{<|H(1,0)|>}\right)^2
            +O(\mathcal{E}_x,|f_{2000}|^2)\\
&\beta_{y,j}=\beta_{y,j}^{(mod)}\left(\frac{|V(0,1)_j|}{<|V(0,1)|>}\right)^2
            +O(\mathcal{E}_y,|f_{0020}|^2)
\end{aligned}\right.
\quad .\quad \label{beta-tuneline7}
\end{eqnarray}

\section{Impact of octupolar-like terms on the tune lines:  an amplitude 
         dependent focusing}
\label{app:2A}
All results of Appendix~\ref{app:1} and Sec.~\ref{today} are 
valid as long as the beam motion remains in the linear regime, 
i.e. nonlinear terms may be neglected. In this 
appendix this assumption is removed and the 
extension of Eq.~\eqref{eq:tunes1}  to include 
higher-order contributions is derived. The result will be a more 
complicated formula with additional terms dependent on the 
initial oscillation amplitudes $(2I_{x,y})$, both in the tune 
line amplitude and phase. These are proportional to octupolar 
fields ($\propto K_3$) and to quadratic functions of sextupole 
strengths ($\propto K_2^2$). 

In Table V-VII of Ref.~\cite{Bartolini1} the list of secondary 
harmonics of the the complex signals $h_x=\tilde{x}-i\tilde{p}_x$ and 
$h_y=\tilde{y}-i\tilde{p}_y$ generated by octupolar-like RDTs 
is presented. It can be seen how their spectral lines 
$H_h(-1,0)$ and $V_h(0,-1)$, which in the linear regime 
are excited only by focusing errors via the two quadrupolar 
RDTs $f_{2000}$ and $f_{0020}$, respectively, receive a 
contribution from several octupolar-like RDTs too. To the 
first order in the RDTs, these lines read
\begin{eqnarray}
\left\{\begin{aligned}
&H_h(-1,0)=\left[-4i\sqrt{2I_x}f_{2000}-6i(2I_x)f_{3100}
                -4i\sqrt{(2I_x)(2I_y)}f_{2011}+O(f^2,I^2)\right]e^{-\tau_x}\\
&V_h(0,-1)=\left[-4i\sqrt{2I_y}f_{0020}-6i(2I_y)f_{0031}
                -4i\sqrt{(2I_x)(2I_y)}f_{1120}+O(f^2,I^2)\right]e^{-\tau_y}\\
&\tau_x=i\left\{2\pi N\left[Q_x^{(mod)}-2h_{1100}-4h_{2200}(2I_x)
                          -2h_{1111}(2I_y)+O(I^2)\right]+\phi_x+\psi_{x0}\right\}\\
&\tau_y=i\left\{2\pi N\left[Q_y^{(mod)}-2h_{0011}-4h_{0022}(2I_y)
                          -2h_{1111}(2I_x)+O(I^2)\right]+\phi_y+\psi_{y0}\right\}
\end{aligned}\right. 
\quad .\quad \label{eq:nonlin1}
\end{eqnarray}
The above expressions are derived from the more general expression 
for the TBT complex signals~\cite{Bartolini1}
\begin{eqnarray}\hspace{-1mm} 
\left\{\begin{aligned}
&\hspace{-1mm} h_{x}(N)=\sqrt{2I_x}e^{i(2\pi Q_xN+\phi_x+\psi_{x0})} - 
2i\sum_{pqrt}{pf_{pqrt}(2I_x)^{\frac{p+q-1}{2}}
                  (2I_y)^{\frac{r+t}{2}}}
 e^{i[(1-p+q)(2\pi Q_xN+\phi_x+\psi_{x0})+
                  (t-r)(2\pi Q_yN+\phi_y+\psi_{y0})]} \\
&\hspace{-1mm} h_{y}(N)=\sqrt{2I_y}e^{i(2\pi Q_yN+\phi_y+\psi_{y0})} -
2i\sum_{pqrt}{rf_{pqrt}(2I_x)^{\frac{p+q}{2}}
                 (2I_y)^{\frac{r+t-1}{2}}}
e^{i[(q-p)(2\pi Q_xN+\phi_x+\psi_{x0})+
                 (1-r+t)(2\pi \nu_yN+\phi_x+\psi_{x0})]}
\end{aligned}\right. \hspace{-1mm} . \nonumber \\ \label{e:TbTComplex}
\end{eqnarray}
The harmonic $H_h(-1,0)$ ($V_h(0,-1)$) is the sum of all terms 
in the above summation such that $1-p+q=-1$ and $t-r=0$ ($q-p=0$ 
and $1-r+t=-1$), i.e. all those oscillating with the opposite 
betatron tune and phase. The first term, scaling with $\sqrt{2I}$, 
is then $f_{2000}$, as $1-2+0=-1$ and $0-0=0$ ($f_{0020}$, 
since $0-0=0$ and $1-2+0=-1$). Following the same logic, the 
next terms in $H_h(-1,0)$ scaling with $(2I_x)$ are $f_{3100}$, 
($1-3+1=-1$ and $0-0=0$) and $f_{2011}$ ($1-2+0=-1$ and 
$1-1=0$). The same rule applied to $V_h(0,-1)$ selects $f_{0031}$ 
and $f_{1120}$. All last four RDTs are normal octupolar-like
($f_{pqrt}$, with $p+q+r+t=4$), with $f_{3100}$ and $f_{0031}$ 
excited by the potential terms $\propto x^4$ and $y^4$, respectively, 
whereas both $f_{2011}$ and $f_{1120}$ stem from the monomial 
$\propto x^2y^2$. Detuning terms $h$ are those 
elements of the complex C-S Hamiltonian
\begin{eqnarray}\label{e:HamComplex}
\tilde{H}=\sum_n{\sum_{pqrt}^{n=p+q+r+t}{h_{pqrt}
            (2I_x)^{\frac{p+q}{2}}(2I_y)^{\frac{r+t}{2}}}
	e^{i[(p-q)(\phi_x+\psi_{x0})+(r-t)(\phi_y+\psi_{y0})]}}\ \ ,
\end{eqnarray}
that do not depend on the betatron phase, i.e. $p=q$ and $r=t$. 
In the above definition $n$ represents the multipole order (normal 
and skew): $n=2$ for quadrupoles, $n=3$ for sextupoles, $n=4$ for 
octupole, etc. The explicit formula for $h_{pqrt}$ reads 
\begin{eqnarray}
h_{pqrt}&=&-\displaystyle
\frac{\bigl[K_{n-1}\Omega(r+t)+
           iJ_{n-1}\Omega(r+t+1)\bigr]}
     {p!\quad q!\quad r!\quad r!\quad 2^{p+q+r+t}}\nonumber 
     \ i^{r+t} \bigl(\beta_{x}\bigr)^{\frac{p+q}{2}}
     \bigl(\beta_{y}\bigr)^{\frac{r+t}{2}},\label{eq:h_Vs_KJ}\\
     \Omega(i)&=&1 \hbox{ if } i \hbox{ is even},\quad
                      \Omega(i)=0 \hbox{ if } i \hbox{ is odd}
     \qquad \ .
\end{eqnarray}
$\Omega(i)$ is introduced to select either the normal or the 
skew multipoles. $K_{n-1}$ and $J_{n-1}$ are the integrated 
magnet strengths of Eq.~\eqref{eq:MADX}. If several sources 
are to be included in the above Hamiltonian, a further 
summation taking into account the relative phase advances 
between magnets and observation point needs to be included 
in Eq.~\eqref{e:HamComplex}, see Appendix A of 
Ref.~\cite{Andrea-arxiv}.

The RDT $f_{pqrt}$ and the detuning terms $h_{pprr}$ introducing 
a dependence on the initial amplitude $(2I)$ in Eq.~\eqref{eq:nonlin1}
are proportional to the octupolar strengths 
($\propto K_3$, generated either by physical octupole magnets 
or by octupolar components of other magnets) and to 
quadratic functions of sextupole strengths ($\propto K_2^2$). 

Nonlinear terms affect the tune lines (of the complex 
signals). According to Eqs.(C29)-(C30) of Ref.~\cite{Andrea-arxiv} 
these read
\begin{eqnarray}
\left\{\begin{aligned}
H_h(1,0)&=\sqrt{2I_x}\left[1+T_{H_x}(2I_x)+T_{H_y}(2I_y)
          +O((2I_{x,y})^2)\right]e^{\tau_x}\\
V_h(0,1)&=\sqrt{2I_y}\left[1+T_{V_x}\hskip0.4mm(2I_x)+T_{V_y}\hskip0.4mm(2I_y)
          +O((2I_{x,y})^2)\right]e^{\tau_y}
\end{aligned}\right. 
\quad ,\quad \label{eq:nonlin2}
\end{eqnarray}
where the complex functions $T$ are quadratic functions of the 
sextupole strengths ($\propto K_2^2$) and the expansion is 
truncated to the first non-zero terms of the invariants. These 
four spectral lines can be conveniently rewritten as 
\begin{eqnarray}
\left\{\begin{aligned}
 H_h(1,0)&=\sqrt{2I_x}\left[1+T_H(K_2^2,I_{x,y})\right]e^{\tau_x}\\
 V_h(0,1)&=\sqrt{2I_y}\left[1+T_V(K_2^2,I_{x,y})\right]e^{\tau_y}\\
H_h(-1,0)&=-i\sqrt{2I_x}\left[4f_{2000}+F_{xx}^*(K_2^2,K_3,I_{x,y}
                )\right]e^{-\tau_x}\\
V_h(0,-1)&=-i\sqrt{2I_y}\left[4f_{0020}+F_{yy}^*(K_2^2,K_3,I_{x,y}
                )\right]e^{-\tau_y}
\end{aligned}\right. 
\quad ,\quad \label{eq:nonlin3}
\end{eqnarray}
where the remainders $O(f^2,I^2)$ have been ignored. The dependence 
of the complex and longitudinally varying functions $T$ and $F$ 
on the magnetic strengths $K_{2,3}$ is 
indicated in the parenthesis. Since the harmonic analysis is 
performed here on the real signals  $\tilde{x}$ and 
$\tilde{y}$, the observable tune lines read 
\begin{eqnarray}
\left\{\begin{aligned}
H(1,0)=\frac{1}{2}[H_h(1,0)+H_h^*(-1,0)]&=\frac{\sqrt{2I_x}}{2}
        \left[1+4if^*_{2000}+iF_{xx}(K_2^2,K_3,I_{x,y})
               +T_H(K_2^2,I_{x,y})\right]e^{\tau_x}\\
V(0,1)=\frac{1}{2}[\ V_h(0,1)+\ V_h^*(0,-1)]&=\frac{\sqrt{2I_y}}{2}
        \left[1+4if^*_{0020}+iF_{yy}(K_2^2,K_3,I_{x,y})
               +T_V(K_2^2,I_{x,y})\right]e^{\tau_y}
\end{aligned}\right. 
.\qquad \label{eq:nonlin4}
\end{eqnarray}
The generalization of the tune line amplitude and phase of 
Eq.~\eqref{eq:tunes2} at a BPM $j$ in the nonlinear 
(amplitude dependent) regime are then 
\begin{eqnarray}
\left\{\begin{aligned}
&|H(1,0)_j|=\frac{\sqrt{2I_x}}{2}|B_{x,j}| \quad , \quad 
\Phi_{H(1,0),j}=\phi_{x,j}^{(mod)}+\psi_{x0}+\hbox{arg}\{B_{x,j}\} 
                -2h_{1100,j}-4h_{2200,j}(2I_x)-2h_{1111,j}(2I_y)\\
&|V(0,1)_j|=\frac{\sqrt{2I_y}}{2}|B_{y,j}| \quad , \quad 
\Phi_{V(0,1),j}=\phi_{y,j}^{(mod)}+\psi_{y0}+\hbox{arg}\{B_{y,j}\} 
                -2h_{0011,j}-2h_{1111,j}(2I_x)-4h_{0022,j}(2I_y) \\
&B_{x,j}=1+i4f^*_{2000,j}+iF_{xx,j}(K_2^2,K_3,
        I_{x,y})+T_{H,j}(K_2^2,I_{x,y})\\
&B_{y,j}=1+4if^*_{0020,j}+iF_{yy,j}(K_2^2,K_3,
        I_{x,y})+T_{V,j}(K_2^2,I_{x,y})
\end{aligned}\right. 
.\qquad \label{eq:nonlin5}
\end{eqnarray}
The linear regime may be then defined as the range of initial 
oscillation amplitudes, i.e. of $2I_{x,y}$, such that 
the functions $F$, $T$ may be ignored along with the 
amplitude dependent detuning terms $h_{2200,j}$, 
$h_{1111,j}$ and $h_{0022,j}$. These are  defined as 
the summation of all octupolar-like sources from the 
beginning of the ring up to the BPM $j$, in the same 
way $h_{1100,j}$ and $h_{0011,j}$ were defined in 
Eq.~\eqref{phase-shift-5} and are non-zero even if 
the global detuning with amplitude is zero or negligible. 
The generalization of observable BPM phase advance of 
Eq.~\eqref{eq:BetaPhase1} eventually reads
\begin{eqnarray}
\begin{aligned}
\Delta\Phi_{H,ij}&=\Delta\phi_{x,ij}^{(mod)}+\hbox{arg}\{B_{x,i}-B_{x,j}\} 
                -2h_{1100,ij}-4h_{2200,ij}(2I_x)-2h_{1111,ij}(2I_y)\\
\Delta\Phi_{V,ij}&=\Delta\phi_{y,ij}^{(mod)}+\hbox{arg}\{B_{y,i}-B_{y,j}\} 
                -2h_{0011,ij}-2h_{1111,ij}(2I_x)-4h_{0022,ij}(2I_y)
\end{aligned}
\qquad,\qquad \label{eq:nonlin6}
\end{eqnarray}
where the the subscript $ij$ in the detuning terms 
$h_{pqrt,ij}$ means that only the summation of 
detuning sources between the two BPMs $i$ and $j$ is 
to be taken into account, as in Eq.~\eqref{phase-shift-9} 
for the linear case. 

In conclusion, if the initial oscillation amplitude $(2I)$ 
is {\sl too large}, the betatron BPM phase advance $\Delta\phi_{ij}$ 
is no longer measurable from the difference of the tune line 
phases $\Delta\Phi_{ij}$, since amplitude dependent focusing, 
octupolar-like, resonant and detuning terms {\sl corrupt} 
the tune line. The same is true for the invariant itself 
$(2I)$, which is no longer measurable from the tune line 
amplitude. The latter is no longer constant along the ring and 
its modulation depends on the invariant itself via the 
functions $F$ and $T$ of Eq.~\eqref{eq:nonlin5}.

\section{Impact of octupolar-like terms on the coupling lines:  an amplitude 
         dependent coupling}
\label{app:2B}
Octupolar-like RDTs do not contribute to the tunes lines only, 
but to several other harmonics of the complex TBT signals, including 
the coupling lines. By applying the same procedure presented 
in Appendix~\ref{app:2A}, the contributions to the harmonic 
$H_h(0,\pm1)$ ($V_h(\pm1,0)$) are all terms in the summation 
of Eq.~\eqref{e:TbTComplex} such that $1-p+q=0$ and $t-r=\pm1$ ($q-p=\pm1$ 
and $1-r+t=0$). The result is
\begin{eqnarray}
&&\left\{\begin{aligned}
H_h(0,\ 1)&=-2i\left[f_{1001}+ 2(2I_x)f_{2101}+(2I_y)f_{1012}
                +O(f^2,I^2)\right]\sqrt{2I_y}e^{\tau_y}\\
H_h(0,-1)&=-2i\left[f_{1010}+ 2(2I_x)f_{2110}+(2I_y)f_{1021}
                +O(f^2,I^2)\right]\sqrt{2I_y}e^{-\tau_y}
\end{aligned}\right. \ , \nonumber\\ \label{eq:NLcoup1} \\\nonumber
&&\left\{\begin{aligned}
V_h(\ 1,0)&=-2i\left[f_{1001}^*+ (2I_x)f_{2101}^*+2(2I_y)f_{1012}^*
                +O(f^2,I^2)\right]\sqrt{2I_x}e^{\tau_x}\\
V_h(-1,0)&=-2i\left[f_{1010}+ (2I_x)f_{2110}+2(2I_y)f_{1021}
                +O(f^2,I^2)\right]\sqrt{2I_x}e^{-\tau_x}
\end{aligned}\right. 
\quad .\quad 
\end{eqnarray}
The amplitude dependent coupling is then generated by skew 
octupolar-like RDTs, with $f_{2101}$ and $f_{2110}$ excited by the 
$x^3y$ potential term, while $f_{1012}$ and $f_{1021}$ originate 
from the $xy^3$ monomial. Note that in the above equations the 
relation $f_{pqrt}=f_{qptr}^*$ is used here for simplicity, though 
it is not strictly true when second order terms are to 
be taken into account: See Appendix A of Ref.~\cite{Andrea-arxiv}.
The coupling lines of the real signals  $\tilde{x}$ and 
$\tilde{y}$, then read
\begin{eqnarray}
\left\{\begin{aligned}
H(0,1)=\frac{1}{2}[H_h(0,1)+H_h^*(0,-1)]&=-i\left[F_{xy}(J_1)+
       T_{xy}(J_3,K_3,K_2^2,J_1,I_{x,y})\right]\sqrt{2I_y}e^{\tau_y}\\
V(1,0)=\frac{1}{2}[\ V_h(1,0)+\ V_h^*(-1,0)]&=-i\left[F_{yx}(J_1)+
       T_{yx}(J_3,K_3,K_2^2,J_1,I_{x,y})\right]\sqrt{2I_x}e^{\tau_x}\\
F_{xy}=f_{1001}  -f_{1010}^*\quad , \qquad &
T_{xy}=2(f_{2101}-f_{2110}^*)(2I_x) + (f_{1012}-f_{1021}^*)(2I_y) \\
F_{yx}=f_{1001}^* -f_{1010}^* \quad , \qquad &
T_{yx}= (f_{2101}^*-f_{2110}^*)(2I_x) +2(f_{1012}^*-f_{1021}^*)(2I_y) 
\end{aligned}\right. 
,\qquad \label{eq:NLcoup2}
\end{eqnarray}
where the remainders $O(f^2,I^2)$ have been ignored. $F_{xy}$ and 
$F_{yx}$ are the same of Eq.~\eqref{e:F_xy} and generate betatron 
coupling, which is amplitude independent. $T_{xy}$ and 
$T_{yx}$ are instead responsible for the amplitude dependent 
coupling and can be excited by several sources. To the first order 
they are generated by skew octupole fields $J_3$, and by 
cross terms $\propto K_3\otimes J_1$ (i.e. between normal octupole 
and skew quadrupole) to the second order. $J_3$ in turn can stem 
from the cross product $\propto K_2\otimes J_2$ (i.e. between normal 
and skew sextupole), with $J_2$ originating from another 
cross term $\propto K_2\otimes J_1$. $K_3$ is also created by a last 
cross term $\propto K_2\otimes K_2$. In summary, the following scaling 
laws may be drafted:
\begin{eqnarray}
J_3&&\ \leftrightarrow\ (K_3)\otimes J_1\ ,\ K_2\otimes [J_2]\ 
   \ \leftrightarrow\ (K_2\otimes K_2)\otimes J_1\ ,\ K_2\otimes [K_2\otimes J_1]
   \ \leftrightarrow\ K_2\otimes K_2\otimes J_1 \ ,\\
K_3\otimes J_1&&\ \leftrightarrow\ K_2\otimes K_2\otimes J_1
\ .\qquad \label{eq:NLcoup3}
\end{eqnarray}
Hence, $T_{xy}$ and $T_{yx}$ are nonzero even in the absence of physical 
normal or skew octupoles and scale quadratically with the sextupole 
strength $K_2$ and linearly with the skew quadrupole field $J_1$. 
Since the betatron coupling terms $F_{xy}$ and $F_{yx}$ scale 
linearly with $J_1$, the overall amplitude dependent modulation 
of the coupling lines scales quadratically with the sextupole fields, 
i.e. with the same order of magnitude of the tune line modulation of 
Eq.~\eqref{eq:nonlin5}.

Amplitude and phase of the coupling lines at a generic BPM $j$ are 
eventually derived from Eq.~\eqref{eq:NLcoup2}, resulting in 
\begin{eqnarray}
\left\{\begin{aligned}
|H(0,1)_j|&=\left|F_{xy,j}(J_1)+T_{xy,j}(J_3,K_3,K_2^2,J_1,I_{x,y})\right|\sqrt{2I_y}\\
|V(1,0)_j|&=\left|F_{yx,j}(J_1)+T_{yx,j}(J_3,K_3,K_2^2,J_1,I_{x,y})\right|\sqrt{2I_x}\\
\hbox{arg}\left\{H(0,1)_j\right\}&=\phi_{x,j}+\psi_{x0}+
                  \hbox{arg}\left\{F_{xy,j}+T_{xy,j}\right\}-\frac{\pi}{2}\\
\hbox{arg}\left\{V(1,0)_j\right\}&=\phi_{y,j}+\psi_{y0}+
                  \hbox{arg}\left\{F_{yx,j}+T_{yx,j}\right\}-\frac{\pi}{2}
\end{aligned}\right. 
.\qquad \label{eq:NLcoup4}
\end{eqnarray}

\end{widetext}




\end{document}